\documentclass{article}
\pdfoutput=1
\usepackage{amsmath,amssymb,graphicx,microtype,hyperref}

\usepackage{framed}
\def\be{\begin{eqnarray}}
\def\ee{\end{eqnarray}}
\def\nn{\nonumber}

\def\p{\partial}

\def\K{K}
\def\l[{\phantom.[}

\def\ttau{p}
\def\ta{\tau}

\def\aa{\alpha}

\hoffset= - 1.0in         

\begin{document}

\title{\vspace{.1cm}{\Large {\bf Explicit examples of DIM constraints
for network matrix models  }\vspace{.2cm}}
\author{
{\bf Hidetoshi Awata$^a$}\footnote{awata@math.nagoya-u.ac.jp},
\ {\bf Hiroaki Kanno$^{a,b}$}\footnote{kanno@math.nagoya-u.ac.jp},
\ {\bf Takuya Matsumoto$^a$}\footnote{takuya.matsumoto@math.nagoya-u.ac.jp},
\ {\bf Andrei Mironov$^{c,d,e,f}$}\footnote{mironov@lpi.ru; mironov@itep.ru},\\
\ {\bf Alexei Morozov$^{d,e,f}$}\thanks{morozov@itep.ru},
\ {\bf Andrey Morozov$^{e,f,g}$}\footnote{andrey.morozov@itep.ru},
\ {\bf Yusuke Ohkubo$^a$}\footnote{m12010t@math.nagoya-u.ac.jp}
\ \ and \ {\bf Yegor Zenkevich$^{d,f,h}$}\thanks{yegor.zenkevich@gmail.com}}
\date{ }
}

\maketitle

\vspace{-6cm}

\begin{center}
\hfill FIAN/TD-09/16\\
\hfill IITP/TH-06/16\\
\hfill ITEP/TH-08/16\\
\hfill INR-TH-2016-011
\end{center}

\vspace{4.3cm}

\begin{center}
$^a$ {\small {\it Graduate School of Mathematics, Nagoya University,
Nagoya, 464-8602, Japan}}\\
$^b$ {\small {\it KMI, Nagoya University,
Nagoya, 464-8602, Japan}}\\
$^c$ {\small {\it Lebedev Physics Institute, Moscow 119991, Russia}}\\
$^d$ {\small {\it ITEP, Moscow 117218, Russia}}\\
$^e$ {\small {\it Institute for Information Transmission Problems, Moscow 127994, Russia}}\\
$^f$ {\small {\it National Research Nuclear University MEPhI, Moscow 115409, Russia }}\\
$^g$ {\small {\it Laboratory of Quantum Topology, Chelyabinsk State University, Chelyabinsk 454001, Russia }}\\
$^h$ {\small {\it Institute of Nuclear Research, Moscow 117312, Russia }}
\end{center}

\vspace{.5cm}

\begin{abstract}
Dotsenko-Fateev and Chern-Simons matrix models,
which describe  Nekrasov functions for SYM theories in different dimensions,
are all incorporated into network matrix models
with the hidden Ding-Iohara-Miki (DIM) symmetry.
This lifting is especially simple for what we call
{\it balanced} networks.
Then, the Ward identities (known under the names
of Virasoro/${\cal W}$-constraints or loop equations
or regularity condition for $qq$-characters)
are also promoted to the DIM level,
where they all become corollaries of a single identity.
\end{abstract}

\vspace{.5cm}


\section{Introduction }

Nekrasov functions, describing instanton corrections in supersymmetric Yang-Mills theories
\cite{SWfirst}-\cite{Nek},
and AGT related conformal blocks \cite{BPZ,CFT} possess rich symmetries that can be separated into
large and infinitesimal.
The former describe dualities between different models, while the latter
define equations on the partition functions in each particular case.
They are also known as "Virasoro constraints" \cite{Vircon} for associated conformal or Dotsenko-Fateev (DF)
matrix models \cite{AGTmamo,MMSha,AGTmamo1}, which are further promoted to network matrix models \cite{MZ,MMZ}, looking like convolutions of refined topological vertices \cite{IKV,AK} and possessing direct topological string interpretation.

As conjectured in a number of papers throughout recent years \cite{Nak}-\cite{Mat1}
and recently summarized in \cite{MMZdim},
in full generality the symmetry underlying the AGT correspondence \cite{AGT}, is the
Ding-Iohara-Miki algebra (DIM) \cite{DI}-\cite{Rmat}, in particular, the infinitesimal Ward identities are controlled by DIM
from which the (deformed) Virasoro and ${\cal W}_{\K}$
emerge as subalgebras in particular representations.
In other words, the full symmetry of the Seiberg-Witten theory
seems to be the {\it Pagoda} triple-affine elliptic DIM algebra (not yet fully studied and
even defined), and particular models (brane patterns or Calabi-Yau toric varieties labeled by integrable systems {\it a la} \cite{GKMMM})
are associated with its particular representations.
The ordinary DF matrix models arise when one specifies "vertical" and "horizontal" directions, then convolutions of topological vertices can be split into vertex operators and screening charges, and the DIM algebra constraints can be attributed in the usual
way \cite{confMAMO1,confMAMO2,UFN23} to commutativity
of screening charges with the action of the algebra in the given representation.
Dualities are associated with the change of the vertical/horizontal splitting, or, more general, with the
choice of the section, where the algebra acts \cite{triality}.
\newpage

\begin{figure}[h!]
  \centering
  \includegraphics[width=0.39\textwidth]{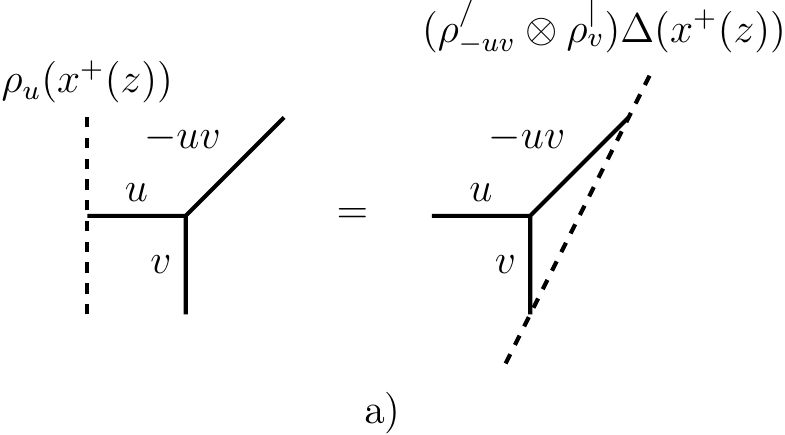}   \includegraphics[width=0.6\textwidth]{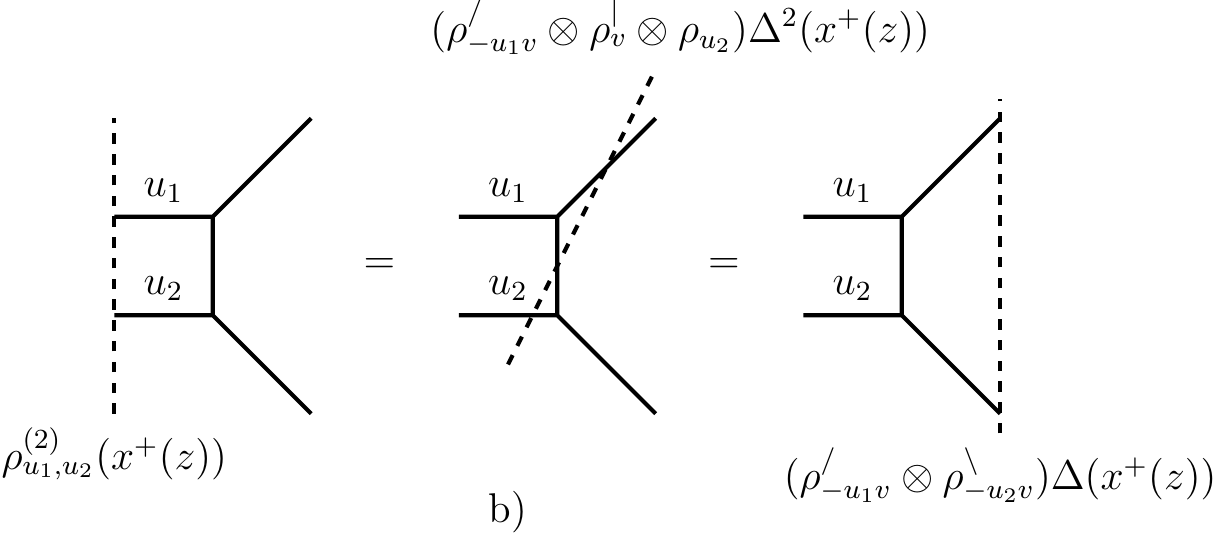}
  \caption{Topological vertex as the intertwiner of DIM
    representations. a) The action of the generator $x^{+}(z)$ on the
    level one Fock representation $\rho_u$ sitting on the horizontal
    leg of the topological vertex (denoted by the dashed line) is the
    same as its action on the product of two representations --- the
    ``vertical'' $\rho_v^{|}$ and ``diagonal'' $\rho_{-uv}^{/}$. b)
    Appropriate contraction of two intertwiners is also an
    intertwiner. This gives the vertex operator of the corresponding
    conformal field theory with deformed Virasoro symmetry,
    corresponding to a single vertical brane in Fig.~\ref{fig:1}.}
  \label{fig:2}
\end{figure}

\begin{figure}[h!]
  \centering
    \includegraphics[width=10cm]{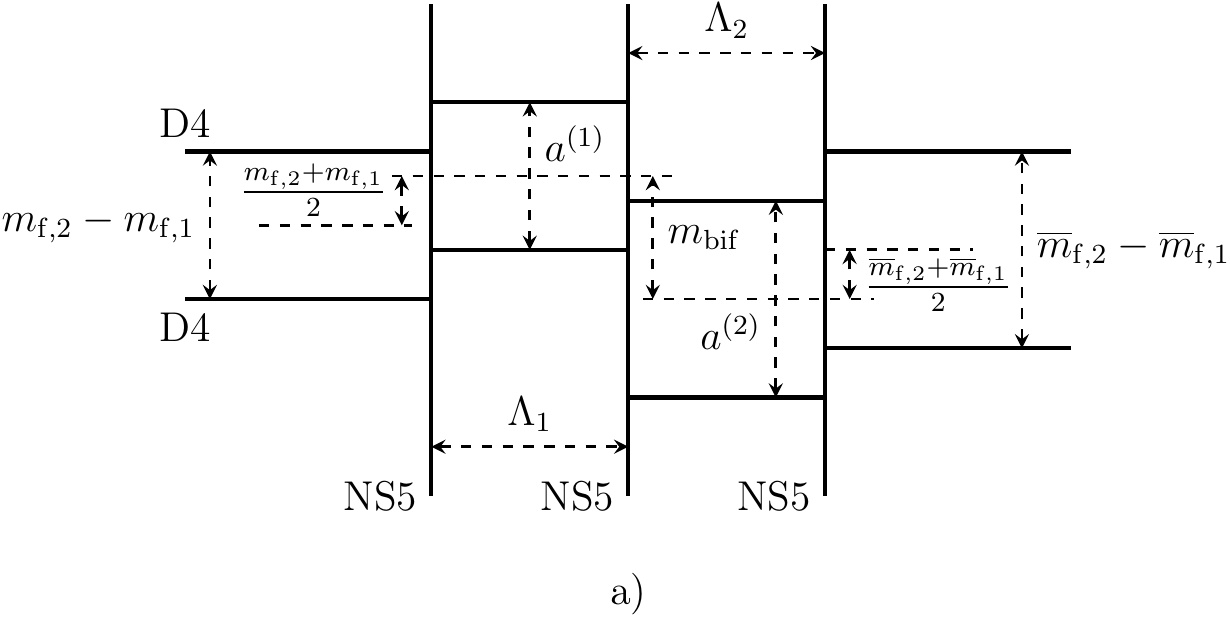}\\
  \includegraphics[width=8cm]{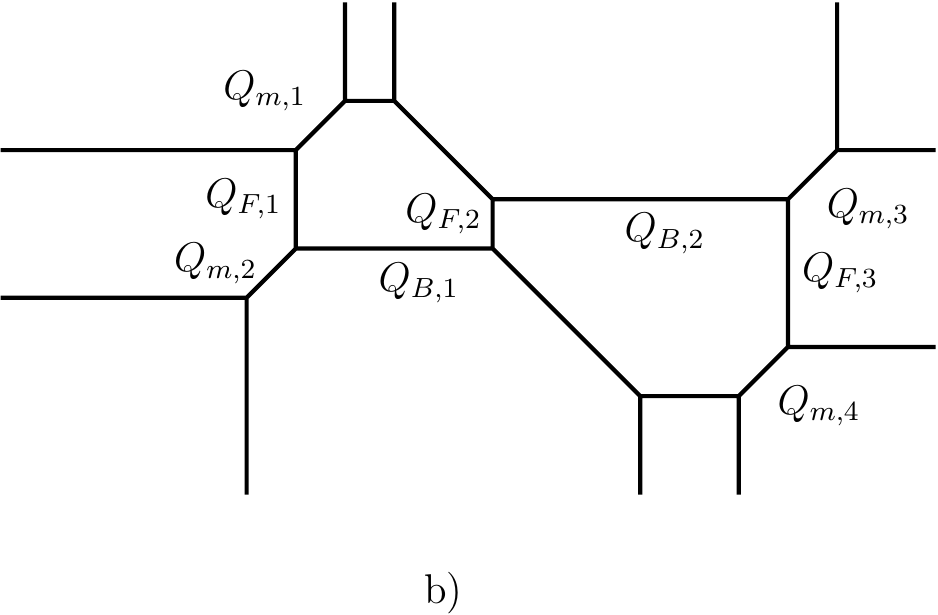} \qquad \includegraphics[width=5cm]{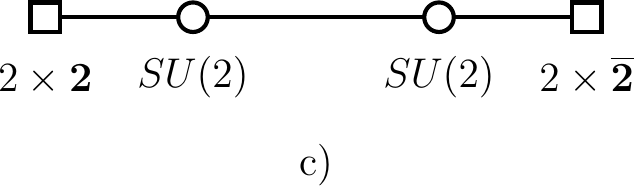}
  \caption{\footnotesize{a) Type IIA brane diagram consisting of
    two horizontal and three vertical intersecting lines representing
    NS5 and D4 branes. The low energy theory in this background is
    4d $\mathcal{N}=2$ gauge theory with $SU(2)^2$ gauge
    group. $\Lambda_i$ are exponentiated complexified gauge couplings,
    $a^{(a)}$ are Coulomb moduli and $m_a$ are the hypermultiplet
    masses. b) The toric diagram of the Calabi-Yau threefold,
    corresponding to the $5d$ gauge theory with the same matter
    content. Edges represent two-cycles with complexified K\"ahler
    parameters $Q_i$, which play the same role as the distances
    between the branes in a). c) The quiver encoding the matter
    content of the gauge theory. $SU(2)$ gauge groups live on each
    node and bifundamental matter on each edge. The squares represent
    pairs of (anti)fundamental matter hypermultiplets.}}
  \label{fig:1}
\end{figure}

All this is illustrated in pictures 1 and 2, which we borrowed from \cite{MMZdim},
and our purpose in this paper is to provide very explicit examples of how these pictures
are converted into formulas.
A great deal of these formulas already appeared in the literature.
Putting them together, we hope to illustrate their general origin and better formulate
the remaining open problems.

The main {\bf scheme} could be formulated as follows:

\begin{itemize}
\item  To build a {\it functor}
\be\label{functor}
\text{rank-}r \text{ Lie algebra } G \ \longrightarrow\
&\text{ quantized double-center  double-loop DIM}(G), \\
& \text{perhaps, } q_{123\ldots}\text{-dependent and elliptic}
\nn
\ee

\item To obtain a non-linear Sugawara construction of stress tensor and other symmetry generators
from a comultiplication $\Delta_{DIM}$.

\item To clarify the interplay between two "orthogonal" ("horizontal" and "vertical")
comultipilcations.

\item To apply the functor (\ref{functor}) to central-extended loop algebras $G$,
starting from $G=\widehat{(gl(1)}$,
to obtain triple-affine Pagoda DIM algebras.
One of the immediate problems is that the known construction of DIM($G$) for non-affine $gl_N$ algebras
\cite{DI,F} already involves the {\it affine} Dynkin diagrams,
thus, for an affine $\widehat{gl_N}$ one can need something more sophisticated.

\item An additional light on the problem can be shed by comparative analysis of
DIM$(gl_2)$, DIM($gl_3$), DIM($so_5$), DIM$(g_2)$ and DIM$(\widehat{gl_1})$,
first four of them being explicitly constructed,
and by studying their various limits including the one to the affine Yangian and further to the standard conformal algebras (coset constructions of conformal field theories, \cite{coset}).
\end{itemize}

Actually, the first three issues are actively studied by various authors (and there has been already achieved a serious progress), and we do not achieve too much in the two last challenging directions in the present paper,
which can be considered as an introduction to the problem.
What we actually do, is search for
a $q,t$-deformed network analogue of the CFT Ward identity
\cite{BPZ}
\be
 \left< \prod_a \hat V_{\alpha_a}(z_a) \cdot \hat {\cal T}_+(z) \cdot \hat {\cal Q}^r \right> \ = 0
\label{CFTwid}
\ee
where $<...>$ denotes the matrix element $< vac|...|vac>$ between two vacua of operators in the fixed chronological order and in the chiral sector \cite{TK}. Here $V_\alpha(z)$ is a primary field (vertex operator) in the free field $c=1$ CFT, ${\cal T}(z)$ is its stress-energy tensor and ${\cal Q}$ is the corresponding screening charge \cite{confMAMO1,confMAMO2}, which is the integral ${\cal Q}=\oint_x {\cal S}(x)$ of the screening current ${\cal S}(x)$.

The order of operators in (\ref{CFTwid}) means that in the conformal correlator
\be
\left<\!\left< \prod_a V_{\alpha_a}(z_a) {\cal T}_+(z) {\cal Q}^r \right>\!\right>
\ee
(where $<<...>>$ denotes the chiral part of the CFT correlator) all $|z_a|>|z|$ and $|z|>|x_i|$, where $x_i$'s lies on the integration contours of the screening currents.

The Ward identity (\ref{CFTwid}) can be manifestly written as
\be
z^2
\oint_{x_i} \left(\underline{\sum_{a,b}{1\over 4}\frac{\alpha_a\alpha_b}{(z-z_a)(z-z_b)}} +\sum_{a,i}\frac{\alpha_a}{(z-z_a)(z-x_i)}
+ \sum_{i,j=1}^r \frac{1}{(z-x_i)(z-x_j)}\right)
\left<\!\left< \prod_a V_{\alpha_a}(z_a)  \prod_{i=1}^r {\cal S}(x_i) \right>\!\right>\  =
\text{Pol}(z)
\label{WI3p}
\ee
\noindent
and the notation Pol($z$) means a power series, i.e. any positive powers of $z$ are allowed. The underlined terms just contribute to Pol($z$) (since $|z_a|>|z|$) and can be omitted giving finally
\be\boxed{
z^2
\oint_{x_i} \left(\sum_{a,i}\frac{\alpha_a}{(z-z_a)(z-x_i)}
+ \sum_{i,j=1}^r \frac{1}{(z-x_i)(z-x_j)}\right)
\left<\!\left< \prod_a V_{\alpha_a}(z_a)  \prod_{i=1}^r {\cal S}(x_i) \right>\!\right>\  =
\text{Pol}(z)}
\label{WI3}
\ee

\bigskip

Though equivalent, (\ref{CFTwid}) and (\ref{WI3}) are in fact very different.
The second one is about field theory {\it correlators},
it is dictated by operator expansions and is especially simple
because a free field formalism is available for conformal theories.
The first one is actually about matrix elements, and the difference
is that it depends on the ordering of operators,
while correlators do not.
Another way to say this is that the projected stress tensor ${\cal T}_+(z)$
does not have a simple operator product expansion (OPE) with other operators, the projection is a non-local
operation and actually depends on the position:
if ${\cal T}_+(z)$ was placed to the left of vertex operators $V(z_a)$,
the matrix element would no longer vanish. At the same time, in this case the underlined terms in (\ref{WI3p}) also contribute (since $|z|>|z_a|$), and they exactly cancel non-zero matrix element leading to the same Ward identity (\ref{WI3}).

\bigskip

These are trivial remarks for the old-fashioned field theory,
where the Ward identities were discovered and treated as sophisticated
recurrence relations between Feynman diagrams,
but in modern CFT we got used to the formalism based on the operator product expansion
and moving the integration contours, which provides a shortcut
for the derivations.
Unfortunately, in the network models,
only the operator approach  is currently available,
and this is the reason why we need to develop the formalism from this starting point.

\begin{figure}[h!]
  \centering
  \includegraphics[width=9cm]{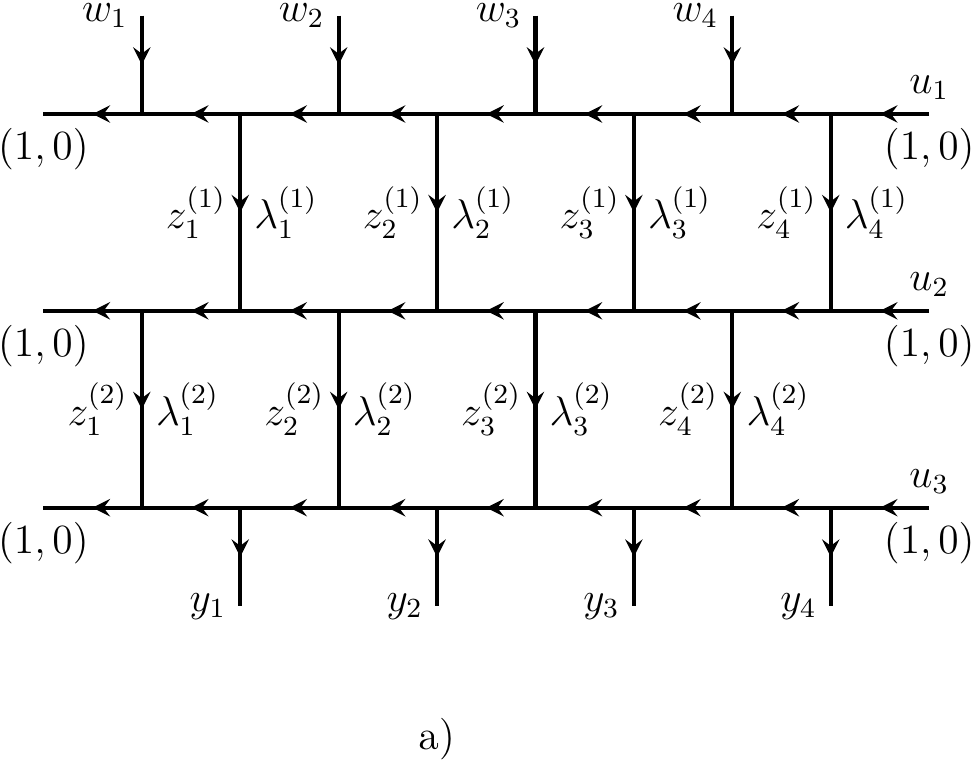}
  \\
  \vspace{0.5cm}
  \includegraphics[width=8.5cm]{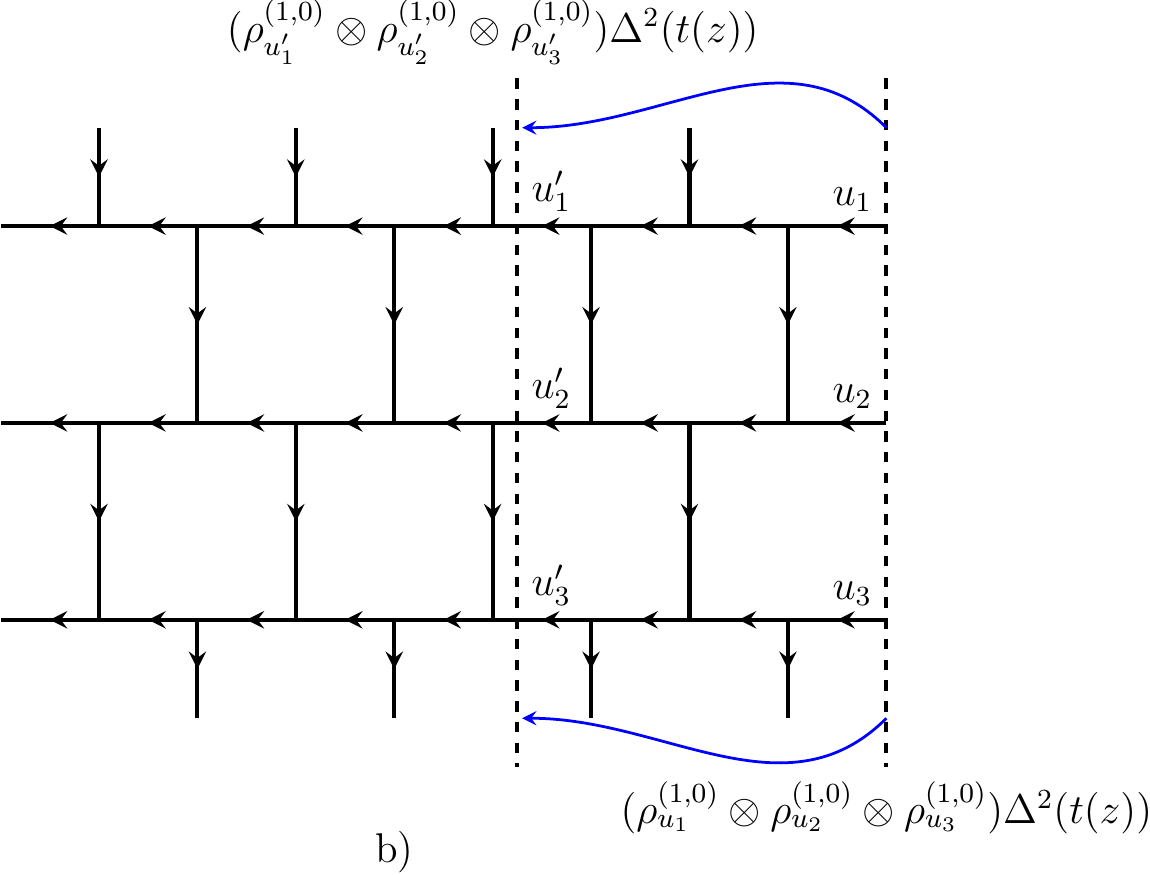}
  \hspace{-2cm}
    \includegraphics[width=10cm]{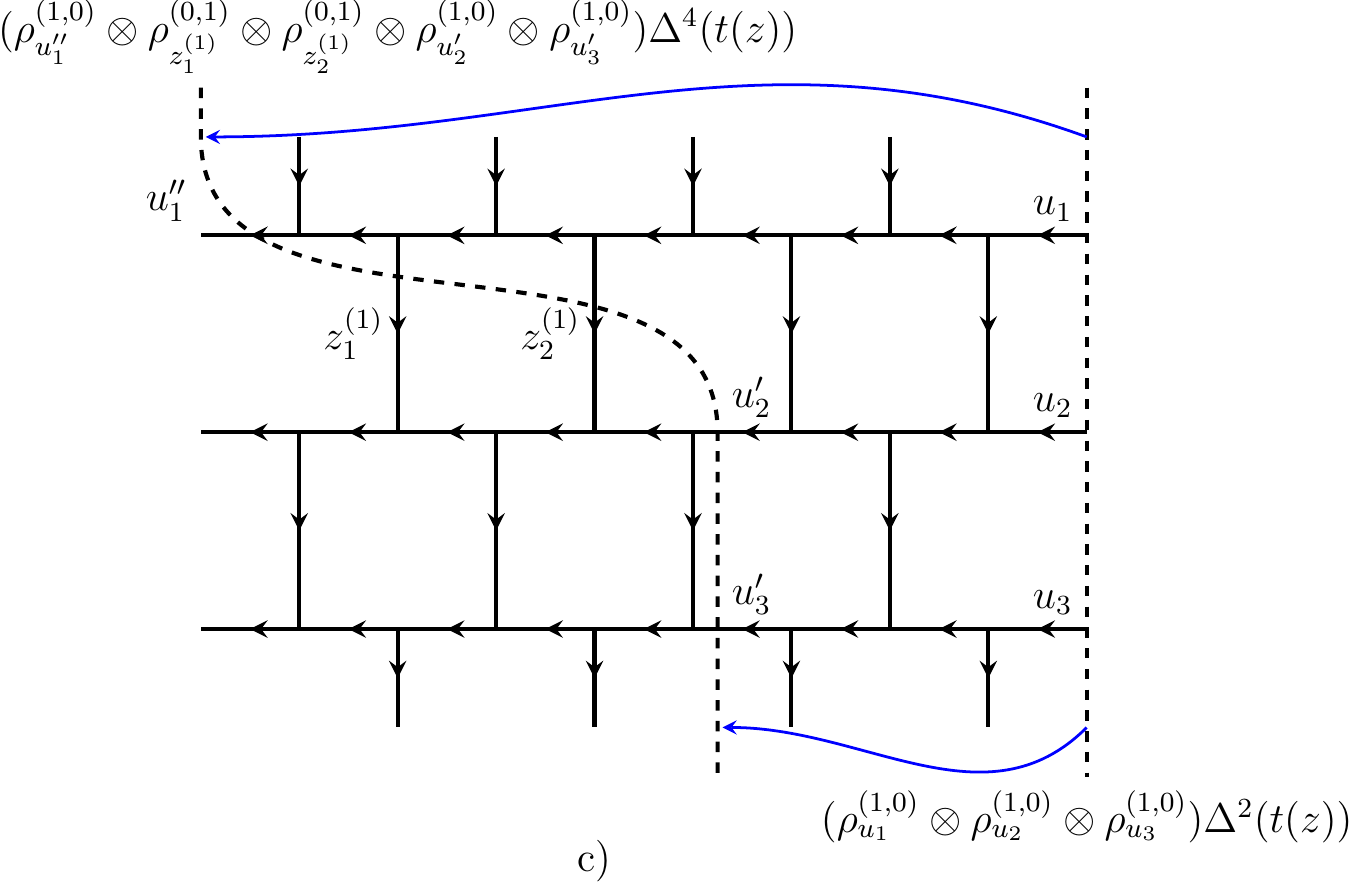}
    \caption{\footnotesize{a) An example of balanced network. Notice
        that the numbers of incoming and outgoing vertical branes are the same in each
        horizontal section. Because of this, the slopes of the
        horizontal branes have the same slopes $(1,0)$ to the left and
        to the right of the diagram. b) The action of a DIM algebra
        element on the section of the diagram.}}
  \label{bane}
\end{figure}

Still, some elements of the free field formalism are already worked out
in particular representations of DIM,
and for a special class of {\it balanced} network models, drawn as a set of horizontal lines with vertical segments
in between, see Fig.~\ref{bane},~a), one has a direct counterpart of (\ref{CFTwid}).
In (extremely) condensed notation it looks like
\be
\left< \prod_a{\Psi}_{\lambda_a}[z_a]\Psi^*_{\mu_a} [z_a^*] \  \
\hat{\cal T}_+(z;u|\xi) \ \ \prod_{b} \left(\sum_\mu {\hat\Psi}_{\mu}\hat\Psi^*_{\mu}\right)\right> \ = 0
\ee
and involves operators like
\be\label{7}
\prod_I \Psi_{\lambda_I}[z_I] \prod_J\Psi^*_{\mu_J}[z_J^*] \ \longrightarrow \
\prod_{I,J} \exp\left( \sum_{n\ne 0}\frac{1}{n}\left(\omega^{|n|}[\lambda_I,z_{I}]_n\mathfrak{a}_n -[\mu_J,z_{J}^*]_n\mathfrak{a}^*_n\right)\right)
\ee
where
\be
[\lambda,z]_n\equiv \hbox{sign}(n)\ \sum_i\left(q^{\lambda_i-1/2}t^{1/2-i}z\right)^n
\ee
are the Miwa variables associated with the Young diagram $\lambda$,
and the Drinfeld-Sokolov operator (generalized stress energy tensor = Miura transformation from $\Lambda_i(z)$)
\be
\hat{\cal T}(z;u|\xi) = z^{1/2\log_{\omega}\xi}\ :\prod_{i=1}^{\K}\Big( \omega^{-2z\p_z}-u_i\Lambda_{i}(z\omega^{2(i-1)})\Big):\ z^{-1/2\log_{\omega}\xi}=\sum_{k=1}^{\K}  \xi^{{\K}-k}\sum_{i_1<...<i_M} \prod_{a=1}^k u_{i_a} :\Lambda_{i_a}(z\omega^{2(a-1)}):
\ee
defining numerous flows, is a linear combination of all ${\cal W}^{(m)}$ with $m\leq {\K}$.
Here $\Lambda_i(z)$ are also
made from the annihilation and creation operators $\hat\aa_{\pm n}$, $\omega=\sqrt{q/t}$ and $\hat{\cal T}(z;u|\xi)$ depends on an additional parameter $\xi$ generating different ${\cal W}^{(m)}(z)$ and on spectral parameters of DIM representation $u_i$.

A counterpart of (\ref{CFTwid}) emerges when the dashed vertical
section in Fig.\ref{bane},~b) is shifted to the left, through {\it
  external} vertical legs, which do {\it not} commute with $\hat{\cal
  T}(z)$.  Moreover, now we can also consider deformations of the
section which do not preserve verticality, like the dotted one in
Fig.~\ref{bane},~c), and everything can still be calculated.  This
should provide a qualitatively new insight into spectral dualities
\cite{specdu} associated with global rotations of the network graph.

\begin{figure}[h!]
  \centering
\includegraphics[width=5cm]{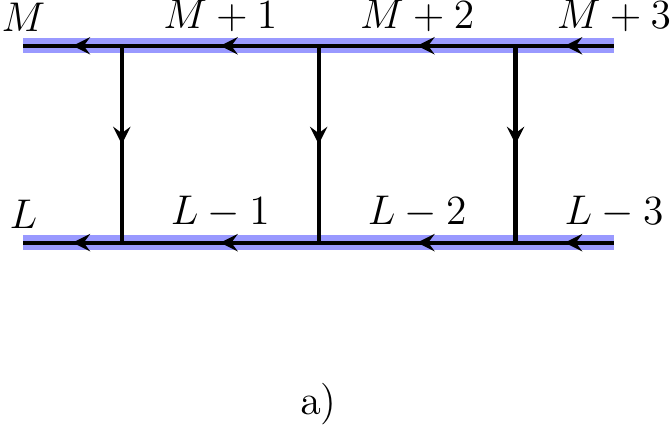} \hspace{3cm} \includegraphics[width=2cm]{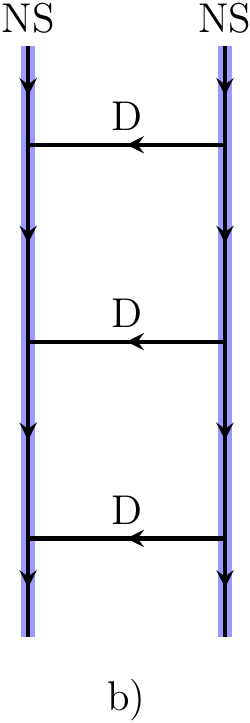}
\caption{\footnotesize{a) An example of non-balanced web with infinite
    ``horizontal'' lines shown in blue. Bending of the ``horizontal''
    lines due to tension from the vertical segments is reflected in
    their slopes marked above them. b) Spectral duality acts by
    rotating the diagram a). After rotation one can identify the
    conventional Hanany-Witten (or brane web/geometric engineering)
    setup with NS5 and D5-branes ().}}
\label{fig:hor}
\end{figure}

Non-balanced networks, where the right-most and left-most branes
in Fig.~\ref{fig:hor} are tilted and the number of operators $\Psi$ differs
from that of $\Psi^*$, can be considered as certain limits of the balanced ones,
but these limits are non-trivial and singular when, say, $q,t\longrightarrow 1$.
From the point of view of representation theory these limits should have
independent description, making use of more complicated intertwiners.
A full-fledged free field description for them comparable to the one in \cite{GMMOS}
for ordinary affine case still needs to be worked out.

Restriction to the balanced networks is a great technical simplification,
but it requires a somewhat lengthy comments on what this means and
whether this really restricts the set of handy physical models.

DIM is a quantization of double loop (double affine) algebras, and the
existing free field formalism, which we are going to expose and
exploit in the present paper, explicitly breaks the symmetry between
the two loops.  Bosonized/fermionized are only the Chevalley
generators, in the case of DIM there are many, still they depend on
one of the two loop parameters, while the other loop is associated
with their multiple commutators and is described very differently: in
terms of Young diagrams parameterizing states in the Fock space. This
breaks the symmetry of the DIM algebra: the
$SL(2,\mathbb{Z})$-automorphisms acting on the square lattice of the
generators and introduces asymmetry between horizontal and vertical
directions in the planar graphs which are used to define the network
models, and makes the spectral dualities interchanging these two
directions highly non-trivial.  In particular, allowed networks look
like infinite "horizontal" lines, connected by vertical segments, see
Fig.~\ref{fig:hor},~a), and not vice versa.  We call these lines {\it
  horizontal}, though they can have varying slopes, however, they have
a non-trivial projection on the horizontal axis, i.e. are strictly
non-vertical.  In the original brane theory interpretation these
horizontal lines depict the $D$-branes, while vertical are the $NS$
branes, from this point of view our description applies only to the
conformal models ($N_f=2N_c$) with definite $N_c=M=\#$ of horizontal
lines.  Quiver models $\otimes SU(N_i)$ with different $N_i$ can seem
excluded, but in fact they appear after application of the spectral
duality: a $90^\circ$ rotation of the graph, see
Fig,~\ref{fig:hor},~b).  After this rotation, the infinite horizontal
lines get associated with the infinite $NS$ branes, while the vertical
segments with $D$-branes between them.  This pattern looks more
relevant from the gauge theory point of view, but we emphasize that
our free fields live on the infinite horizontal lines, the
three-valent vertices (the DIM algebra intertwiners $\Psi$ and $\Psi^*$, also
known as topological vertices) act as operators in the Fock spaces
horizontally, while the third vertical edge carries a Young-diagram
label, not converted into operator language.  In result these vertices
can look like $\bot$ or $\top$, but not like $\vdash$ or $\dashv$.

All these restrictions can be lifted by switching from Fock to
MacMahon modules, which are representations of DIM spanned by 3d
partitions, but such a description is only combinatorial so far, no
generalization to the full-fledged double-loop free field formalism is
available yet.  This is what makes tedious the consideration of dotted
sections in Fig.~\ref{bane},~c).  We briefly touch this issue at the
very end of this text, but detailed presentation is postponed to the
future work.  Our main purpose here is to describe the powerful free
field formalism for the {\it balanced} network as a straightforward
generalization of that for the ordinary conformal theories, and
explain how the DIM algebra becomes the symmetry of generic Nekrasov
functions generalizing the Virasoro/W symmetry of the ordinary
conformal blocks and Dotsenko-Fateev matrix models.

In the next section 2, we explain how the elementary theory of a harmonic oscillator
can be straightforwardly developed and lifted to description of
generic networks, i.e. of generic Nekrasov functions.
In section 3, in the simplest examples we demonstrate the actual formalism in full detail.
It is important that most complications come from sophisticated notation,
which are largely no more than a change of variables (normalization of creation
and annihilation operators).
The really big change comes in section 4, when one looks at the {\it symmetry}: it is indeed
essentially deformed.
But this deformation actually simplifies things, reducing all the symmetries to the action
of the DIM generators, while the Sugawara construction of Virasoro and W-operators
and of their sophisticated $q$-deformations is no more than the simple comultiplication rule.
At last, at section 5 we briefly discuss the spectral duality
action on symmetry generators.
Finally, the Appendix contains further details about various DIM algebras and
their representations.
At present stage of development, different parameters are treated as providing
different algebras, but further studies can promote them to parameters of different
representations of a single unified algebra (like the triple-affine elliptic Pagoda DIM algebra
anticipated in \cite{MMZdim}).

\paragraph{Notation.} Throughout the text we use the notation
\be
\boxed{\omega\equiv\sqrt{q\over t}}
\ee

\section{Basic example: Theme with variations}

We assume some familiarity with \cite{MMZdim} and do not repeat the general logic, leading to
Ward identities like (\ref{CFTwid}) in DF and network matrix models.

\setcounter{subsection}{-1}

\subsection{The main theme
\label{mathe}}

Screening charge $\hat Q$, acting on the Fock space
${\cal F}_\alpha =\Big\{{\rm Pols}(\ta_n)\Big\}\cdot e^{\alpha T_0}$, is
$$
\hat Q = \oint \hat S(x)dx = \hbox{res}_{x=0}\ \hat S(x),
$$
\be
\hat S(x) = \ :e^{\sqrt{2}\hat \phi(x)}: \ =
\underbrace{\exp\left(\sum_{n>0} \frac{\ta_nx^n}{n}\right)}_{\sum_n x^{n}\chi_n\{\ta\}} e^{T_0} x^{2\p_0}
\exp \left(-\sum_{n>0} \frac{2}{x^n}\frac{\p}{\p \ta_n}\right)
\label{scree}
\ee
where $\chi_n\{\ta\}$ are the characters of symmetric representations $[n]$ of $sl$ algebras
(the Schur polynomials in this particular case).
Applied to a highest-weight state
(i.e. the one annihilated by all negative modes
$\hat a_{-n} = -\sqrt{2}n\frac{\p}{\p \ta_n}$)
with negative half-
    integer $\alpha$
\be
\Big|m+1\Big>\ = e^{-{1\over 2}(m+1) T_0}
\ee
it gives
\be
\hat Q \ \Big|m+1\Big>\  =   \chi_m\{\ta\}\, \Big|m-1\Big>\
\ee
Residue is non-vanishing, because $x^{2\p_0}$ converts $|m+1>$ into $x^{-m-1}$.
Similarly
\be
\hat Q^2\  \Big|m+2\Big>\  =  \chi_{[mm]}\{\ta\}\,\Big|m-2\Big>\
\ee
where the calculation involves
\be\label{Q2}
-{1\over 2}\sum_{m_1,m_2} \chi_{m_1}\chi_{m_2} \oint\oint dx_1dx_2 x_1^{m_1-m-2}x_2^{m_2-m-2} (x_1-x_2)^2
= \chi_{m}^2-\chi_{m+1}\chi_{m-1} = \chi_{[m,m]}
\ee
and
\be
\boxed{
\hat Q^r\ \Big|m+r\Big>\ = \chi_{[m^r]}\{\ta\}\,\Big|m-r\Big>\
}
\label{Qronvac}
\ee
i.e. the power of $\hat Q$ acts as a character of rectangular Young diagram.
This is the old result by \cite{spCFT,mamoCFT1,mamoCFT2}.
The rectangular diagrams arise from the Cauchy formula
\be
\prod_{i=1}^r \exp\left(\sum_{n>0} \frac{\ta_nx_i^n}{n}\right)
=  \exp\left(\sum_{n>0} {1\over n}\ta_n \sum_{i=1}^r x_i^n\right) =
\sum_\lambda \chi_\lambda\{\ta\} \chi_\lambda[\vec x]
\ee
with a sum over {\it all} Young diagrams $\lambda$
(actually, with no more than $r$ lines)
after the Vandermonde projection
\be
\prod_{i=1}^r \oint \frac{dx_i}{x_i^{m+r}}\, \Delta(\vec x)^2\, \chi_\lambda[\vec x]
\sim  \delta_{\lambda,[m^r]}
\ee
which is a direct generalization of (\ref{Q2}).

\bigskip

Since the screening charge commutes
\be
\boxed{
\l[ \hat L_n, \hat Q ] = 0
}
\label{LQcomm}
\ee
with the Virasoro generators
\be
\hat L_n = \sum_{k} (k+n)\ta_k\frac{\p}{\p \ta_{k+n}}
+ \sum_{k=1}^{n-1}k(n-k)\frac{\p^2}{\p \ta_k\p \ta_{n-k}}
+ 2n\frac{\p^2}{\p \ta_n \p T_0}, \ \ \ \ n>0
\label{Virpos}
\ee
one has
\be
\hat L_n \hat Q^r\, \Big|m+r\Big>\ = \hat Q^r \hat L_n\,\Big|m+r\Big>\ = 0
\ \ \ {\rm for \ } n> 0
\ee
In application to (\ref{Qronvac}), this gives
\be
\boxed{
\hat L_n \chi_{[m^r]} = n(m-r)\frac{\p \chi_{[m^r]}}{\p \ta_n}
}
\ \ \ \ \  \ n>0
\label{Lchi}
\ee
while the action of
\be
\hat L_0 = \sum_{k} k\ta_k\frac{\p}{\p \ta_{k}} + \frac{\p^2}{\p T_0 \p T_0}
\label{Vir0}
\ee
gives just the size of the Young diagram:
\be
\hat L_0\chi_{[m^r]}=mr\cdot \chi_{[m^r]}
\label{L0chi}
\ee
In the Miwa parametrization $\ta_n = \sum_i X_i^n$,
this turns into the statement about the Calogero eigenfunctions.
Also $Q^r\Big|m+r\Big>$ are singular vectors in Verma modules and (\ref{Lchi})
can be considered as the simplest version of BPZ equations for correlators with
degenerate fields, \cite{BPZ}.

\bigskip

Equation (\ref{Lchi}) provides a simple example of the Ward identity for
the state $\hat Q^r|m+r>$, which can be promoted to identity for
the matrix element in conformal field theory, i.e. in the abstract Fock module and corresponding Sugawara energy-momentum tensor (which we denote by Gothic letters),
$\chi_{[m^r]} =<m-r|\hat C \hat {\cal Q}^r|m+r>$
by additional insertion of the intertwining operator, see below.
We are now ready to formulate the main theme of the present paper:

\bigskip

\begin{framed}
A trivial symmetry property (\ref{LQcomm}) gives rise to a non-trivial equation for the matrix element
(\ref{Lchi}), provided one can calculate (\ref{Qronvac}).
\end{framed}

\bigskip

\noindent
In what follows we extend this simple example to matrix elements
of an arbitrary network of intertwining operators,
what allows to reveal in a rather explicit form the hidden DIM symmetry of
the Seiberg-Witten/Nekrasov theory.

We continue in this section with {\it variations} on the main theme,
developing it at conceptual level.
Next sections will describe technical details of the story.

\subsection{Variation I: Matrix elements in the free-field theory}

Actually, in theory of free field $\phi(z)$, the bra vacuum state
is annihilated by all the negative mode operators $\hat a_{-n}=\tau_n/\sqrt{2}$,\ $n>0$,
i.e. contains $\prod_{n>0} \delta(a_{-n})$ in holomorphic representation.
Thus, one can not simply convert (\ref{Qronvac}) into a statement that
$\chi_{[m^r]}\{\ta\}$  is equal to $<m-r|\,Q^r\,|m+r>$:
this matrix element would not depend on $\ta$ at all. The way out is to introduce a special intertwining operator
\be
\hat C \{p\} = \exp\left(\sum_{n>0} \frac{p_n\hat a_{n}}{n}\right)
\ee
which converts the bra vacuum into the coherent state
\be
\langle m | \ \longrightarrow \ \langle m | \,\hat C\{p\}
\ee
with the property
\be
\langle m |\, \hat C\{p\}\, \hat a_{-n} = p_n \cdot \langle m | \, \hat C\{p\}
\ee
This allows us to rewrite (\ref{Qronvac}) as
\be
\chi_{[m^r]}\{\sqrt{2}p_n\} = \ \left<m-r\Big|\,\hat C\{p\}\, \hat {\cal Q}^r\,\Big|m+r\right>
\ee
Among many complications as compared with (\ref{Qronvac}), there is $\sqrt{2}$,
which reflects the fact that the character is extracted here from the screening charge
in a single field ("current") realization.
A more adequate kind of formulas arise within the fermionic realization
(see sec.3.2 of \cite{confMAMO2} and sec.\ref{VOP} below) which involves two scalar fields,
and $\sqrt{2}$ is a result of basis rotation to their symmetric combination.

\subsection{Variation II: Generating functions}

We can make from particular Virasoro generators $\hat L_n$ a single operator (stress tensor)
\be
\hat T(z) = \sum_{n\in \mathbb{Z}} \frac{\hat L_n}{z^{n+2}}
\ee
Positive and zero modes with $n\geq 0$ are given by (\ref{Virpos}) and (\ref{Vir0}) respectively,
negative modes are:
\be
\hat L_{-n} = \sum_k k\ta_{k+n}{\p\over\p \ta_k}+\ta_n{\p\over\p T_0}+{1\over 4}\sum_{k=1}^{n-1} \ta_k\ta_{n-k}
\ee
so that
\be
\l[\hat L_n,\hat L_m] = (n-m)\hat L_{n+m} + \frac{n(n^2-1)}{12}\delta_{n+m,0}
\ee
for the properly regularized sum $\ \sum_{n>0} n = -\frac{1}{12}$.

\bigskip

Symmetry (\ref{LQcomm}) actually holds for all $n\in Z$.

\bigskip

\noindent
We will also need a "current"
\be
\hat J(z) = \p_z\hat\phi(z)=\sum_{n\in Z} \frac{\hat J_n}{z^{n+1}} \nn \\
{\rm with} \ \ \ \
\hat J_{-n} = {\ta_n\over\sqrt{2}}, \ \ \ \hat J_0 = \sqrt{2}{\p\over\p T_0}, \ \ \ \hat J_{n} = \sqrt{2}n\frac{\p}{\p \ta_n}
\ee
and
\be
\l[\hat J_n,\hat J_m] = n\delta_{n+m,0}\nn \\
\l[\hat L_n, \hat J_m] = -mJ_{n+m}
\ee
The two operators are related by the Sugawara relation
\be
\hat T(z) = \ :\frac{1}{2}\hat J(z)^2:
\ee
where normal ordering puts all $p$-derivatives to the right of all $p$'s
(in each term of the formal series).

\noindent
The generating functions satisfy the commutation relations
\be
\l[\hat J(z),\hat J(w)] = \delta'(w/z)  \nn \\
\delta(x)=\sum_{n\in\mathbb{Z}}x^{n}
\ee

\bigskip

In terms of generating functions, the Ward identity (\ref{Lchi}), i.e.
the corollary of symmetry (\ref{LQcomm}) becomes
\be
\left[z^2\ \hat T(z) - {m-r\over\sqrt{2}}\ z\hat J(z)\right]_-\cdot\chi_{[m^r]} = 0
\ee
or, in other words, a regularity constraint
\be
\boxed{
\left(z^2\ \hat T(z) - {m-r\over\sqrt{2}}\ z\hat J(z)\right)\cdot \chi_{[m^r]} = {{\rm Pol}(z) }
}
\label{chiwid}
\ee
This will be the typical form of Ward identities (regularity condition for $qq$-characters)
for network Nekrasov functions $Z$ generalizing the simple character $\chi_{[m^r]}$.

\subsection{Variation III:  DF model
\label{DFmod}}

Expressions (\ref{Qronvac}) and (\ref{scree}) together imply the integral representation
of the matrix element
$$
\chi_{[m^r]}\{\ta\}
= \ \left<m-r\Big| \ \hat C\{\ta_n/\sqrt{2}\}\, \hat {\cal Q}^r \ \Big|m+r\right> \
= -{1\over r!}\underbrace{\oint\ldots\oint}_r
\left(\prod_{i=1}^r  \frac{\,{\cal G}\{\ta|x_i\}\, dx_i}{ x_i^{m+r}} \right)
\prod_{i<j} (x_i-x_j)^2  =
\ \Big<\  1\  \Big>_{\rm{DF}_{m,r}}
$$
\vspace{-0.5cm}
\be
{\cal G}\{\ta|x\} \ = \ \exp\left(\sum_{n=1}^\infty \frac{\ta_nx^n}{n}\right) \  =\
\sum_{n=0}^\infty x^{n}\chi_n\{\ta\}
\label{intDF}
\ee
which is the archetypical example of DF or conformal matrix model \cite{confMAMO1,confMAMO2,mamoCFT1,mamoCFT2}.

Ward identity (\ref{chiwid}), which is a trivial corollary of commutativity (\ref{LQcomm})
looks now like a not-so-obvious set of integral identities:
\be
\left(z^2\ \hat T(z) - {m-r\over\sqrt{2}}\ z\hat J(z)\right)\Big< \ 1 \ \Big>_{\rm{DF}_{m,r}} = \
\left< \sum_{k,i}{\tau_kx_i^{k+1}\over z-x_i}+\sum_{i,j=1}^r \frac{x_ix_j}{(z-x_i)(z-x_j)}-(m-r)\sum_i{x_i\over (z-x_i)} \right>_{\rm{DF}_{m,r}}
= {\rm Pol}(z)
\label{DFwid}
\ee
Actually there are two standard ways to derive the l.h.s.:

(1) by using bosonization, which is the simplest version of free-field (FF) formalism,
i.e. the Wick rule for decomposition of correlators into pair ones,
\be
\hat T(z)\Big< \ 1 \ \Big>_{\rm{DF}_{m,r}} = \
\left< \hat C(\tau_n/\sqrt{2})\cdot\underbrace{{\cal T}(z)}_{\frac{1}{2}:\p\varphi(z)^2:}
\cdot \prod_{i=1}^r \oint e^{\sqrt{2}\varphi(x_i)}dx_i \right>_{\rm{FF}_m}
\label{OPEwid}
\ee
where the index $m$ refers to a special way of handling the zero mode of $\varphi$ and $\varphi(z)$
refers to the scalar field acting in the abstract Fock module, \ and

(2) by a change of integration variables $\delta x_i = \epsilon x_i^{n+1}$
in the multiple integral (\ref{intDF}), \cite{Vircon,UFN23}:
in this case we get the identities in a slightly different form:
\be
\left<\sum\limits_{i<j}^r 2\frac{x_i^{n+1}-x_j^{n+1}}{x_i-x_j}+\sum_{i,k}\tau_k x_i^{k+n}+\sum\limits_i^r(n+1-m-r)x_i^n\right>_{\rm{DF}_{m,r}}=0,\ \ \ \ \ \ n>0
\label{widDFchava}
\ee
In this paper we actually need an outdated and tedious third way:

(3) the operator formalism based on an explicit calculation of commutators arising when the
stress tensor is carried from the left to the right through the screening operators:
this is what we are now doing, starting from sec.\ref{mathe} and this is what in the simplest case
brought us to the Ward identity in the form (\ref{DFwid}).

Both the OPE-based and change-of-integration-variables/total-derivative approaches should also
work in the network model context, but they still need to be developed.

\subsection{Variation IV: Multi-field case
\label{multifmod}}

The network matrix models can be considered as associated with networks of branes (brane-webs \cite{web5d}),
which being projected onto the $4-5$ plane look like segments with different slopes.
From the point of view of Yang-Mills theories, interpretation of the different slopes is different.
Surprisingly or not, it is also different at the present level of understanding of the DIM symmetry. Throughout the section, we distinguish only between the horizontal and vertical segments, while intermediate slopes appear in this section only in ss.\ref{net1} and \ref{net2}.
Our next variations introduce and describe the associated notions.

The first one is {\it horizontal branes}.
These are associated with different free fields.
Generalization of the DF model to ${\K}$-field case provides ${\cal W}_{{\K}}$ constraints
for models with ${\K}$ horizontal branes.
An additional procedure can be applied to separate a "center-of-mass" field: this explains
why in the previous subsection \ref{DFmod} the number of fields was one rather than two.

The multi-field conformal model \cite{confMAMO2} is defined as
\be
\ \left<\vec m-\vec r\left| \ \prod_{a=1}^{{\K}-1}
\hat C_a\{\ta_n^{(a)}/\sqrt{2}\}\, \hat {\cal Q}_a^{r_a} \ \right|\vec m+\vec r\right> \
=\left< 1 \right>_{{\rm DF}_{\vec m,\vec r}} \ee where the screening
charges now carry additional indices labeled by ${\K}-1$ simple
roots $\vec\alpha_a$ of $sl_{\K}$.  They are actually associated
with segments of the {\it vertical} branes ending on two adjacent
horizontal branes, Fig.~\ref{bane},~a), in accordance with the
decomposition $\vec\alpha_a = \vec e_{a+1}-\vec e_a$.  In other words,
a better labeling of ${\cal Q}$ is by pairs of indices $ab$, each
corresponding to a particular horizontal (in fact, any non-vertical,
see s.\ref{CSmod}) brane.\footnote{To avoid possible confusion, note
  that in \cite{MMZdim} an "orthogonal" labeling rule was used,
  treating horizontal edges of the network as segments between the
  vertical ones.}  Now the matrix model partition function depends on
${\K}$ sets of times, one of which is associated with the "center
of mass" and actually decouples in the DF model (\ref{intDF}), thus it
was actually suppressed in that formula.  However, this is not always
true: the decoupling will not take place already in the Chern-Simons
deformation of (\ref{intDF}) in sec.\ref{CSmod}, and all the ${\cal
  M}$ sets of times will be relevant in generic DIM considerations.
This phenomenon is familiar in the CFT approach to Nekrasov functions,
where relevant is the $Heis+Virasoro$ symmetry and its generalizations
rather than the $Virasoro$ alone.  This is also reflected in
appearance of $"1"$ in the popular notation ${\cal W}_{1+\infty}$.

\bigskip

Algebraically, the multi-field generalization is controlled by the comultiplication $\Delta_{DIM}$,
which builds all the symmetry generators from a single element
of DIM:
\be
\text{current algebra}\nn \\
\downarrow \nn \\
\text{Virasoro} \nn \\
\downarrow \nn \\
{\cal W}_3 \nn \\
\downarrow \nn \\
\ldots \nn \\
\downarrow \nn \\
{\cal W}_{\K} \nn \\
\downarrow \nn \\
\ldots
\ee
This comultiplication adds new scalar fields, and non-linearity of the usual $4d$ Sugawara formulas
is mostly due to elimination of the center-of-mass field; what makes this possible
is the exponential form of symmetry generators beyond $4d$.
Somewhat symbolically,
the Sugawara formulas for the stress tensor (at the second level of DIM) arise from the expansion of characters (in fact, $q$-characters)
\be
{\K}=2: & {\cal T}_{sl_2} = \frac{1}{2}\Big(e^J + \underline{e^{-J}}\Big)  = 1 + \frac{1}{2}J^2 + \ldots
\nn \\
{\K}=3: & {\cal T}_{sl_3} = \frac{1}{3}\Big(e^{J_1}+ e^{J_2-J_1} + \underline{e^{-J_2}}\Big) =
1 + \frac{1}{3}(J_1^2 - J_1J_2+ J_2^2) + \ldots \nn \\
\ldots \nn \\
{\K}: & {\cal T}_{sl_{\K}}= 1 + \frac{1}{{\K}!} \sum_{a,b=1}^{{\K}-1} C_{ab}J_aJ_b + \ldots \nn\\
\ldots \nn \\
{\K}=\infty: & {\cal T}_{sl_\infty} = 1 + {\rm const}\cdot \int(\nabla J)^2 + \ldots
\ee
underlined in the first two lines are terms appearing due to the center-of-mass reduction
\be
\sum_{a=1}^{\K} J_a = 0
\ee
$C_{ab}$ is the Cartan matrix for $sl_{\K}$, which the ${\K}=\infty$ limit describes a difference Laplace operator
$\nabla^2$.
Other ${\cal W}$-operators made from higher powers of $J$ arise in the same way
at higher levels of DIM, i.e. after several applications of the comultiplication $\Delta_{DIM}$,
e.g. at ${\K}=3$ the second generator of the $W_3$-algebra is
\be
{\cal W}^{(3)}_{sl_3}=\frac{1}{3}\Big(e^{-J_1}+ e^{J_1-J_2} + \underline{e^{J_2}}\Big)
\ee
so that the standard $W_3$-generator is a difference
\be
{\cal T}_{sl_3}-{\cal W}^{(3)}_{sl_3}\sim \frac{1}{3}J_1J_2(J_1-J_2)+ \ldots
\ee

\subsection{Variation V:  Chern-Simons (CS) model
\label{CSmod}}

The brane slopes show up in a specially designed $4d$ limit as additional square-logarithmic terms
$(\log x_i)^2$ in the action of the DF matrix model (\ref{intDF}), giving rise to what is often called
the CS matrix model \cite{CSmod,CSknots,BEMT,CSknotmore}:
\be
\Big<\  1 \  \Big>_{\rm{CS}_{r}} =
{1\over r!}\prod_{i=1}^r \oint dx_i{\cal G}\{\tau|x_i\}e^{\gamma (\log x_i)^2}  \prod_{i<j} (x_i-x_j)^2
\label{intCS}
\ee
The parameter $\gamma$ controls the brane slope, it vanishes for the horizontal branes,
while for the vertical ones it becomes infinite and the story gets a separate twist,
see s.\ref{VOP} below.

From the point of view of DIM symmetry of the network model, the
Virasoro/Ward constraints should look similar with and without these logarithmic terms,
in the sense that they should be always dictated by the Wick theorem
hidden in the algebraic structures of DIM. There is, however, a crucial difference: in this case,
the $U(1)$-mode should not decouple for non-trivial slopes, and two sets of times survive (see s.\ref{VOP}).
This is reflected in the fact that one needs to consider ${\cal G}\{\ta|x\}$ depending on $\tau_{n>0}$ and $\tau_{n<0}$ in (\ref{intCS}),
\be
{\cal G}\{\ta|x\} \ = \ \exp\left(\sum_{n\in\mathbb{Z}}^\infty \frac{\ta_nx^n}{n}\right)
\ee
in order to construct the Ward identities. Then, a counterpart of (\ref{widDFchava})
for (\ref{intCS}) looks somewhat different \cite{BEMT,Slept,Dubinkin}:
\be
\left<(n-r+1)\sum\limits_i^r x_i^n+\sum\limits_i^r x_i^n \left(\log\theta(x_i|q)\right)' +2\sum\limits_{i<j}\frac{x_i^{n+1}-x_j^{n+1}}{x_i-x_j}+
\sum\limits_{k,i}\tau_k x_i^{n+k}\right>=0
\label{widDFchava1}
\ee
where $q=\exp({1\over 2\gamma})$ and $\theta(x|q)=\sum_{\nu=-\infty}^{\infty}q^{\nu^2/2}x^\nu$.

\subsection{Variation VI: Correlators with vertex operators
\label{VOP} }

The vertical branes are associated with insertions of vertex operators into the DF and CS models.
A particular instance of the vertex operator is the screening current.
As already mentioned in s.\ref{multifmod}, screening charges
are segments of vertical branes between the two neighbour horizontal ones,
and they can be considered as contractions of two vertex operators attached to these
two branes.
However, the relevant operators are special, namely, they are $e^{\alpha\phi}$ with $\alpha=\pm 1$:
a kind of "fermion
vertices" (in fact, intertwining operators) $\psi^\pm =e^{\pm\phi}$.
Accordingly, the screening charges should be associated with bilinears
$\psi_a^+(x)\psi_b^-(x)$, "non-local" in the vertical direction:
\be\label{fsc}
\hat Q_{ab} = \oint \psi^+_a(x) \psi^-_b(x) dx
\ee
This non-locality explains, among other things, why the screening currents are "naturally"
exponentials rather than $\p\phi$-like currents, as well as the emergency of peculiar
$\sqrt{2}$ in (\ref{scree}) coming from the $45^\circ$ rotation of the basis $\phi_1,\phi_2$ into
$\frac{\phi_1\pm\phi_2}{\sqrt{2}}$.

In general, fermion operators (peculiar intertwiners in DIM) carry a Young-diagram label $\lambda$
instead of $x$ and the screening charge is a convolution of these indices (see s.3.2 of \cite{confMAMO2}
for details).
Interchanging of $+$ and $-$ labels changes the screening charge to the dual one (in algebraic terms, this corresponds to using instead of a positive root the corresponding negative one):
as usual in conformal matrix models \cite{confMAMO1,confMAMO2}, the use of dual charges is unnecessary. In fact, one can connect every screening charge with a simple root: one can associate with each end of leg $a$ a basis vector $\vec e_a$, then, the screening charge $Q_{a,a+1}$ corresponds to a simple root $\vec\alpha_a=\vec e_{a+1}-\vec e_a$.

\bigskip

In operator formalism the correlator  of vertex operators is just a matrix element of an ordinary product
of linear operators.
A generic vertex operator is constructed from the primary field $V_\alpha (x)$ and is labeled by the Young diagram $\lambda$:
\be
\hat V^{\lambda}_\alpha = \hat L_{-\lambda}V_\alpha (x)
\ee
with $\hat L_{-\lambda} = \prod_i \hat L_{-\lambda_i}$. The conjugation with $\hat L_{-1}$ moves it to an arbitrary point $z$:
\be
\hat V_\alpha^\lambda(x+z) = e^{z\hat L_{-1}}\hat V^{\lambda}_\alpha (x) e^{-z\hat L_1}
\ee
However, in CFT the positions of operators does not matter:
they can be considered as located at points in the complex $z$-plane,
or, more generally, on a Riemann surface (in the latter case same traces
need to be taken in operator formalism).

Still, location of the stress-tensor insertion does matter:
in the Riemann surface picture, it is associated with a choice of a contour encircling the vertex operator insertions,
and correlator depends on the homology class of this contour.
Changing the class is equivalent to commutation of $T(z)$ with the vertex operator, which is read off the commutation relations
\be
[L_n,V_\alpha(x)]=x^{n+1}V'_\alpha(x)+\alpha^2 (n+1)x^nV_\alpha(x)
\ee
and those of the Virasoro algebra. This is what we did in the derivation of (\ref{WI3}) placing the stress-tensor to the left, and to the right of vertex operators.

\paragraph{Central-charge-preserving comultiplication $\Delta_{MS}$.}
The action of Virasoro algebra is provided by
the Moore-Seiberg comultiplication $\Delta_{MS}$, which is given by the ordinary Leibnitz rule on the negative modes $T_-$, but the positive modes
act differently:
\be\label{MS}
\Delta_{MS} (L_n) R_{1} \otimes R_2  = \left(\sum\limits_{k=0}^{\infty}z^{n+1-k}\left(n+1 \atop k\right)L_{k-1}R_1\right)\otimes R_2  + R_1 \otimes  L_nR_2
\ee
This comultiplication can be read off the conformal Ward identities, \cite{MS} and celebrates two important properties:
\begin{itemize}
\item It is {\it parameterized} by an arbitrary parameter $z$,
\item it does not change the central charge, in contrast with the comultiplication in the DIM algebra that we use below.
\end{itemize}

\subsection{Variation VII. Nekrasov functions}

We define the Nekrasov function as partition function of the DF/CS network matrix model depending on parameters
$\vec\alpha_i$, $z_i$ and $N_{a}$, associated respectively with external legs
(assumed vertical), horizontal and vertical edges  of
the graph $\Gamma$: schematically,
\be
Z_\Gamma = \ \left<  \prod_{i=1}^{4} \hat V_{\vec\alpha_i}(z_i) \exp\left(\sum_{a=1}^{{\K}-1}
\hat Q_{a,a+1}\right)    \right>_{\mathrm{DF}_{N_a}}
\ee
and this partition function describes the $A_1$-quiver with obvious modifications for more sophisticated quivers, \cite{MMZdim} (changing the number of vertex operators and adding more screening charges that differ by the choice of the integration contours).
The right numbers of screening charges are automatically selected from the series expansion
of the exponential by zero mode conditions.

On the gauge theory side, this data describes the theory with the gauge group $SU({\K})$ and $2{\K}$ fundamental matter hypermultiplets (i.e. zero $\beta$-function). Here the numbers $N_a$ are the Coulomb moduli, the hypermultiplet masses are parameterized by the vertex operator parameters $\vec\alpha_i$ and the positions of vertices (rather their double-ratio) control the instanton expansion in the gauge theory. Note that this theory is characterized by zero $\beta$-function, all other cases are obtained by evident degeneration. The case of adjoint matter hypermultiplets is described by the elliptic DIM algebras\footnote{By DIM algebras in this paper we mean both DIM and its limits like affine Yangian \cite{4d,Mat1,Pro}.} \cite{ellDIM} and is out of scope of the present paper. The other quiver theories, say $A_k$ are described, on the physical side, by a product of $k$ gauge groups: $\prod_i^k SU(n_i)$ with $\kappa_i=2n_i-n_{i-1}-n_{i+1}$ bifundamental hypermultiplets for each $i$ transforming under the gauge groups $SU(n_i)$ and $SU(n_{i+1})$. There are also $\kappa_0$ and $\kappa_k$ fundamental hypermultiplets that are transformed under $SU(n_1)$ or $SU(n_k)$ (we put $n_0=n_{k+1}=0$). These theories have also zero $\beta$-functions, other cases can be obtained by a degeneration of hypermultiplet masses. Note that the Nekrasov network partition functions typically contain additional singlet fields, which corresponds to $U({\K})$ instead of $SU({\K})$ group. The contribution of this singlet factorizes out and reduces just to a simple multiplier in the Nekrasov function.

While exponentiation of bosonized screenings $Q=\oint e^\phi$ can look somewhat artificial,
the same procedure is very natural in the fermionic version $Q = \oint \psi^+\psi^-$:
this adds $\psi$-bilinear terms to the free fermion action, i.e. leaves it quadratic.
This is the reason for integrability, and in bosonized version this is reflected in
integrable properties of Toda like systems with exponential actions.

Exponentiation of fermionic screenings makes a new interesting
twist after the $q$-deformation in sec.\ref{sec:variation-x:-q}, see eq.(\ref{expofeS}) below.

\subsection{Variation VIII: Network model level. Network as a Feynman diagram\label{net1}}

\begin{figure}[h!]
  \centering
  \includegraphics[width=3cm]{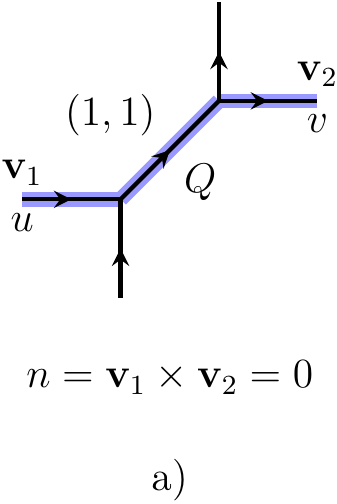} \hspace{3cm}   \includegraphics[width=7cm]{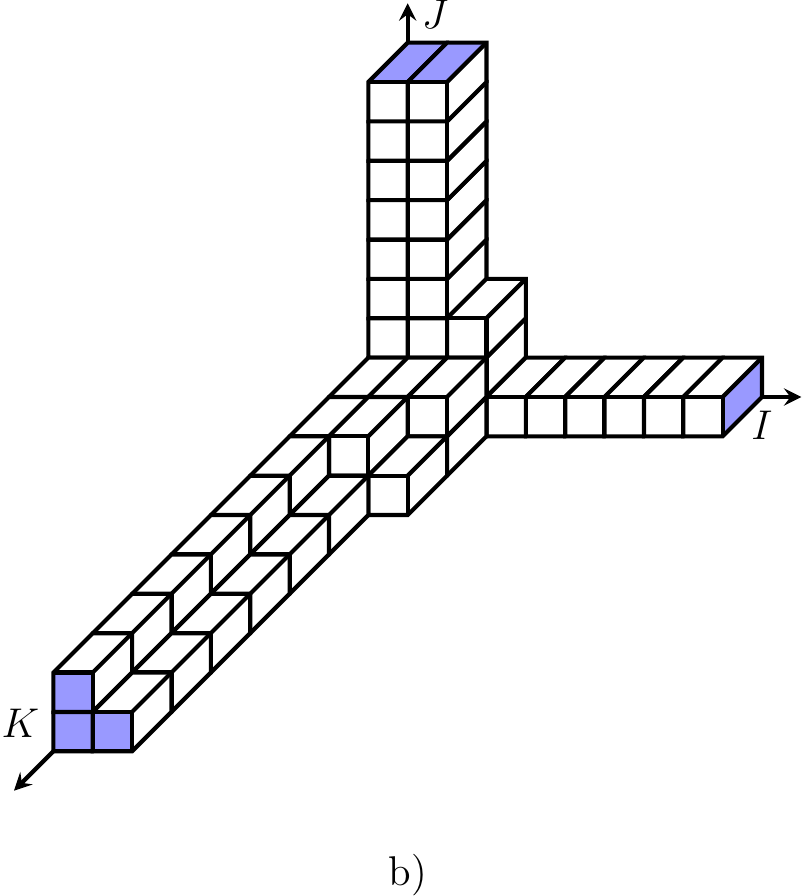}
  \caption{a) Simplest toric diagram. The intermediate edge has slope
    $(1,1)$ and length $Q$ and framing factor $n = \mathbf{v}_1 \times
    \mathbf{v}_2 = 0$. The ``horizontal'' line with spectral
    parameters $u$ and $v$ is shown in blue. The length of the
    intermediate edge is determined by the ratio of the spectral
    parameters on the adjacent edges, $Q = \frac{v}{u}$. b) An example
    of a 3d Young diagram which contributes to the vertex
    $C_{[1],[2],[2,1]}$. The vertex $C_{[1],[2],[2,1]}$ is given by
    the weighted sum over all 3d Young diagrams with three fixed
    asymptotics shown in blue.}
  \label{fig:toric}
\end{figure}

Network model is defined for a planar 3-valent graph $\Gamma$ with edges parameterized by slopes and lengths.
Slopes are given by pairs of numbers $(X_1,X_2)$,
see Fig.\ref{fig:toric}, and lengths by parameters $Q$. The
$2$-component vectors $\vec X$ are conserved at the vertices of $\Gamma$: $\vec X_v\,'+\vec X_v\,'' + \vec X_v\,'''=0$
at each vertex $v$; this is a stability condition for the brane-web.
The graph $\Gamma$ with this structure
describes {\it a la} \cite{GKMMM} the tropical spectral curve of the underlying integrable system,
but for our purposes it can be considered just as a Feynman diagram with cubic vertices
and momenta $Q\vec X$ on the edges, associated with some effective Chern-Simons-type field theory.
Expressions $Z_\Gamma$ for this Feynman diagram
(Nekrasov partition function or generalized conformal block) is build by convolution of
vertices $C_{IJK}(\vec X\,',\vec X\,'',\vec X\,'''|q)$ and propagators $\Pi^{IJ}(Q)$, where
indices $I,J,K$ are Young diagrams, and $C_{IJK}$ are, in turn,
"(refined) topological vertices" \cite{IKV,AK} given \cite{3dvertex} by sums over $3d$ (plane) partitions
with three boundary conditions described by three ordinary Young diagrams $I,J,K$,
see Fig.~\ref{fig:toric},~b).

In the generic network matrix model, the exponentials of screening charges no longer turn into
exponential of "fermions": it produces an elementary 3-valent vertex (=refined topological vertex)
providing the true DIM intertwiner.
Automatic is now not only adjustment of the number of screenings, but also
matching between their $\psi^+$ and $\psi^-$ constituents.

$\bullet$  Screening charges are substituted by vertical lines between
pairs of horizontal brains,
$\oint\exp(\vec\alpha_{ij}\vec\phi)$, involving two free fields associated with
the corresponding branes.

$\bullet$ Slopes of the horizontal branes enter the matrix model description through
$(\log x_i)^2$ terms in the action, see (\ref{eq:70}) in s.3.
The coefficient is made out of the skew product (see
Fig.~\ref{fig:toric}~a))
\be
\vec X_{v_1}\wedge\vec X_{v_2}
\ee
where $\vec X_{v_1}$, $\vec X_{v_2}$ are associated with the external horizontal lines, one incoming, the other one outgoing. In the case with several horizontal lines, see e.g. (\ref{eq:18}), one has to consider $\vec X_{v_1}$, $\vec X_{v_2}$ for different horizontal lines, and the answer in this case does not depend on the concrete choice of these lines.

\bigskip

We described in this subsection a generic network model. One can consider its particular case: the model that gives rise to the quiver gauge theory (as described in the previous subsection). In this case (for any quiver gauge theory), one can construct a K-theoretic version of the Nekrasov functions, $Z_\Gamma$, \cite{KNekr1,KNekr2}. They coincides \cite{AK,Nektv} with the refined partition functions in the corresponding geometry, which can be constructed via the refined topological vertex.

Another possibility is to consider the quiver theories with zero $\beta$-functions (so that all other can be obtained via various limiting procedures from these) and all gauge groups coinciding, $n_i=n$ $\forall i$.
These theories are associated with so called balanced networks and can be immediately described within the representation theory of DIM algebras, and the requirement of all gauge groups having the same rank is implied by a possibility of immediate extension of DIM to the elliptic DIM: this latter describes the quiver gauge theories with adjoint matter, where the condition $n_i=n$ is inevitable. We discuss the issue of balanced networks in the next subsection.

\subsection{Variation IX: Balanced network model\label{net2}}

As usual, the $q,t$-deformation leads to overloaded formulas,
but in fact it drastically simplifies them by providing a very
clear and transparent interpretations and unifying seemingly different ingredients.
Namely, everything gets controlled by the DIM symmetry:
the edges of graph carry DIM representations,
the topological vertices $C$ become their intertwiners, and symmetries
(stress-tensor and its ${\cal W}$-counterparts)
are just the generators of DIM acting in tensor products of representations
and thus defined by powers of the comultiplication $\Delta_{DIM}$ (which is different from $\Delta_{MS}$).

An exhaustive description of the network models depends on development of representation
theory for the double affine algebra DIM, and it is not yet brought to the generality level of \cite{GMMOS}
for ordinary affine algebras. In particular, at the moment, it is not immediate to describe within the DIM framework an arbitrary DF or CS matrix model. However, among the DF matrix models there is a subclass that is directly lifted to rather peculiar
networks, which we call {\it balanced} which are controlled by an analogue of
the level one representations of Kac-Moody algebras and allow a drastically simplified
bosonization and even fermionization. As we already mentioned the balanced networks correspond to special quiver gauge theories with zero $\beta$-functions.

We provide the details in section \ref{calculus} below, and devote the rest of this subsection
to a bird's eye view survey which makes use of an oversimplified, almost symbolic notation.
One can find the exact formulas in s.\ref{calculus}.

\bigskip

The network basically is a constructor with the main building block being a (refined) topological vertex, which is a matrix element of an intertwining operator that intertwines three representations, hence, the topological vertex is associated with three legs.

The balanced network is defined by three requirements:

(a) Consider a class of representations of DIM such that each leg is parameterized by a pair of integers $\vec M=(M_1,M_2)$ (DIM central charges) and a Young diagram $Y$. Then, the integers are subject to the conditions: $\sum_i^3\vec M_i=0$ and $|\vec M_i\wedge\vec M_j|=1$ for any pair of legs in the vertex. \footnote{In terms of topological strings, these are the Calabi-Yau and smoothness conditions. } As in s.\ref{net1}, we associate every vector $(M_1,M_2)$ with an edge of the network, and parameterize slopes as ratios $M_2/M_1$.

(b) Assume one of the legs of vertices is always vertical, $\vec M=(0,M)$. This implies that $M=\pm 1$ and that the two other vertices are $(\pm 1,L\pm 1)$.
From a general network with rational slopes one can make this minimal (i.e. that with all vertices having a vertical edge) by a sequence of {\it resolutions}, introducing new edges and triple vertices.
Reversing, a general rational network arises from a minimal one,
when some edges are shrunk to a point while others "fattened" (i.e. described
by $M_1$ and $M_2$ which are not coprime, this can be needed to keep vertices
three-valent).

We represent such a {\it minimal rational} network
(Fig.~\ref{fig:hor}, a))
by a set of ${\K}$ horizontal lines connected by
vertical segments (for planar graph, only adjacent lines can be connected),
which can also be as external vertical legs (to the lowest and highest horizontal lines).
Horizontal segments are also labeled by slopes: in other words, we draw all non-vertical edges
horizontal, but keep the slopes as labels.

(c) {\it Balanced} is the {\bf minimal rational} network where all {\it external}
legs are either vertical or
horizontal, i.e. either $(\pm 1,0)$ or $(0,\pm 1)$.

\bigskip

Partition functions for non-balanced networks have singular limit $t\longrightarrow 1$, $q/t=$fixed
and thus do not directly reduce to a DF model in $4d$.
Also the $U(1)$ center-of-mass field does not split from the Virasoro and other symmetries in this case.
However, maybe not these two issues are the main drawbacks, the real problem is a more sophisticated
representation theory needed to lift any of the three above restrictions:
balance, minimality and rationality (in the order of complexity).

\bigskip

From now on, we draw all networks  on the square lattice: the vertical lines $(0,1)$ are vertical, while all the lines with slopes $(1,M)$ are horizontal and just carry the charges $(1,M)$.

\bigskip

The partition function for the balanced network is a contraction of just two types of vertices:
the generalized "fermions" $\Psi$ and $\Psi^*$, which intertwine the DIM representations:
$(1,M)\otimes (0,1) \longrightarrow (1,M+1)$ and $(1,M+1)\otimes (0,-1)\longrightarrow (1,M)$.
These intertwiners can be described in terms of free field, which acts as an operator
in "horizontal" direction, i.e. converts the Verma module $V_{(1,M)}$ into $V_{(1,M\pm 1)}$.
Thus, of the three Young diagrams $\Psi$ depends explicitly only on one: that
sitting on the vertical leg, while those on two other, horizontal legs parameterize the
states in the Fock space, but not the operator.
Instead $\Psi$ depends on the spectral parameter $u$, as well as on the position.
The position is described by a continuous coordinate $z$ along the horizontal line and by the discrete number $a$ labelling the horizontal line itself.
Actually, all $\Psi_a$ and $\Psi^*_a$ with a given $a$ depend on the free field $\phi_a$:
there are ${\K}$ independent free fields for ${\K}$ horizontal lines.
It remains to provide explicit formulas for the $\Psi$-operators, slightly symbolically
\be
\prod_I \Psi_{\lambda_I}[z_I] \prod_J\Psi^*_{\mu_J}[z_J^*] \ \longrightarrow \ \boxed{
\prod_{I,J} \exp\left( \sum_{n\ne 0}\frac{1}{n}\left(\omega^{|n|}[\lambda_I,z_{I}]_n\mathfrak{a}_n -[\mu_J,z_{J}^*]_n\mathfrak{a}^*_n\right)\right)}
\ee
and details can be found in the next section \ref{calculus} (see especially s.\ref{3.5}).

Clearly, this description of balanced networks is as asymmetric w.r.t. vertical/horizontal
symmetry as only possible.
Thus, it does not respect most of interesting dualities, which appear as non-trivial
properties of the answers.
Instead, it is extremely simple and very close to conventional matrix model techniques.
In particular, it provides a very simple description of infinitesimal symmetries
(Ward identities), and this is some compensation for non-transparency of large
invariances (dualities).
Moreover, as mentioned in the Introduction, the Ward identities are now labeled by sections of the network.
The description is simple when the sections are pure vertical, but they can be easily
deformed to include horizontal pieces, and the study of such cases can bring us
closer to description of spectral dualities, even in this asymmetric formalism.

\subsection{Variation X: $q$-deformation\label{sec:variation-x:-q}}

The main new thing at this level is Jackson discretization of integrals:
\be
\int_0^z f(x) dx \longrightarrow (1-q) \sum_{n>0} zq^nf(q^nz)
\ee
It can seem that there is a problem here, because the screening charges would require integrals along closed contours, and one may think the Jackson integral is not their good counterpart.
What makes this deformation possible is the fact that the screening charges in the DF matrix models
of \cite{MMSha,AGTmamo1} are actually defined along {\it open} contours between
ramification points.

The most important result of discretization is the Young diagram expansion for
exponentiated screening in fermionic realization (\ref{fsc}) (an avatar of the Cauchy formula):
\be
\exp(\hat Q) = \exp\left(\oint \psi^+_a(x)\psi^-_{a+1}(x) dx\right) \
\longrightarrow \
\exp\left((1-q)\sum_{n>0}^\infty q^n\psi^+_a(q^n)\psi^-_{a+1}(q^n) \right) =
\sum_{\lambda} {q^{|\lambda|}\over \zeta_{\lambda}}\Psi^+_{a,\lambda} \Psi^-_{a+1,\lambda}
\label{expofeS}
\ee
where $\lambda = \{\lambda_1\geq\lambda_2\geq\ldots \geq \lambda_{l(\lambda)}>0\}$
is the Young diagram with
$|\lambda|=\sum_{i=1}^{l(\lambda)}\lambda_i$ boxes, $\Psi_\lambda = \prod_i \psi(q^{\lambda_i})$ and $\zeta_\lambda=\prod_r m_r!$, where $m_r$ is a number of times $r$ appears in the partition $\lambda$.
This formula is a simple avatar of the Cauchy expansion.

Operators of the type $\psi_\lambda$ play a crucial role in
building particular network models:
they are intertwiners of peculiar representations of DIM
and their matrix elements are the topological vertices (perhaps, refined) within the topological string framework.
Since elements of some Verma modules of double affine algebra DIM($\mathfrak{gl}_1$) are
labeled by 3d Young diagrams
(just like Verma modules of affine Virasoro by ordinary Young diagrams),
topological vertices are naturally expressed as sums
over plane partitions.

\subsection{Variation XI: $q,t,\ldots$-deformations}

Everything, what we surveyed above is straightforwardly deformed,
at least from Schur to the Macdonald level, or, in group theory terms,
from ordinary and affine (current) to double affine algebras DIM.
Moreover, one can expect a topicality of the elliptic and further Kerov deformations,
and, perhaps, even further, to triple-affine {\it Pagoda} algebras of \cite{MMZdim},
at least, to those corresponding to the double elliptic systems.

A short list of algebraic deformations is (in accordance with the columns: dimension$|$deformation \\parameters$|$symmetric polynomials$|$algebra of symmetry):
\be
\begin{array}{cccc}
4d & t=1,\ q=1 & \hbox{Schur} & \hbox{Virasoro}/{\cal W}_{1+\infty} \\
4d & t=q^{\beta},\ q \to 1& \hbox{generalized Jack} & \hbox{affine Yangian}\\
5d & t=q& \hbox{Schur}& q-\hbox{Virasoro} \\
5d &t,q& \hbox{generalized Macdonald} & \hbox{DIM}\\
 \end{array}
\ee
From the gauge theory/string perspective, the deformation parameters are associated with
compactification radius of the fifth dimension $R_5$:
\be
q=e^{\epsilon_1R_5},\ \ \ \ \ t=e^{-\epsilon_2R_5}
\ee
One naturally expects more parameters: the $q,t$ probably can be lifted to a three-parameter deformation associated with F-theory compactified on an elliptically fibred Calabi-Yau four-fold. Some evidence that the Seiberg-Witten/Nekrasov theory
survives in a nice form beyond the Macdonald $q,t$-deformation is provided
by the double elliptic studies on integrability side \cite{dell} and
by reinterpretation of the Seiberg duality for $N_f>2N_c$ \cite{seiberg} in terms of topological strings \cite{NfNc}.
It remains to repeat once again that potential of the DIM algebras is also far from being
exhausted by the $q,t$-deformation.

From CFT perspective, the most natural is the $\beta$-deformation, $t=q^\beta$,
$\beta = \sqrt{\epsilon_1/\epsilon_2}$,
which shifts the Virasoro central charge away from unity and other integer values in
the multi-field case.
As to the $q$-like deformations, they are long known to be natural for
hypergeometric series and their generalizations, which CFT is really about.
One of the main new things is that the stress tensor and more general ${\cal W}_{\K}$
generators are now unified: they are all combinations of primary vertex operators,
form a closed subalgebra
and possess a non-vanishing centralizer so that
one can consider models with the corresponding symmetry.

Another interesting point is a drastic increase of applicability domain for
fermionization: after discretization of screening integrals, it continues to work
in many representations beyond $c=1$,
moreover, the fermionic intertwiners in DIM are actually the refined topological
vertices from topological string theory.

The most impressive result of deformation is clear unification of a huge variety
of notions and phenomena, which appeared in different branches of science.
It gets clear that they were describing the same things, just in different
interpretations and limits,  about one and the same object: the network matrix model,
which is no more than a generic DIM-symmetric partition function on graphs.

\bigskip

In the last part of this section, we briefly consider the peculiarities
of the simplest deformation, $t=q^\beta,\ q\to 1$.

\subsection{Variation XII: $\beta$-deformation to non-unit Virasoro central charge}
The main new thing at non-unit $\beta=\log t/\log q$ as compared with subsection 2.0 is that the Vandermonde determinants in the matrix model measure
are raised to power $2\beta$ instead one $2$, i.e. the matrix models
are lifted to the $\beta$-ensembles \cite{beta,mamoCFT2,MMSha,AGTmamo1}, what leads to a temporal loss of connection
to integrability theory (which is presumably restored after the $q$-deformation).
Anyhow, technically most formulas are obtained by analytical continuation from
integer values of $\beta$.
The possibility to do so (unambiguously) comes from $\beta$-polynomiality of the
Selberg integrals, which define most correlators in the DF $\beta$-ensembles.

In the conformal field theory representation \cite{mamoCFT2}, the $\beta$-ensemble corresponds to theory with non-unit central charge.
As already mentioned, for non-integer Virasoro central charge $c$ one can expect problems
with fermionization: only bosonization is straightforward.
However, an appropriate substitute of fermionized formulas actually survives all the deformations,
all the way to DIM, at least in some representations (not restricted to $\beta=1$).

Screening charge $\hat Q$, acting on the Fock space
${\cal F}_\alpha =\Big\{{\rm Pols}(\ta_n)\Big\}\cdot e^{\alpha T_0}$, is
$$
\hat Q = \oint \hat S(x)dx = {\rm{res}}_{x=0}\ \hat S(x),
$$
\be
\hat S(x) = \ :e^{\sqrt{2\beta}\phi(x)}: \ =
\underbrace{\exp\left(\sum_{n>0} \frac{\sqrt{\beta} \ta_nx^n}{n}\right)}_{\sum_n x^{n}\chi_n\{\ta\}} e^{\sqrt{\beta}T_0} x^{2\sqrt{\beta}\p_0}
\exp \left(-\sum_n \frac{\sqrt{\beta}}{nx^n}\frac{\p}{\p \ta_n}\right)
\label{screebeta}
\ee
where $\chi_n\{p\}$ are the characters of symmetric representations $[n]$ of $sl$ algebras
(the Jack polynomials in this particular case).
Applied to the highest-weight state
\be
\Big|m+1\Big>\ = e^{-\alpha_{r,s} T_0} , \ \ \
\alpha_{r,s} = (1+r)\frac{\sqrt{\beta}}{2} - (1+s)\frac{1}{2\sqrt{\beta}}
\ee
it gives
\be
\hat Q \ \Big|\alpha_{-1,m}\Big>\  =   \chi_m\{\ta\}\, \Big|\alpha_{1,m}\Big>\
\ee
Similarly
\be
\hat Q^r\ \Big|\alpha_{-r,m}\Big>\ = \chi_{[m^r]}\{\ta\}\,\Big|\alpha_{r,m}\Big>\
\label{Qronvacbeta}
\ee

\bigskip

These screening charge commutes
\be
\l[ \hat L_n, \hat Q ] = 0
\label{LQcommbeta}
\ee
with the Virasoro generators
\be
\hat L_n = \sum_{k} (k+n)\ta_k\frac{\p}{\p \ta_{k+n}}
+ \sum_{k=1}^{n-1}k(n-k)\frac{\p^2}{\p \ta_k\p \ta_{n-k}}
+ 2n\sqrt{\beta}\frac{\p^2}{\p \ta_n \p T_0}-n(n+1)\mathfrak{Q}{\p\over\p\ta_n}, \ \ \ \ n>0
\label{Virposbeta}
\ee
where $\mathfrak{Q}=\sqrt{\beta}-{1\over\sqrt{\beta}}$. Then, one obtains
\be
\hat L_n \chi_{[m^r]} = 2n\sqrt{\beta}\alpha_{r,m}\frac{\p \chi_{[m^r]}}{\p \ta_n}
\ \ \ \ \  \ n>0
\label{Lchibeta}
\ee
while the action of
\be
\hat L_0 = \sum_{k} k\ta_k\frac{\p}{\p \ta_{k}} + \beta\frac{\p^2}{\p T_0 \p T_0}+(1-\beta){\p\over\p T_0}
\label{Vir0beta}
\ee
still gives the size of the Young diagram:
\be
\hat L_0\chi_{[m^r]}=mr\cdot \chi_{[m^r]}
\label{L0chibeta}
\ee
The negative modes are:
\be
\hat L_{-n} = \sum_k k\ta_{k+n}{\p\over\p \ta_k}+\sqrt{\beta}\ta_n{\p\over\p T_0}+{1\over 4}\sum_{k=1}^{n-1} \ta_k\ta_{n-k}+{n-1\over 2}\mathfrak{Q}\ta_n
\ee
so that
\be
\l[\hat L_n,\hat L_m] = (n-m)\hat L_{n+m} + \frac{n(n^2-1)}{12}\left(1-6\mathfrak{Q}^2\right)\delta_{n+m,0}
\ee
and the current modes are now
\be
\hat J_{-n} = {\ta_n\over\sqrt{2}}, \ \ \ \hat J_0 = \sqrt{2\beta}{\p\over\p T_0}, \ \ \ \hat J_{n} = \sqrt{2}n\frac{\p}{\p \ta_n}
\ee
while the Sugawara relation is
\be
\hat T(z) = \ :\frac{1}{2}\hat J(z)^2:+{\mathfrak{Q}\over\sqrt{2}}\p_z J(z)
\ee
In terms of generating functions, the Ward identity (\ref{Lchibeta}), i.e.
the corollary of symmetry (\ref{LQcommbeta}) becomes
\be
\left[z^2\ \hat T(z) - \sqrt{2\beta}\alpha_{r,m}\ z\hat J(z)\right]_-\cdot\chi_{[m^r]} = 0
\ee
or
\be
\left(z^2\ \hat T(z) - \sqrt{2\beta}\alpha_{r,m}\ z\hat J(z)\right)\cdot \chi_{[m^r]} = {{\rm Pol}(z) }
\label{chiwidbeta}
\ee

\bigskip

Now similarly to obtaining (\ref{intDF}), we can get the matrix element that is given by the integral ($\beta$-ensemble) representation. It looks like
\be
\chi_{[m^r]}\{\ta\}
= \ \left<\alpha_{r,m}\Big| \ \hat C\{\ta_n/\sqrt{2}\}\, \hat {\cal Q}^r \ \Big|\alpha_{-r,m}\right> \
= -{1\over r!}\underbrace{\oint\ldots\oint}_r
\left(\prod_{i=1}^r  \frac{\,{\cal G}\{\ta|x_i\}\, dx_i}{ x_i^{m+r}} \right)
\prod_{i<j} (x_i-x_j)^{2\beta}  =
\ \Big<\  1\  \Big>_{\rm{DF}_{m,r}}
\ee
However, the symmetric function $\chi_{[m^r]}\{\ta\}$ is now not the Schur, but the Jack polynomial.

The Ward identity (\ref{DFwid}) is now substituted by
\be
\left(z^2\ \hat T(z) - \sqrt{2\beta}\alpha_{r,m}\ z\hat J(z)\right)\Big< \ 1 \ \Big>_{\rm{DF}_{m,r}} = \ {\rm Pol}(z) =\\=
\left< \sum_{k,i}{\tau_kx_i^{k+1}\over z-x_i}+\sum_{i,j=1}^r \frac{x_ix_j}{(z-x_i)(z-x_j)}-\mathfrak{Q}\sum_i{z^2\over (z-x_i)^2}-2\sqrt{\beta}\alpha_{r,m}\sum_i{x_i\over (z-x_i)} \right>_{\rm{DF}_{m,r}}
\label{DFwidbeta}
\ee

\bigskip

One has to get two important points from considering this $\beta$-deformation:

\begin{itemize}
\item The deformation preserves the structure of equations and the vertex operators, moderately changing only the screening charges (the change that can be removed to a rescaling of the Heisenberg algebra operators), while the main change is due to changing the Sugawara relation, i.e. the construction of the Virasoro/W algebra.
\item Matrix models partition functions are also changed moderately, basically with only the Vandermonde determinant being deformed (hence, changing the Ward identities).
\end{itemize}

These two properties will persist in the generic $q,t$-case, as we demonstrate in the next sections.

In fact, one could repeat this matrix model consideration in the deformed case with non-unit $q$, following the lines of \cite{AY,5dAGTmamo}. However, the actual symmetry in this case becomes much larger than the Virasoro algebra: it is the DIM algebra, and we start its general description in the next section.

\section{DIM calculus for balanced network model
\label{calculus} }

In this section, we demonstrate how to deal with the balanced network model
by methods of the DIM algebra, which is a development based on the previous
consideration in \cite{Z,MZ,Pestun,MMZ,MMZdim}.
It is rather special from the algebraic perspective: only the DIM($gl_1$) algebra with special
values of central charges and rather
peculiar representations allowing straightforward bosonization and
even fermionization  is considered, however,
this covers almost all what is presently known about Nekrasov partition functions.

Details on various DIM algebras and their simplest representations are provided in the Appendix,
which can be useful for further development of the theory.

\subsection{DIM algebra\label{sec:dim-algebra}}

Let us first remind the definition of the
DIM algebra $U_{q,t}(\widehat{\widehat{\mathfrak{gl}}}_1)$.
It looks like a deformation of the affine quantum algebra
$U_q(\widehat{\mathfrak{gl}}_2)$ with the four Drinfeld currents: the positive/negative root
generators $x^\pm(z)=\sum_{n\in\mathbb{Z}}x^\pm_n z^{-n}$, two exponentiated Cartan generators
$\psi^+(z)$ and $\psi^-(z)$, which are power series in $z^{-1}$ and $z$ correspondingly, and the central element $\gamma$.

Commutation relations are
\begin{gather}
  G^{\mp}(z/w)\,x^\pm(z)\,x^\pm(w) \ =\  G^\pm(z/w)\, x^\pm(w)\, x^\pm(z)\nn \\
  \l[x^+(z),\, x^-(w)]\ =\
  \frac{(1-q)(1-t^{-1})}{1-q/t}\,\Big(\delta(\gamma^{-1}z/w)\,\psi^+(\gamma^{1/2}w)
  \ -\ \delta(\gamma z/w)\,\psi^-(\gamma^{-1/2}w)\Big)\nn \\
  \nn \\
  \psi^\pm(z)\,\psi^\pm(w) \ = \ \psi^\pm(w)\,\psi^\pm(z) \label{eq:24} \\
  \psi^+(z)\,\psi^-(w)\ = \ \frac{g(\gamma w/z)}{g(\gamma^{-1}w/z)}\, \psi^-(w)\psi^+(z) \nn \\
  \psi^+(z)\, x^\pm(w) \ =\  g(\gamma^{\mp 1/2}w/z)^{\mp 1}\, x^\pm(w)\,\psi^+(z) \nn \\
  \psi^-(z)\, x^\pm(w) \ =\ g(\gamma^{\mp 1/2}z/w)^{\pm 1}\,
  x^\pm(w)\,\psi^-(z)\nn\\
  \mathop{\mathrm{Sym}}\limits_{z_1, z_2, z_3} z_2 z_3^{-1} [x^{\pm}(z_1),
  [x^{\pm}(z_2),x^{\pm}(z_3)]]=0\nn
\end{gather}
The DIM algebra is a Hopf algebra with comultiplication
\begin{gather}
  \Delta\Big(\psi^\pm(z)\Big) \ = \ \psi^\pm(\gamma^{\pm 1/2}_2z)\,\otimes\, \psi^\pm(\gamma^{\mp 1/2}_1z)\nn\\
  \Delta\Big(x^+(z)\Big)\ = \ \psi^-(\gamma^{ 1/2}_1
  z)\,\otimes\,x^+(\gamma_1 z) \ + \ x^+(z)\,\otimes\, 1 \label{eq:23} \\
  \Delta\Big(x^-(z)\Big)\ = \ 1\,\otimes\, x^-(z)\ + \ x^-(\gamma_2
  z)\,\otimes\,\psi^+(\gamma^{ 1/2}_2 z)\nn
\end{gather}
where $\ \gamma_1^{\pm 1/2} = \gamma^{\pm 1/2}\otimes 1, \ \ \ \
\gamma_2^{\pm 1/2} = 1\otimes \gamma^{\pm 1/2}$ and the functions $g(z) =
\frac{G^+(z)}{G^-(z)}$ is restricted by the associativity requirement
$g(z)^{-1}=g(z^{-1})$.
We omit expression for the counit and antipode, since we will not need them.

This data allows one to construct the universal $R$-matrix \cite{Rmat}.

In these relations, $\gamma^{\pm 1/2}$ and $\psi_0^+\equiv\psi^+(z=\infty)$, $\psi^-_0\equiv\psi^-(z=0)$ are the central elements. Parameterizing their values as
\be
\gamma=\omega^{-M_1},\ \ \ \ \ \ \psi_0^\pm=\omega^{\pm M_2},\ \ \ \ \omega\equiv \sqrt{q\over t}
\ee
we reproduce the $(M_1,M_2)$ pairs of integers enumerating representations in s.\ref{net2}. The action of this comultiplication increases the central charges, in contrast with the Moore-Seiberg comultiplication $\Delta_{MS}$ (\ref{MS}). This is why the number of free fields is also increased by action of the comultiplication. In particular, starting from one free field (Kac-Moody level), we produce the Virasoro by acting with comultiplication, which adds yet another free field etc. Of the two integers $M_1$ and $M_2$, the first one is a counterpart of the Kac-Moody algebra level so that the refined topological vertex is a matrix element of the operator intertwining the level one representations, i.e. it can be realized by one free field. We explain this construction manifestly in the next subsections.

The structure of the algebra is encoded in the function
$G(z)$
which is often chosen to be {\it cubic} in $z$
with additional restriction $q_1q_2q_3=1$:
\begin{equation}
  G^\pm(z) = (1 -q_1^{\pm 1} z) (1 -q_2^{\pm 1} z) (1 -q_3^{\pm 1} z) = \Big(1-q^{\pm 1}z\Big)\Big(1-t^{\mp 1}z\Big)\Big(1-(q/t)^{\mp 1}z\Big)
\end{equation}
Without any harm to commutation relations and comultiplication,
it can be further promoted to
unrestricted $q_{1,2,3}$ and more general Kerov deformations,
and even to the elliptic function, though details of bosonization procedure below
should still be worked out in these cases.

\subsection{Bosonization in the case of special slopes}

Explicit expressions for $C_{IJK}$ are currently known only for
particular slopes: $\vec s'' = (0,1)$ and $\vec s'=(1,M)$, $\vec s'''=(1,M\pm 1)$, see Fig.\ref{fig:toric}.  According to
\cite{AFS}, they can be expressed in terms of the following
bosonization:
\begin{gather}
  \phi(z) = \sum_{n>0}
\left(\frac{1-t^n}{1-q^n}\frac{z^n
    a_{-n}}{n} -\frac{1-t^n}{1-q^n}\frac{a_n}{nz^n}\right),
\nn \\
[ a_n, a_m] =
n\,\frac{1-q^{|n|}}{1-t^{|n|}}\,\delta_{m+n,0} \label{eq:64}
\end{gather}
From this free field we can construct pre-vertex operators
depending on infinitely many time-variables $p_n$:
\begin{gather}
C\{p\} = \exp\left(\sum_{n>0}\frac{1-t^n}{1-q^n}\frac{a_n}{n}p_n\right) \nn \\
\bar C\{p\} = \exp\left(-\sum_{n>0}\frac{1-t^n}{1-q^n}\frac{a_n}{n}p_n\right)\nn \\
C^\dagger \{p\} = \exp\left(\sum_{n>0}\frac{1-t^n}{1-q^n}\frac{a_{-n}}{n}p_n\right) \label{eq:60}\\
\bar C^\dagger \{p\} =
\exp\left(-\sum_{n>0}\frac{1-t^n}{1-q^n}\frac{a_{-n}}{n}p_n\right)\nn
\end{gather}
with
\begin{equation}
  C^\dagger\left\{p_n =\sum_i
    z_i^n\right\}\bar C\left\{p_n = \sum_i z_i^{-n}\right\} \ = \ :\prod_i
  e^{\phi(z_i)}:
\end{equation}
These operators can be used to define the main vertex operators for
the above-mentioned particular slopes:
\begin{gather}
\Psi_{\lambda}(z) = \Psi\left[ \begin{picture}(45,20)(-20,3) \put(15,0){\vector(-1,0){15}}
    \put(0,0){\vector(-1,0){15}} \put(0,20){\vector(0,-1){20}}
    \put(-7,23){\mbox{{\footnotesize $z,\lambda$}}}
    \put(-20,-9){\mbox{{\footnotesize $-uz$}}}
    \put(10,-9){\mbox{{\footnotesize $u$}}}
    \put(-16,3){\mbox{{\footnotesize $M$}}}
    \put(5,3){\mbox{{\footnotesize $M\!-\!1$}}}
\end{picture} \right] =
 \frac{(-z)^{M|\lambda|}}{c_\lambda (f_\lambda)^M}\
C^\dag\Big[t^{-1}q^\lambda t^{\rho+1/2}z\Big]\ \bar
C\Big[q^{-\lambda}t^{-\rho-1/2}q z^{-1}\Big]
\times (-uz)^{|\lambda|}q^{n(\lambda^T)}
\nn \\ \nn \\ \nn \\
\Psi^{*}_{\lambda}(z) = \Psi\left[ \begin{picture}(45,20)(-24,9)
\put(15,20){\vector(-1,0){15}}
\put(0,20){\vector(-1,0){15}}
\put(0,20){\vector(0,-1){20}}
\put(-7,-9){\mbox{{\footnotesize $z,\lambda$}}}
\put(-25,23){\mbox{{\footnotesize $-v/z$}}}
\put(10,23){\mbox{{\footnotesize $v$}}}
\put(-20,12){\mbox{{\footnotesize $L$}}}
\put(5,12){\mbox{{\footnotesize $L\!+\!1$}}}
\end{picture} \right] =
 \frac{(-z)^{L|\lambda|}\,(f_\lambda)^L}{c_\lambda} \
\bar C^\dag\Big[q^{\lambda}t^{\rho} q^{-1/2} z \Big]\
C\Big[q^{-\lambda}t^{-\rho}q^{1/2} z^{-1}\Big]
\times (v/q)^{-|\lambda|}q^{n(\lambda^T)}\label{86}
\end{gather}
Here only one Young diagram $\lambda$ is shown explicitly, the two
others label matrix elements of the operator. The operator acts on the Fock space
$\mathcal{F}_u$, in which the basis vectors are labelled by Young
diagrams (e.g.\ the Schur functions provide a basis, $\chi_Y(a_{-n})|u,
\varnothing \rangle$). The edge parameters/lengths $Q$ are encoded in
the spectral parameters $u$ and $z$. More precisely, edge lengths are
given by \emph{ratios} of the spectral parameters between the
parallel lines, as shown in Fig.~\ref{fig:toric},~a). Notice that the vertices
$\Psi$, $\Psi^{*}$ depend only on two spectral parameters, the third
one being determined by the momentum conservation condition. This
condition follows from the requirement that the vertices intertwine
the action of the zero mode of the generator $x^{+}(z)$. An additional
notation is:
\begin{equation}
  f_\lambda = \prod_{(i,j)\in\lambda} (-)\cdot
  q^{j-1/2} t^{1/2-i}, \ \ \ \ \ \ \ \ c_\lambda = \prod_{(i,j)\in
    \lambda} \Big(1-q^{\lambda_i-j}t^{\lambda^T_j-i+1}\Big), \ \ \ \ \
  \ \ \ \ n(\lambda^T)=\sum_{(i,j)\in\lambda} (j-1)
\end{equation}
where $\lambda^T_j$ are row lengths of transposed Young diagram.
Finally,
\begin{equation}
  C[q^\lambda t^{\rho+1/2}] = C\left\{p_n =\sum_{i\geq
      1}(q^{n\lambda_i}-1)t^{n(1-i)}+\frac{1}{1-t^{-n}}\right\} \
  \stackrel{|t|>1}{=}\ C\left\{p_n=\sum_{i\geq
      1}q^{n\lambda_i}t^{n(1-i)}\right\}
\end{equation}
Here the requirement $|t| > 1$ is needed for convergence of the
sum. However, the result is analytic, and thus valid for any complex
$t \neq 1$.

The Feynman diagram is made from horizontal lines and vertical segments
between them. Operators along the horizontal lines are simply
multiplied, but each horizontal line depends on its own free field,
i.e.\ with ${\K}$-line diagram we associate operators acting in the
${\K}$-th tensor power of the single field Fock space,
$\mathcal{F}_{u_1}\otimes \cdots \otimes \mathcal{F}_{u_{\K}}$.  Sum over the
Young diagrams, $\lambda$ on vertical segments is performed with the
simple weight, which is independent of the edge length $Q$ : all
$Q$-dependent factors are already included in the definitions of
$\Psi$. One can understand this procedure as cutting the propagators
$\Pi^{IJ}$ in two halves (taking a ``square root'') and attaching the
resulting stubs to the corresponding adjacent vertices.

\subsection{Relation to topological vertex}

The operator $C\{p\}$ defined in Eq.~\eqref{eq:60} switches between the Fock
space and the time variables: for the vacuum state annihilated by all
operators $a_{-n}$ with $n>0$, $\langle 0 | a_{-n} = 0$ we have
\begin{gather}
\langle 0|  C\{p\} \,  a_{-n}  = p_n\, \langle 0|  C\{p\} \nn \\
\langle 0| C\{p\} \,  a_{n}  =  n\,\frac{1-q^n}{1-t^n}\,\frac{\p}{\p p_n}\,
\langle 0|  C\{p\}
\end{gather}
This $\langle 0 | C\{p\}\ $ is a $p$-dependent set of common coherent
eigenstates of all the annihilation operators $a_{-n}$.  Accordingly,
one can use the Macdonald polynomials $M_{\lambda}\{p\}$ to define
``Macdonald states'':
\begin{equation*}
  M_{\lambda}\{p\} = \langle 0 | C\{p\}|M_{\lambda} \rangle
\end{equation*}
and their involutions
\begin{equation*}
  \bar M_{\lambda}\{p\} = M_{\lambda} \{-p\} = \langle 0 |
  \bar{C}\{p\} | M_{\lambda} \rangle
\end{equation*}
The skew characters are given by the matrix elements
\be
M_{\lambda/\mu}\{p\} = \frac{\langle M_\mu |  C\{p\} | M_\lambda\rangle}
{\langle M_\mu | M_\mu\rangle}
\ee

The matrix elements of the intertwiners $\Psi$, $\Psi^{*}$ in the
basis of the Macdonald states give the standard expression for the AK
version of the refined topological vertex~\cite{AK}
\begin{gather}
  \langle \bar M_\mu
| \Psi\left[ \begin{picture}(42,20)(-20,3)
      \put(15,0){\vector(-1,0){15}} \put(0,0){\vector(-1,0){15}}
      \put(0,20){\vector(0,-1){20}} \put(-7,23){\mbox{{\footnotesize
            $z,\lambda$}}}
\end{picture}\right] | \bar M_\nu\rangle \ =
||M_{\lambda}||^2 ||M_{\nu}||^2 \left( - \frac{t^{1/2} u}{q (-z)^M}
\right)^{|\lambda|} f_{\lambda}^{-M} (t^{-1/2} z)^{|\mu|-|\nu|} f_{\nu}^{-1} C^{\mu\lambda}_\nu (q,t)
\\
\langle \bar M_\nu | \Psi\left[ \begin{picture}(45,20)(-23,6)
\put(15,20){\vector(-1,0){15}}
\put(0,20){\vector(-1,0){15}}
\put(0,20){\vector(0,-1){20}}
\put(-7,-9){\mbox{{\footnotesize $z,\lambda$}}}
\end{picture}\right]
| \bar M_\mu\rangle \
=   ||M_{\mu}||^2 \left( -\frac{q(-u)^L}{t^{1/2} z} \right) f_{\lambda}^L
(t^{-1/2}u)^{-|\mu| + |\nu|} f_{\nu} C_{\mu\lambda}{}^\nu(q,t)
\end{gather}
where
\begin{gather}
  C_{\mu\lambda}{}^\nu(q,t) = M_\lambda^{(q,t)}[t^\rho]\sum_\sigma \bar
  M_{\mu^T/\sigma^T}^{(t,q)}[t^{-\lambda^T}q^\rho]\,
  M_{\nu/\sigma}^{(q,t)}[q^\lambda t^\rho] (q/t)^{\frac{|\sigma|-|\nu|}{2}}f_\nu^{-1}(q,t),\\
  C^{\mu \lambda}{}_{\nu} (q,t) = (-1)^{|\lambda|+|\mu|+|\nu|}
  C_{\mu^T \lambda^T}{}^{\nu^T}(t,q)
\end{gather}

The IKV vertices~\cite{IKV} arise in another basis: for the
$q,t$-independent Schur states $|s_{\lambda}\rangle$ and their
$q,t$-dependent duals $\langle S_{\lambda} |$ w.r.t.\ to the Macdonald
scalar product.

\subsection{Building screening charges and vertex operators}
\label{sec:build-scre-charg}
The screening charges and vertex operators of the Virasoro or
$W_{\K}$-algebra arise as combinations of intertwiners $\Psi$,
$\Psi^*$. The screening charges should commute with the Virasoro
generators, and since the Virasoro algebra is generated by an element of
DIM algebra, the intertwiners of DIM are the natural candidates for
the screening charges. We will see in the next section that one can interpret the commutation
graphically. The Virasoro generators act on the horizontal lines, and the
screening charges are segments of the vertical lines between the
horizontal ones. There are also external vertical lines, which
correspond to the Virasoro vertex operators. These \emph{do not} commute
with the Virasoro algebra, because the corresponding intertwiner contains an
extra representation, the \emph{vertical} one. The action of
energy-momentum tensor on this additional representation gives extra
terms, making the commutation rules nontrivial.

\paragraph{Screenings charges.} Let us start by building the screening
charges. They correspond to internal vertical lines in the web. The
minimal example contains two intertwiners, which are contracted with
each other to form a vertical segment between the adjacent horizontal
lines. The whole procedure resembles the free fermion construction of
the screening currents from sec.~\ref{VOP}. Each intertwiner plays the
role of a free fermion, so that their contraction gives rise to fermion
bilinears, i.e.\ the screening currents of dimension one. The integral
of the currents is replaced by the sum over intermediate states in the
vertical representation as in sec.~\ref{sec:variation-x:-q}.

The product of intertwiners is given by
\begin{multline}
  \label{eq:13}
  \parbox{3cm}{\includegraphics[width=3cm]{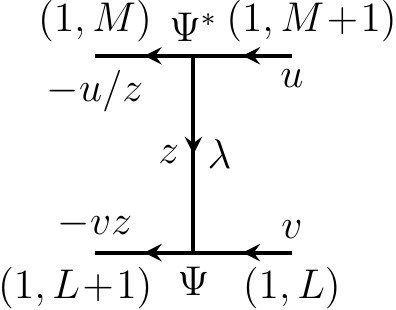}} = \sum_{\lambda} ||M_{\lambda}||^{-2} \Psi^{*}_{\lambda}(z) \otimes \Psi_{\lambda}(z) =\\
  = \sum_{\lambda} \left( \frac{qv}{u} (-z)^{M-L+1}
  \right)^{|\lambda|} \frac{f_{\lambda}^{M-L-1} q^{2
      n(\lambda^{T})}}{c_{\lambda} c'_{\lambda}} \exp \left\{ -
    \sum_{n \geq 1} \frac{1}{n} \frac{1-t^n}{1-q^n} \left( 1 + \left(
        \frac{q}{t} \right)^n \right) p_n(q^{\lambda}t^{\rho}
    q^{-1/2}z)
    \tilde{\alpha}_{-n} \right\} \times\\
  \times \exp \left\{ \sum_{n \geq 1} \frac{1}{n} \frac{1-t^n}{1-q^n}
    \left( 1 + \left( \frac{q}{t} \right)^n \right)
    p_n(q^{-\lambda}t^{-\rho} q^{1/2}z^{-1})
    \tilde{\alpha}_{n} \right\} =\\
  =\sum_{\lambda} \left( \frac{qv}{u} (-z)^{M-L+1} \right)^{|\lambda|}
  \frac{f_{\lambda}^{M-L-1} q^{2 n(\lambda^{T})}}{c_{\lambda}
    c'_{\lambda}} :\prod_{i \geq 1} S(q^{\lambda_i} t^{\rho_i}
  q^{-1/2} z):
\end{multline}
where $c'_{\lambda} = \prod_{(i,j)\in \lambda} (1 - q^{\lambda_i - j
  +1} t^{\lambda^T_j - i})$ and
\begin{equation}
  \label{eq:21}
  S(x) = \exp \left\{ - \sum_{n \geq 1} \frac{1}{n}
  \frac{1-t^n}{1-q^n} \left( 1 + \left( \frac{q}{t} \right)^n \right) x^n
  \tilde{\alpha}_{-n} \right\} \exp \left\{  \sum_{n \geq 1} \frac{1}{n}
  \frac{1-t^n}{1-q^n} \left( 1 + \left( \frac{q}{t} \right)^n \right) x^{-n}
  \tilde{\alpha}_{n} \right\}
\end{equation}
We see that the contraction of two intertwiners depends on a
particular (``Virasoro'') combination of the bosonic oscillators acting on the
two horizontal Fock representations\footnote{We conform with the
  notations of~\cite{MMZdim}.}:
\begin{align}
  \label{eq:14}
  \tilde{\alpha}_n &= \frac{1}{1 + \omega^{2|n|}} (a_n^{(1)} - \omega^{|n|}
  a_n^{(2)}),\ \ \ \ \ \ \ \ \omega\equiv\sqrt{q\over t}
\end{align}
where $a^{(1)}_n = a_n \otimes 1$ and $a_n^{(2)} = 1 \otimes a_n$.
The generators $\tilde{\alpha}_n$ are normalized differently from the
original Heisenberg generators $a_n$ (cf.\ Eq.~\eqref{eq:64}):
\begin{equation}
  \label{eq:15}
  [\tilde{\alpha}_n, \tilde{\alpha}_m] = n \frac{1 - q^{|n|}}{(1 -
    t^{|n|})(1 + \omega^{2|n|})}  \delta_{n+m,0}.
\end{equation}

The contraction of intertwiners provides us with an \emph{indefinite}
number of screening currents, since the product in the last line of
Eq.~\eqref{eq:13} is infinite. This corresponds to the \emph{exponential}
of the screening charge and fits well with the picture where the
\emph{pair} of intertwiners gives fermion bilinear screening current:
\begin{equation}
  \label{eq:19}
  \sum_\lambda ||M_\lambda||^{-2} \Psi_\lambda^*(z) \otimes
  \Psi_\lambda(z) \sim \exp \left( \oint S(x) dx \right) =  \sum_N \frac{1}{N!} \oint
  \prod_{i=1}^N S(x_i) d^Nx
\end{equation}
According to the $q$-deformation prescription from
sec.~\ref{sec:variation-x:-q}, the positions of the screening currents
are discrete and parameterized by the Young diagrams $\lambda$:
\begin{equation}
  \label{eq:20}
  x_i = q^{\lambda_i} t^{\rho_i} q^{-1/2} z,
\end{equation}
so that the contour integral in Eq.~\eqref{eq:19} is replaced by the
sum over $\lambda$\footnote{One can understand this recipe in
  different ways: either as the Jackson integral, or as a sum over
  residues of the normal ordered operator expression. The final result
  for the sum over Young diagrams is the same in the both approaches.}. To
get a \emph{definite} number of screenings one should put some vertex
operators and external states into the system. Then, the selection rules
automatically provide one with a necessary number of screening
charges. We will see this effect below, when discussing the vertex
operators.

The operator product expansion of two screening operators immediately defines the
corresponding matrix model measure. We have
\begin{equation}
  \label{eq:66}
  S(x) S(y)  \sim \frac{\left( \frac{x}{y}; q \right)_{\infty} \left(
      \frac{y}{x}; q \right)_{\infty} }{\left(t \frac{x}{y}; q
    \right)_{\infty} \left(t \frac{y}{x}; q \right)_{\infty}} :S(x) S(y):
\end{equation}
where $(x;q)_{\infty} = \prod_{k\geq 1} (1 - q^k x)$. This means that
the matrix model is of $(q,t)$-deformed type, with the measure given
by
\begin{equation}
  \label{eq:67}
  \Delta^{(q,t)}(x) = \prod_{i \neq j} \frac{\left( \frac{x_i}{x_j}; q
    \right)_{\infty}}{\left( t \frac{x_i}{x_j}; q \right)_{\infty}}
\end{equation}
It is known \cite{triality,MZ} that such a matrix model explicitly
computes the Nekrasov partition function and the $\mathsf{Vir}_{q,t}$
conformal block.

Of course, the expression for the intermediate vertical segment
between any two adjacent horizontal lines (e.g.\ $i$-th and
$(i+1)$-th) has the same form as Eq.~\eqref{eq:13}. The only
difference is that the Heisenberg generators are $a^{(i)}_n$ and
$a_n^{(i+1)}$ instead of $a_n^{(1)}$ and $a_n^{(2)}$. On the tensor
product of ${\K} \geq 3$ Fock representations acts the $W_{\K}$-algebra and
the intermediate segments correspond to $({\K}-1)$ different screening
charges commuting with this algebra. The combinations of the
differences between the adjacent bosonic oscillators correspond to the
\emph{roots} of the $A_{{\K}-1}$ algebra.

\paragraph{Vertex operators.} As we have already mentioned, vertex
operators should be built from the intertwiners with external vertical
legs. Again, a minimal example contains a pair of intertwiners on two
horizontal lines, which are, however, not contracted in this
case. Their product now essentially depends on the \emph{both} horizontal
oscillators. This corresponds to a composite vertex operator having
two parts: the Virasoro part depending on $\tilde{a}_n$ and the
Heisenberg part depending on the orthogonal linear combination of
the oscillators, $\bar{a}_n$. This is exactly as prescribed by the AGT
relation \cite{AY,5dAGT,5dAGTmamo}, where the Nekrasov functions for the gauge group $U(N)$
correspond to the conformal block of the algebra $\mathsf{Vir}_{q,t}
\otimes \mathsf{Heis}_{q,t}$.

We have the following result:
\begin{equation}
  \label{eq:16}
   \parbox{3cm}{\includegraphics[width=3cm]{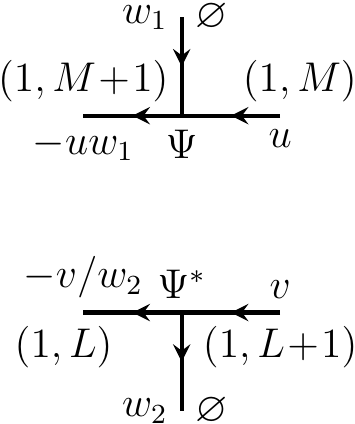}} = \Psi_{\varnothing}(w_1,v) \otimes \Psi^{*}_{\varnothing}(w_2,u) =
  \widetilde{V}_{w_1/w_2}^{\mathrm{Vir}} \left( ( w_1 w_2 )^{1/2}\right)  V_{w_1/w_2}^{\mathrm{Heis}} \left( ( w_1 w_2 )^{1/2}\right)
\end{equation}
where the indices denote the Liouville-like momenta of the vertex
operators
\begin{align}
  \label{eq:17}
  \widetilde{V}_{P}^{\mathrm{Vir}} \left( z \right) &= \exp \left\{ - \sum_{n \geq 1}
  \frac{P^{n/2} + P^{-n/2}}{n(1-q^n)} z^n \tilde{\alpha}_{-n} \right\} \exp \left\{ - \sum_{n \geq 1}
  \frac{q^n(P^{n/2} + P^{-n/2})}{n(1-q^n)} z^{-n} \tilde{\alpha}_n \right\} ,\\
  V_P^{\mathrm{Heis}} \left( z\right) &=\exp \left\{ - \sum_{n \geq 1}
  \frac{\omega^{-n}((\omega^2 P)^{n/2} - (\omega^2 P)^{-n/2})}{n(1-q^n)} z^n \bar{\alpha}_{-n} \right\} \exp \left\{ - \sum_{n \geq 1}
  \frac{(qt)^{n/2} ((P/\omega^2)^{-n/2} - (P/\omega^2)^{n/2})}{n(1 - q^n)} z^{-n} \bar{\alpha}_n \right\} .
\end{align}
where $\tilde{\alpha}_n$ are defined in Eq.~\eqref{eq:14} and
\begin{gather}
  \label{eq:65}
  \bar{\alpha}_n = \frac{\omega^{|n|}}{1 + \omega^{2|n|}} \left( \omega^{|n|}
    a_n^{(1)} + a_n^{(2)} \right)
\end{gather}
Notice that the momenta in the $U(1)$ part are slightly different (by
$t^2/q^2$) for the positive and negative modes which matches the AGT
prescription~\cite{AGTa,AFLT}.

The vertex operator~\eqref{eq:17}, though it depends on the right
combination of the oscillators $\tilde{\alpha}_n$ is not the full
Virasoro vertex operator (in particular, it does not have a smooth
limit for $t,q \to 1$). The same comment actually applies to the
exponential of the screening charge~\eqref{eq:13}. The reason for this
behavior is that both~\eqref{eq:13} and~\eqref{eq:16} are not
\emph{balanced}. This means that either incoming or outgoing
representations are not horizontal. To get the balanced combination,
one should consider the product of~\eqref{eq:13}
and~\eqref{eq:16}, i.e.\ the partial contraction of four
intertwiners:
\begin{multline}
  \label{eq:47}
  \parbox{4cm}{\includegraphics[width=4cm]{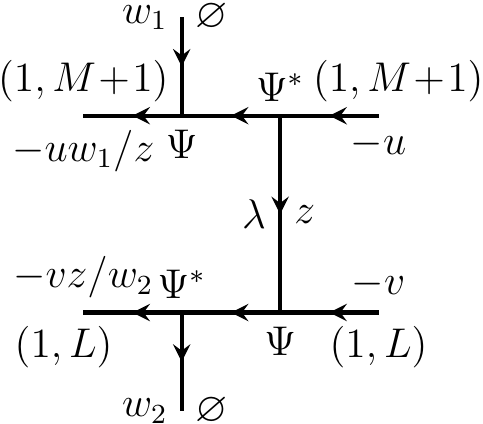}} =  \sum_{\lambda} ||M_{\lambda}||^{-2} \Psi_{\varnothing}(w_1)
  \Psi^{*}_{\lambda}(z) \otimes \Psi^{*}_{\varnothing}(w_2)
  \Psi_{\lambda}(z) = \\
  = V^{\mathrm{Heis}}_{w_1/w_2}\left((w_1 w_2)^{1/2}\right)
  \oint \prod_{i=1}^N  S(x_i) d^Nx\, V_{(tw_2)/(qw_1)}^{\mathrm{Vir}} \left( \left(
      w_1 w_2 q/t \right)^{1/2} \right),
\end{multline}
where $t^N = \sqrt{\frac{t}{q}} \frac{z}{w_1}$ and
\begin{equation}
  \label{eq:51}
  V^{\mathrm{Vir}}_P(x) = \exp \left( - \sum_{n \geq 1} \frac{1}{n}
    \frac{x^n}{1 - q^n}  \tilde{\alpha}_{-n} \left( P^{n/2} - P^{-n/2}
    \right) \right) \exp \left( - \sum_{n \geq 1} \frac{1}{n}
    \frac{q^n}{1 - q^n} x^{-n} \tilde{\alpha}_n \left( P^{-n/2} - P^{n/2} \right) \right)
\end{equation}
Of course, one can change $w_1$ to $w_2$ and vice versa in all the
formulas. The balanced combination of the operators automatically
fixes two problems: it determines the number of screening charges $N$ and
gives the correct expression for the $\mathsf{Vir}_{q,t}$ vertex
operator~\eqref{eq:51} in terms of free fields~\cite{AY}.

\paragraph{Other combinations of four intertwiners.} In this paragraph
we give an exhaustive list of webs, both balanced and unbalanced,
obtained from combinations of four intertwiners on two horizontal
lines. The first possibility is given by Eq.~(\ref{eq:47}), the second
we describe below in~(\ref{eq:18}), when we discuss conformal
blocks. Here we consider two more variations.

\begin{enumerate}
\item \textbf{Two antiparallel lines.}
  \begin{multline}
  \label{eq:94}
  \parbox{5cm}{\includegraphics[width=5cm]{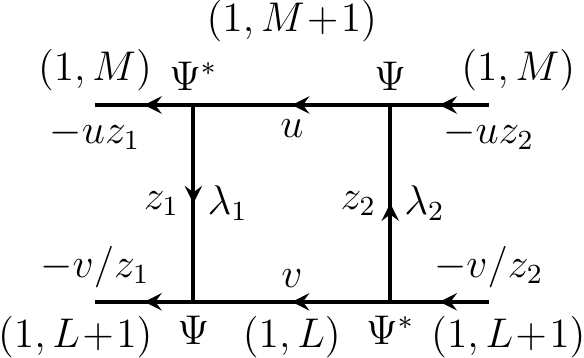}} =
  \sum_{\lambda_1, \lambda_2} \left( \frac{qv}{u} (-z_1)^{M-L+1}
  \right)^{|\lambda_1|} \left( \frac{qu}{v} (-z_2)^{L-M-1}
  \right)^{|\lambda_2|}\times\\
  \times \frac{f_{\lambda_1}^{M-L-1}f_{\lambda_2}^{M-L-1}
    q^{2n(\lambda_1^T)+2n(\lambda_2^T)}}{c_{\lambda_1} c_{\lambda_1}'
    c_{\lambda_2} c_{\lambda_2}'} :\prod_{i \geq 1} S_2(q^{\lambda_{2,i}} t^{\rho_i} q^{-1/2}
  z_2):\, :\prod_{j \geq 1} S_1(q^{\lambda_{1,j}} t^{\rho_j} q^{-1/2}
  z_1):
\end{multline}
Notice that here $S_2$ depends on the combination of the
oscillators corresponding to the \emph{affine} (imaginary) root of the
algebra $\widehat{A}_1$:
\begin{align}
  \label{eq:92}
    \tilde{\alpha}_n^{(2)} &= \frac{1}{1 + \omega^{2n}} \Big(a_n^{(2)} - \omega^{n}
  a_n^{(1)}\Big), \quad n \geq 1\\
  \tilde{\alpha}_{-n}^{(2)} &= \frac{1}{1 + \omega^{2n}} \left(
    a_{-n}^{(2)} - \omega^{n} a_{-n}^{(1)} \right), \quad n \geq 1\notag
\end{align}

This diagram is balanced and corresponds to a particular case of the
compactified toric diagram. The two antiparallel vertical lines should
be understood as living on the two sides of the cylinder. We will give
a more general ``quasi-periodic'' version of this diagram in s.\ref{affine}, where we describe the affine $(q,t)$-matrix model.

\item \textbf{Horizontal cut.} This strange variation is obtained by
  adding two ``internal'' lines ending at empty diagrams:
  \begin{multline}
    \parbox{5cm}{\includegraphics[width=5cm]{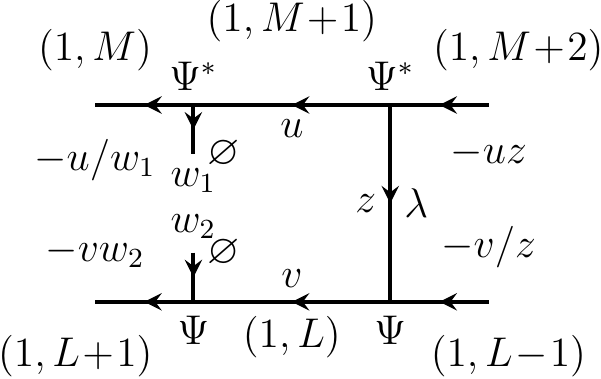}} =
    \sum_{\lambda} \left( \frac{qv}{u} (-z)^{M-L+1}
    \right)^{|\lambda|} \frac{f_{\lambda}^{M-L-1}
      q^{2n(\lambda^T)}}{c_{\lambda}
      c_{\lambda}'}\times\\
    \times \widehat{V}^{\mathrm{Heis}}_{w_1/w_2}(\sqrt{w_1w_2})
    \widehat{V}^{\mathrm{Vir}}_{w_1/w_2}(\sqrt{w_1w_2}) :\prod_{j =
      1}^{\tilde{N}} S(q^{\lambda_j} t^{\rho_j} q^{-1/2} z):
  \end{multline}
  where $t^{\tilde{N}} = \frac{z}{w_1}$ and
  \begin{align}
  \widehat{V}_P^{\mathrm{Vir}} \left( z \right) &= \exp \left\{ -
    \sum_{n \geq 1}
  \frac{\omega^{n}(P^{n/2} - P^{-n/2})}{n(1-q^n)} z^n \tilde{\alpha}_{-n} \right\} \exp \left\{ - \sum_{n \geq 1}
  \frac{q^n \omega^{n}(P^{-n/2} - P^{n/2})}{n(1-q^n)} z^{-n} \tilde{\alpha}_n \right\} ,\\
  \widehat{V}_P^{\mathrm{Heis}} \left( z\right) &=\exp \left\{ - \sum_{n \geq 1}
  \frac{\omega^{-n}(P^{n/2} - P^{-n/2})}{n(1-q^n)} z^n \bar{\alpha}_{-n} \right\} \exp \left\{ - \sum_{n \geq 1}
  \frac{\omega^{-n} (P^{-n/2} - P^{n/2})}{n(1 - q^n)} z^{-n} \bar{\alpha}_n \right\}
\end{align}
This network is unbalanced and, hence, produces wrong vertex
operators $\widehat{V}^{\mathrm{Vir}}$, i.e.\ those which do not
satisfy the usual commutation relations with the $q$-deformed Virasoro
energy-momentum tensor. Notice also that the Heisenberg vertex
operator $\widehat{V}^{\mathrm{Heis}}$ is not the required
Carlsson-Okounkov vertex operator \cite{CO}, i.e. the momenta are not shifted for
the positive and negative modes (see also \cite{AFLT}).
\end{enumerate}

\subsection{Network partition function\label{3.5}}

Now we have all ingredients necessary for constructing network partition functions. It schematically has the form
\be\label{npf}
\left< \prod_a{\Psi}_{\lambda_a}[z_a]\prod_b\Psi^*_{\mu_b} [z_b^*] \prod_{c} \left(\sum_\mu {\Psi}_{\mu}\Psi^*_{\mu}\right)\right>
\ee
where the first product describes the external vertex operators, and the second one the "internal" screening operators. We denoted the vertex attached to brane $a$ by $\Psi_a$.

As we already mentioned at the end of the previous section, the deformation does not influence much the screening and vertex operators. This means that one can straightforwardly construct (\ref{npf}). Indeed, one can choose the normalization of the Heisenberg algebra operators in such a way that the pre-vertex operators become very simple:
\be
C\{p\} = \exp\left(\sum_{n>0} \frac{{\mathfrak{a}}_{n}p_n}{n}\right)  &&
\bar C\{p\} = \exp\left(-\sum_{n>0} \frac{{\mathfrak{a}}_{n} p_n}{n}\right)\nn \\
C^\dagger \{p\} = \exp\left(\sum_{n>0} \frac{{\mathfrak{a}}_{-n}p_n}{n} \right) &&
\bar C^\dagger \{p\} = \exp\left(-\sum_{n>0} \frac{{\mathfrak{a}}_{-n}p_n}{n} \right)
\ee
and the screening currents (\ref{eq:21}) get the non-deformed form
\begin{equation}\label{newsc}
  S(x) = \exp \left\{ - \sum_{n \geq 1} \frac{1}{n}
  x^n
  \tilde{\mathfrak{a}}_{-n} \right\} \exp \left\{  \sum_{n \geq 1} \frac{1}{n}
 x^{-n}
  \tilde{\mathfrak{a}}_{n} \right\}
\end{equation}
In this simplified notation, the first part of formula (\ref{npf}), the external vertex operators, can be rewritten in the form (we are using equation (\ref{86}) with the rescaled Heisenberg algebra)
\be\label{ev}
\prod_I \Psi_{\lambda_I}[z_I] \prod_J\Psi^*_{\mu_J}[z_J^*] \ \longrightarrow \
\prod_{I,J} \exp\left( \sum_{n\ne 0}\frac{1}{n}\left(\omega^{|n|}[\lambda_I,z_{I}]_n\mathfrak{a}_n -[\mu_J,z_{J}^*]_n\mathfrak{a}^*_n\right)\right)
\ee
where all incoming vertex operators (labeled by the index $I$) are associated with the horizontal brane described by the Heisenberg operators $\mathfrak{a}_n$, while those outgoing ones (labeled by the index $J$) correspond to the Heisenberg operators $\mathfrak{a}^*_n$.
Here the symbol
\be
[\lambda,z]_n\equiv \hbox{sign}(n)\ \sum_i\left(q^{\lambda_i-1/2}t^{1/2-i}z\right)^n
\ee
introduces the Miwa variables. This is exactly the formula (\ref{7}).

Formulas (\ref{newsc}) and (\ref{ev}) give simple expressions for the ingredients of (\ref{npf}), thus providing a description of the network partition functions.

As we already explained, in variance with vertex operators, the Virasoro/W-algebra non-trivially changes with deformation. We shall discuss this phenomenon in the next two sections, and here give a few examples of conformal blocks (calculated in terms of the non-rescaled Heisenberg algebras).

\subsection{Examples of conformal blocks}

\paragraph{The simplest conformal block $\mathcal{B}_{\mathrm{PG}}$.}
The simplest possible contraction corresponding to a nontrivial
conformal block includes four intertwiners. It gives a peculiar ``pure
gauge'' limit of the four-point Virasoro conformal block
$\mathcal{B}_4(P_1, P_2, P_3, P_4, P,
\mathfrak{x})$, which, in the gauge theory language, corresponds to the
pure $SU(2)$ gauge theory partition function. In this
limit~\cite{MMS-BGW}, the dimensions $P_i$ of all the external fields
become infinite, and simultaneously the points $0$ and $\mathfrak{x}$ merge in a very particular way:
\begin{equation}
  \label{eq:68}
  P_i \to \infty, \qquad \mathfrak{x} \to 0, \qquad \mathfrak{x}
  P_1 P_2 P_3 P_4 = \Lambda^4 = \mathrm{fixed}
\end{equation}
Only two parameters, $\Delta$ and $\Lambda$ remain finite, so that
$\mathcal{B}_{\mathrm{PG}} = \mathcal{B}_{\mathrm{PG}} (P,
\Lambda)$.

The corresponding web partition function is equal to
\begin{multline}
  \label{eq:18}
  \parbox{5cm}{\includegraphics[width=5cm]{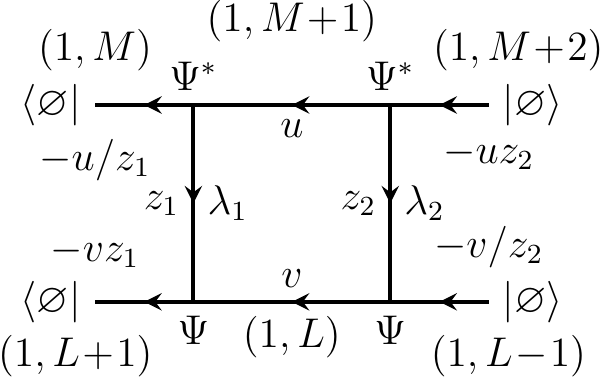}} = \langle
  P_1 | \exp \left( \oint_{\mathcal{C}_1} S(x) dx \right) \exp \left(
    \oint_{\mathcal{C}_2} S(x) dx \right) |P_4 \rangle =
  \mathcal{B}_{\mathrm{PG}} (P, \Lambda) =\\
  =\sum_{\lambda_1, \lambda_2} \left( \frac{qv}{u}
    (-z_1z_2)^{\frac{M-L+1}{2}} \right)^{|\lambda_1|+|\lambda_2|}
  \left(-\frac{z_1}{z_2}\right)^{\frac{M-L+1}{2}
    (|\lambda_1|-|\lambda_2|)} \frac{f_{\lambda_1}^{M-L-1}
    f_{\lambda_2}^{M-L+1} q^{2 n(\lambda_1^{T}) + 2
      n(\lambda_2^{T})}}{c_{\lambda_1} c_{\lambda_2} c'_{\lambda_1}
    c'_{\lambda_2}} \times\\
  \times
  \begin{smallmatrix}
    \langle \varnothing, -u / z_1 |\\
    \otimes\\
    \langle \varnothing,
  -v z_1 |
  \end{smallmatrix}
:\prod_{i \geq 1} S(q^{\lambda_{1,i}} t^{\rho_i} q^{-1/2}
  z_1):\, :\prod_{i \geq 1} S(q^{\lambda_{2,i}} t^{\rho_i} q^{-1/2}
  z_2):
  \begin{smallmatrix}
    |\varnothing, -u z_2\rangle\\
    \otimes\\
    |\varnothing, - v / z_2\rangle
  \end{smallmatrix}
\end{multline}
Here $\mathcal{C}_1$ and $\mathcal{C}_2$ are the contours encircling
the two points $1$ and $\Lambda$. $P_1$ and $P_4$ denote the momenta
of the fields at points $0$, $\infty$. These momenta are actually
infinite in the pure gauge limit. However, the infinite charges at zero
and at infinity are compensated by the infinite number of screening
charges coming from the two exponentials, so that the dimension of the
field in the intermediate channel is \emph{finite} and equal to
$P$. In our formalism, $P$ is related to $\frac{u}{v}$ (since $u$
and $v$ are dimensions associated with the intermediate segments of the
horizontal lines) and the $\Lambda = \frac{z_1}{z_2}$.

Using Eq.~\eqref{eq:66}, we can evaluate the matrix element of the two
normal ordered factors in the last line of~\eqref{eq:18} to obtain the
Vandermonde determinant:
\begin{equation}
  \label{eq:22}
  \langle \varnothing|\otimes \langle \varnothing | :\prod_{i \geq 1} S(q^{\lambda_{2,i}}
  t^{\rho_i} q^{-1/2} z_2):\, :\prod_{i \geq 1} S(q^{\lambda_{1,i}}
  t^{\rho_i} q^{-1/2} z_1): |\varnothing\rangle \otimes
  |\varnothing\rangle \sim \Delta^{(q,t)} (\{ x \}, \{ y \})
\end{equation}
where
\begin{equation}
  \label{eq:69}
  x_i = q^{\lambda_{1,i}}
  t^{-i} z_1, \qquad y_i = q^{\lambda_{2,i}}
  t^{-i} z_2.
\end{equation}
Substituting the Vandermonde determinant back to Eq.~\eqref{eq:18}, one can verify
that what is left is a particular limit of $(q,t)$-matrix model with the
Chern-Simons terms:
\begin{multline}
  \label{eq:70}
  \mathcal{B}_{\mathrm{PG}}(P, \Lambda) =\\
  =\lim_{N_{1,2} \to \infty} \oint \oint d^{N_1} x\, d^{N_2} y\,
  \Delta^{(q,t)}(x)\Delta^{(q,t)}(y)\Delta^{(q,t)}(\{x\},\{y\})
  \prod_{i=1}^{N_1} x_i^{\alpha_1} e^{(M-L+1)\frac{(\ln x_i)^2}{\ln
      q}} \prod_{j=1}^{N_2} y_j^{\alpha_2} e^{(M-L+1)\frac{(\ln
      y_j)^2}{\ln q}}
\end{multline}
where $q^{\alpha_{1,2}}= \frac{qv}{u} z_{1,2}^{M-L+1} $. The parameter
$\Lambda = \frac{z_1}{z_2}$ is hidden inside the definition of the
contour integrals $\mathcal{C}_{1,2}$. Notice that the Chern-Simons
coupling constants depend on the relative slope of the two
``horizontal'' lines and, in particular, vanish for $L= M+1$, when we
also have $\alpha_1 = \alpha_2$.

Let us also give a spectral dual gauge theory interpretation for this
conformal block. In the AGT correspondence, this limit of the
conformal block corresponds to the pure $SU(2)$ gauge theory, with
$\Lambda$ being the instanton counting parameter (coupling constant)
and $P$ being related to the Coulomb modulus $a$. After applying the spectral
duality, however, we have a different interpretation: the coupling
constant $\Lambda$ and the Coulomb modulus $q^{2a}$ are
exchanged. This spectral dual approach is directly applicable to
Eq.~\eqref{eq:18}. If we simplify the infinite products~\eqref{eq:18}
(or equivalently in the Vandermonde factors in Eq.~\eqref{eq:70}) we
get:
\begin{equation}
  \label{eq:71}
  \mathcal{B}_{\mathrm{PG}} (P, \Lambda) = \sum_{\lambda_1,
    \lambda_2} \left( \frac{qv}{u} z_1^{M-L+1} \right)^{|\lambda_1|}
  \left( \frac{qv}{u} z_2^{M-L+1} \right)^{|\lambda_2|} (f_{\lambda_1}
  f_{\lambda_2})^{M-L+1} \frac{1}{z_{\mathrm{vec}}\left(
      \frac{z_1}{z_2}, \lambda_1, \lambda_2 \right)}
\end{equation}
where $z_{\mathrm{vec}}$ is the standard Nekrasov factor. Notice that
the whole sum becomes the Nekrasov function for the pure $SU(2)$ theory
(with additional ``framing'' factors in the case of general slopes $L
\neq M+1$). However, the instanton counting parameter and the Coulomb
modulus are related to $\frac{u}{v}$ and $\frac{z_1}{z_2}$
respectively, while, following the AGT duality it should be vice
versa. Thus, what we write in Eq.~\eqref{eq:71} is actually the
spectral dual of the AGT dual Nekrasov function corresponding to the
pure gauge limit of the conformal block.

Though this example is very simple in the gauge theory, as well as for
the webs of intertwiners, from the point of view of the CFT it looks a
bit contrived. The reason is that the corresponding diagram is not
balanced. Let us describe a more regular example of a balanced diagram
corresponding to a general four-point conformal block.

\paragraph{More general Virasoro conformal blocks.} To get the general
four-point conformal block $\mathcal{B}_4$, we should combine the two
balanced building blocks from Eq.~\eqref{eq:47}:

\begin{multline}
  \label{eq:72}
\parbox{7cm}{\includegraphics[width=7cm]{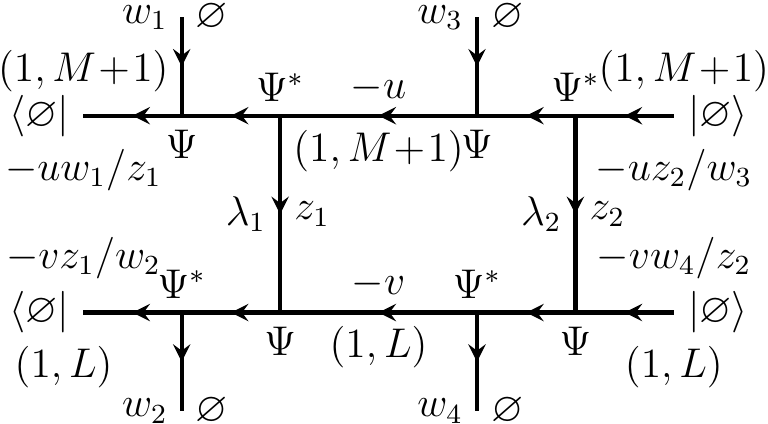}} = \langle P_1 | \exp \left( \oint_{\mathcal{C}_1} S(x) dx
  \right) V_{P_2}(\mathfrak{x}) \exp \left( \oint_{\mathcal{C}_2} S(x)
    dx \right) V_{P_3}(1) |P_4 \rangle =\\
  = \mathcal{B}_4 (P_1, P_2, P_3, P_4, P, \mathfrak{x})
  =\sum_{\lambda_1, \lambda_2} \left( \frac{qv}{u} (-z_1)^{M-L+1}
  \right)^{|\lambda_1|} \left(\frac{qv w_4 w_3}{u}(-z_2)^{M-L-1}\right)^{|\lambda_2|} \frac{f_{\lambda_1}^{M-L-1} f_{\lambda_2}^{M-L+1}
    q^{2 n(\lambda_1^{T}) + 2 n(\lambda_2^{T})}}{C_{\lambda_1}
    C_{\lambda_2} C'_{\lambda_1}
    C'_{\lambda_2}} \times\\
  \times
  \begin{smallmatrix}
    \langle \varnothing, -u z_1/w_1 |\\
    \otimes\\
    \langle \varnothing, -v w_2 / z_1 |
  \end{smallmatrix}
 V^{\mathrm{Heis}}_{w_1/w_2}\left((w_1 w_2)^{1/2}\right) V^{\mathrm{Heis}}_{w_3/w_4}\left((w_3 w_4)^{1/2}\right)
  :\prod_{i = 1}^{N_1} S(q^{\lambda_{1,i}} t^{\rho_i} q^{-1/2}
  z_1):\times\\
  \times V_{(tw_2)/(qw_1)}^{\mathrm{Vir}} \left( \left(
      w_1 w_2 q/t \right)^{1/2} \right) :\prod_{i = 1}^{N_2} S(q^{\lambda_{2,i}} t^{\rho_i} q^{-1/2}
  z_2): V_{(tw_4)/(qw_3)}^{\mathrm{Vir}} \left( \left(
      w_3 w_4 q/t \right)^{1/2} \right)
  \begin{smallmatrix}
    |\varnothing, -u w_3/ z_2\rangle\\
    \otimes\\
    |\varnothing, - v w_4 / z_2\rangle
  \end{smallmatrix}
\end{multline}
Here $t^{N_1} = \sqrt{\frac{t}{q}} \frac{z_1}{w_1}$ and $t^{N_2} =
\sqrt{\frac{t}{q}} \frac{z_2}{w_3}$. Notice that the Heisenberg vertex
operators commute with the Virasoro ones and also with the screening operators, so
that their contribution factorizes and adds the standard ``$U(1)$
factor'' to the conformal block.  Employing the scaling invariance
argument, one can consider only the conformal blocks in which the
position of the last Virasoro vertex operator is the identity, so that
$w_3w_4 q/t =1$ and $w_1w_2 q/t =\mathfrak{x}$. The dimensions of the
primary fields are given by
\begin{gather}
  \label{eq:73}
  P_1 = \frac{qv}{u},\qquad P_2 = \frac{tw_2}{q w_1}, \qquad P_3 =
  \frac{w_4}{w_3}, \qquad P_4 = \frac{u z_2^2}{qvw_4w_3} \\
  P = \frac{t^{N_1+1}P_2 }{q P_1} = \frac{t^{N_2} P_4}{P_3}
\end{gather}
The corresponding matrix model is of the Penner type with the additional
Chern-Simons terms:
\begin{multline}
  \label{eq:74}
  \mathcal{B}_4 (P_1, P_2, P_3, P_4, P, \mathfrak{x}) =\\
  =\oint \oint d^{N_1} x\, d^{N_2} y\,
  \Delta^{(q,t)}(x)\Delta^{(q,t)}(y)\Delta^{(q,t)}(\{x\},\{y\})
  \prod_{i=1}^{N_1} x_i^{\alpha_1} e^{(M-L+1)\frac{(\ln
      x_i)^2}{\ln q}} \frac{\left(q^{1-\alpha_2} x_i/\mathfrak{x}; q
    \right)_{\infty}}{\left( x_i/\mathfrak{x}; q \right)_{\infty}} \frac{\left(q^{1-\alpha_3} x_i; q
    \right)_{\infty}}{\left( x_i; q \right)_{\infty}} \times\\
  \times
  \prod_{j=1}^{N_2} y_j^{\alpha_4} e^{(M-L+1)\frac{(\ln y_j)^2}{\ln
      q}} \frac{\left( q^{1-\alpha_3} y_j; q \right)_{\infty}}{\left(
      y_j; q \right)_{\infty}} \frac{\left( q^{1-\alpha_2} y_j/\mathfrak{x}; q \right)_{\infty}}{\left(
      y_j/\mathfrak{x}; q \right)_{\infty}}
\end{multline}
where $q^{\alpha_i} = P_i$.

The five-point conformal block can be obtained by putting three building
blocks like~\eqref{eq:47} together. This gives a product of three
Virasoro vertex operators, three Heisenberg vertex operators and three
groups of screening charges. Schematically, one has:
\begin{multline}
  \label{eq:75}
  \parbox{9cm}{\includegraphics[width=9cm]{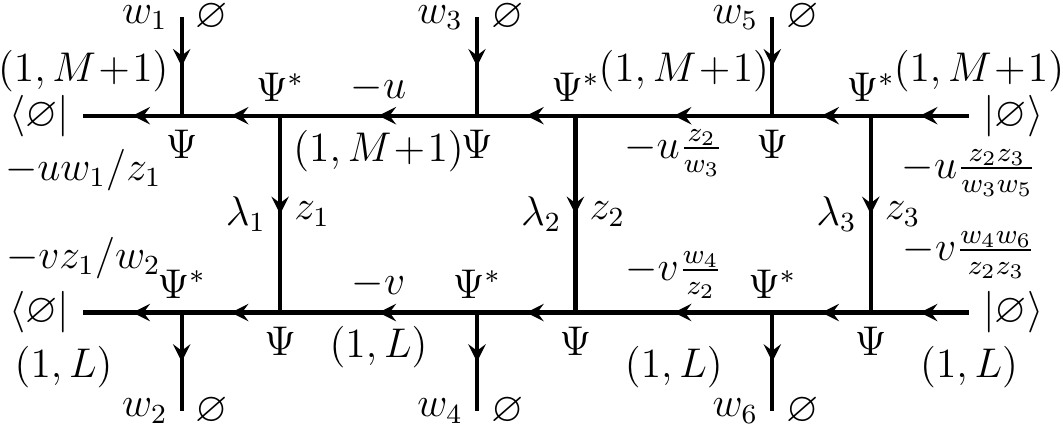}} = \\
  =\langle P_1 | \exp \left( \oint_{\mathcal{C}_1} S(x) dx
  \right) V_{P_2}(\mathfrak{x}_1) \exp \left( \oint_{\mathcal{C}_2} S(x)
    dx \right) V_{P_3}(\mathfrak{x}_2) \exp \left( \oint_{\mathcal{C}_3} S(x) dx
  \right) V_{P_4}(1) |P_5 \rangle
\end{multline}
where $V_P$ includes both the Virasoro and Heisenberg parts.

\bigskip

\paragraph{Conformal blocks of $W_{\K}$-algebra.} Generalizing our
formalism in another direction, we consider the $W_3$ algebra
conformal block. In this case, there are three horizontal lines and two
different types of screening currents $S_1$ and $S_2$, which
correspond to the vertical segments between the first and second or the second
and third lines respectively:
\begin{equation}
  \label{eq:77}
    S_i(x) = \exp \left\{ - \sum_{n \geq 1} \frac{1}{n}
  \frac{1-t^n}{1-q^n} \left( 1 + \omega^{2n} \right) x^n
  \tilde{\alpha}_{-n}^{(i)} \right\} \exp \left\{  \sum_{n \geq 1} \frac{1}{n}
  \frac{1-t^n}{1-q^n} \left( 1 + \omega^{2n} \right) x^{-n}
  \tilde{\alpha}_n^{(i)} \right\}
\end{equation}
where
\begin{align}
  \label{eq:78}
  \tilde{\alpha}_n^{(i)} &= \frac{1}{1 +\omega^{2n}} \Big(a_n^{(i)} - \omega^{n}
  a_n^{(i+1)}\Big), \quad n \geq 1\\
  \tilde{\alpha}_{-n}^{(i)} &= \frac{1}{1 + \omega^{2n}} \left(
    a_{-n}^{(i)} - \omega^{n} a_{-n}^{(i+1)} \right), \quad n \geq 1\notag
\end{align}
The simplest example is the pure gauge
limit of the four-point block, which is given by the following web
diagram:
\begin{multline}
  \label{eq:76}
  \parbox{7cm}{\includegraphics[width=7cm]{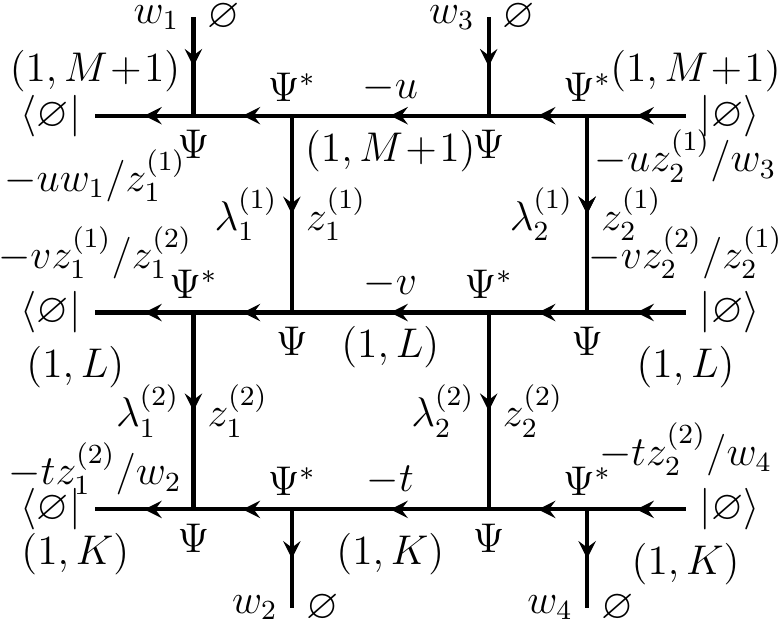}} =\\
  = \langle
  \vec{P}_1 | \exp \left( \oint_{\mathcal{C}_1} S_1(x) dx \right) \exp \left( \oint_{\mathcal{C}_1} S_2(x) dx \right) \exp \left(
    \oint_{\mathcal{C}_2} S_1(x) dx \right) \exp \left( \oint_{\mathcal{C}_2} S_2(x) dx \right) |\vec{P}_4 \rangle
\end{multline}

\subsection{Compactified network and the affine screening operator\label{affine}}

Let us also give an expression for the compactified diagram describing
the affine quiver gauge theory. The compactification identifies the
vertical line going down the lower edge of the diagram with the line
coming from the upper edge. Moreover, to get the general diagram, one
should add one more ingredient, the shift in the spectral
parameter. There is an automorphism of the DIM algebra, which
multiplies the spectral parameters of all lines (and all elements of
the algebra) by a constant. In general, the vertical compactification,
i.e.\ the trace over vertical representation can contain a ``twist''
by this automorphism, which does not spoil the nice intertwining
properties of the whole diagram. Taking the twist into account, one
arrives at the ``quasiperiodic'' compactification, where the lines
wrapping the compactification cylinder have their spectral parameters
shifted. The whole picture now looks as follows:
\begin{multline}
  \label{eq:91}
  \parbox{4cm}{\includegraphics[width=4cm]{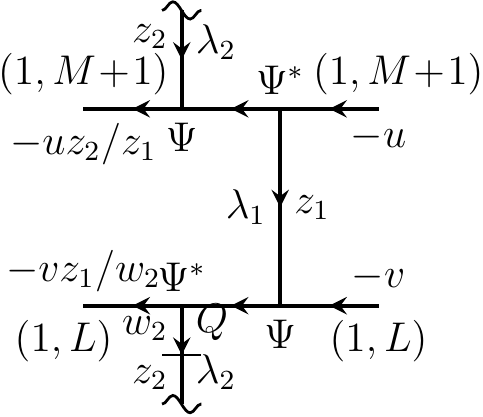}} =
  \sum_{\lambda_1, \lambda_2} \left( \frac{qv}{u} (-z_1)^{M-L+1}
  \right)^{|\lambda_1|} \left( \frac{qu w_2}{v z_2} Q (-z_2)^{L-M-1}
   \right)^{|\lambda_2|}
  \frac{f_{\lambda_1}^{M-L-1}f_{\lambda_2}^{M-L-1}
    q^{2n(\lambda_1^T)+2n(\lambda_2^T)}}{c_{\lambda_1} c_{\lambda_1}'
    c_{\lambda_2} c_{\lambda_2}'}\times\\
  \times:\prod_{i \geq 1}
  \widetilde{S}_2(q^{\lambda_{2,i}} t^{\rho_i} q^{-1/2} z_2):\, :\prod_{j \geq 1}
  S_1(q^{\lambda_{1,j}} t^{\rho_j} q^{-1/2} z_1):
\end{multline}
Here the wavy lines denote the identification of two vertical edges
and the shift automorphism is marked by a short horizontal line. The
automorphism shifts the spectral parameter of the line passing through
it by $\frac{w_2}{z_2}$ and simultaneously adds $Q$ to the length of
the corresponding edge. As in Eq.~\eqref{eq:94} one has two types of
screening operators $S_1$ and $\widetilde{S}_2$ corresponding to two
simple roots of $\widehat{A}_1$, however because of the shift, the
definition of the set of ``root'' oscillators inside the second
screening is different:
\begin{align}
    \tilde{\alpha}_n^{(2)} &= \frac{1}{1 + \omega^{2n}} (a_n^{(2)} - \omega^{n} (z_2/w_2)^n
  a_n^{(1)}), \quad n \geq 1\\
  \tilde{\alpha}_{-n}^{(2)} &= \frac{1}{1 + \omega^{2n}} \left(
    a_{-n}^{(2)} - \omega^{n} (w_2/z_2)^n a_{-n}^{(1)} \right), \quad n \geq 1\notag
\end{align}

Taking an average, i.e.\ using the Wick theorem, one arrives at the
affine $q$-Selberg matrix model~\cite{MMZdim}:
\begin{equation}
  \label{eq:93}
  \oint d^{N_1}x \, d^{N_2}y  \frac{\Delta^{(q,t)}(x)
    \Delta^{(q,t)}(y)}{\Delta^{(q,t)}(\{x\},\{y\})
    \Delta^{(q,t)}(\{y\}, \{\tilde{t}x\})} \prod_{i=1}^{N_1} x_i^{\alpha_1} e^{(M-L+1)\frac{(\ln
      x_i)^2}{\ln q}}  \prod_{j=1}^{N_2} y_j^{\alpha_2} e^{-(M-L+1)\frac{(\ln
      y_j)^2}{\ln q}}
\end{equation}
where the parameter of compactification $\tilde{t} =
\sqrt{\frac{q}{t}} \frac{z_2}{w_2}$, $q^{\alpha_1} = \frac{q
  v}{u}$, $q^{\alpha_2} = Q \frac{q
  u}{v}$. Notice the characteristic combination of the Vandermonde
factors in the measure, which is determined by the Cartan matrix of
the affine algebra $\widehat{A}_1$. This matrix model can also be
understood as the refined version of the ABJM matrix model \cite{CSmod}, in
particular, the level of the two Chern-Simons terms are opposite to
each other. On the other hand, the spectral dual of this network is
described by the elliptic DIM algebra (see Appendix A2).

If one can cuts the diagram (\ref{eq:91}) along the vertical
compactified line, one arrives at the regular Virasoro vertex
operator~(\ref{eq:47}). This is equivalent to the decompactification
limit $Q \to \infty$, since in this case only $\lambda_2 =
\varnothing$ contributes.

\section{The action of Virasoro and DIM$(\mathfrak{gl}_1)$\label{ref}}

There is a simple way~\cite{FHSSY-kernel} to build $q$-deformed
Virasoro or $W_{\K}$-algebras from DIM generators. To this end, one
considers the \emph{dressed} current $t(z)$:
\begin{equation}
  \label{eq:48}
  t(z) = \alpha(z) x^{+}(z) \beta(z)
\end{equation}
where
\begin{equation}
  \label{eq:27}
  \alpha(z) = \exp \left(  - \sum_{n \geq 1} \frac{1}{\gamma^n -
      \gamma^{-n}} b_{-n} z^n \right), \qquad \beta (z) = \exp \left( \sum_{n \geq 1} \frac{1}{\gamma^n -
      \gamma^{-n}} b_n z^{-n} \right)
\end{equation}
and $b_n$ are the modes of the $\psi^{\pm}$ generators:
\begin{equation}
  \label{eq:28}
  \psi^{\pm}(z) = \psi_0^{\pm} \exp \left( \pm \sum_{n \geq 1} b_{\pm
      n} \gamma^{n/2} z^{\mp n} \right).
\end{equation}
The dressing is needed to kill the extra Heisenberg part of the
algebra. The element $t(z)$ acts as a Virasoro current in the ${\K}$-fold
tensor product of Fock modules $\mathcal{F}_{u_1} \otimes \cdots
\otimes \mathcal{F}_{u_{\K}}$. One has:
\begin{equation}
  \label{eq:52}
  \rho_{u_1, \ldots, u_{\K}} (t(z)) = \sum_{i=1}^{\K} u_i \Lambda_i (z)
\end{equation}
where
\begin{equation}
  \label{eq:53}
  \Lambda_i(z) = \exp \left( \sum_{n \geq 1} \frac{1 - t^{-n}}{n} z^n
    \hat{\alpha}^{(i)}_{-n} \right) \exp \left( - \sum_{n \geq 1} \frac{1 - t^n}{n} z^{-n} \hat{\alpha}^{(i)}_n \right)
\end{equation}
The oscillators $\hat{\alpha}_n^{(i)}$ are defined as
\begin{equation}
  \label{eq:54}
  \hat{\alpha}_n^{(i)} = \hat{\hat{\alpha}}_n^{(i)} - \bar{\alpha}_n
\end{equation}
where (compare with \cite[s.4.2]{confMAMO2}
\begin{align}
  \label{eq:55}
  \hat{\hat{\alpha}}_n^{(i)} &= a_n^{(i)} \omega^{(i-1)n},\qquad n \geq
  1,\\
  \hat{\hat{\alpha}}_{-n}^{(i)} &= (1-\omega^{-2n}) \left( a_{-n}^{(1)} +
  a_{-n}^{(2)} \omega^{-n} + \ldots + a_{-n}^{(i-1)}
  \omega^{(2-i)n}\right) + a_{-n}^{(i)} \omega^{(1-i)n},\qquad n \geq
  1
\end{align}
and the Heisenberg part oscillators are given by
\begin{align}
  \label{eq:56}
  \bar{\alpha}_n &= \frac{1 - \omega^{-2n}}{1 - \omega^{-2n {\K}}} \left(a_n^{(1)} +
    a_n^{(2)} \omega^{-n} + \ldots + a_n^{({\K})} \omega^{(1-{\K})
      n} \right),\qquad n \geq
  1,\\
  \bar{\alpha}_{-n} &= \frac{1 - \omega^{-2n}}{1 - \omega^{-2n {\K}}}
  \left(a_{-n}^{(1)} + a_{-n}^{(2)} \omega^{-n} + \ldots +
    a_{-n}^{({\K})} \omega^{(1-{\K}) n} \right),\qquad n \geq 1,
\end{align}
Here $a^{(i)}_n$ acts in the $i$-th Fock module. Notice that we can
obtain the ``root'' bosons $\tilde{\alpha}_n^{(i)}$ from
$\hat{\alpha}_n^{(i)}$:
\begin{align}
  \label{eq:79}
  \tilde{\alpha}_n^{(i)} &= \frac{\omega^{(i+1)n}}{1 + \omega^{2n}}
  ( \hat{\alpha}^{(i)}_n -  \hat{\alpha}^{(i+1)}_n ),\\
  \tilde{\alpha}_{-n}^{(i)} &= \frac{\omega^{(1-i)n}}{1 + \omega^{2n}}
  ( \hat{\alpha}^{(i)}_{-n} -  \hat{\alpha}^{(i+1)}_{-n} ) \notag
\end{align}
Another useful property of the oscillators $\hat{\alpha}_n^{(i)}$ is
that they commute with the ``$U(1)$'' oscillators $\bar{\alpha}_n$:
\begin{equation}
  \label{eq:85}
  [\hat{\alpha}_n^{(i)}, \bar{\alpha}_m] = 0.
\end{equation}
The $W_{\K}$ algebra is built from the generators $\Lambda_i(z)$ by the
Miura transform:
\begin{equation}
  \label{eq:57}
  W_{K}(z) = \sum_{i_1 < \ldots < i_{K-1}} u_{i_1}\cdots u_{i_{K-1}}
  :\Lambda_{i_1}(z) \Lambda_{i_2}(\omega^2 z) \cdots \Lambda_{i_{K-1}}(\omega^{2(K-2)} z):
\end{equation}

The screening charges are built from the contractions of the DIM
intertwiners. Thus, they commute with any element of the DIM algebra,
e.g.\ with $t(z)$ by construction. This returns us to the definition of the
Virasoro algebra as the centralizer of the screening
charges~\cite{MMZdim}. Any element of the DIM algebra acts in the tensor
products of some of the representations corresponding to the lines of
the network. This can be described as an action in a particular section of
the diagram (see Fig.~\ref{fig:2}). The DIM element acts in the tensor
product of Fock modules associated with the legs intersected by the
dotted line. The section can be brought through the intertwiners, so that
eventually the element of the algebra acts on the external
lines. These external lines correspond to the vertex operators, and
the commutation with the intertwiners leads to the Ward identities for the
corresponding CFT or matrix model.

One should always be careful to include all the spaces, which are
intersected by the section. Let us give an example of commutation of
the DIM element with the contraction of two vertices. Pictorially we
have:
\begin{equation}
  \label{eq:84}
  \parbox{8cm}{\includegraphics[width=8cm]{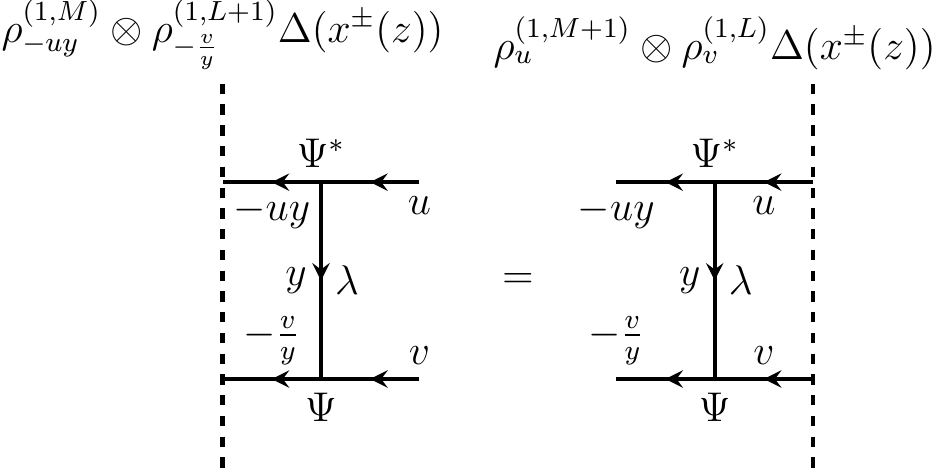}}
\end{equation}
This can be written out as follows:
\begin{multline}
  \label{eq:80}
\left[ \rho_{u_1} \otimes \rho_{u_2}( \Delta (
    x^{\pm}(\omega z ))),\sum_{\lambda} ||M_{\lambda}||^{-2}
    \left(\begin{smallmatrix}
      \Psi^{*}_{\lambda}(y)\\
      \otimes\\
      \Psi_{\lambda}(y)
    \end{smallmatrix}\right)
      \right]
  =\left[ \rho_{u_1}(x^{\pm}(\omega^{\xi_{\pm}}z))\otimes \rho_{u_2}(\Xi^{-\xi_{\mp}}(z)),\sum_{\lambda}
    ||M_{\lambda}||^{-2} \left(\begin{smallmatrix}
      \Psi^{*}_{\lambda}(y)\\
      \otimes\\
      \Psi_{\lambda}(y)
    \end{smallmatrix}\right)\right] +\\
+ \left[ \rho_{u_1} (\Xi^{\xi_{\pm}}(z))\otimes
  \rho_{u_2}(x^{\pm}(\omega^{\xi_{\mp}}z)),\sum_{\lambda} ||M_{\lambda}||^{-2}
  \left( \begin{smallmatrix}
      \Psi^{*}_{\lambda}(y)\\
      \otimes\\
      \Psi_{\lambda}(y)
    \end{smallmatrix}\right) \right] =\\
  =(\omega/q)^{\xi_{\mp}}\sum_{\lambda} \frac{q}{||M_{\lambda}||^2} \sum_{i=1}^{l(\lambda)}
  A_{\lambda,i}^{-} \delta (x_i  ty/(qz))\left( \begin{smallmatrix}
      \rho_{u_1}
  (\Xi^{\xi_{\pm}}(z))\Psi^{*}_{\lambda - \xi_{\pm}\cdot 1_i}(y)\\
      \phantom{\rho_{u_1}
  (\psi^{-} ((q/t)^{1/4}z))} \otimes\\
      \Psi_{\lambda-\xi_{\mp}\cdot 1_i}(y)\rho_{u_2}
  (\Xi^{-\xi_{\mp}}(z))
\end{smallmatrix}\right) +\\
+ (\omega/q)^{\xi_{\mp}}\sum_{\lambda} \frac{1}{||M_{\lambda}||^2} \sum_{i=1}^{l(\lambda)+1}
  A_{\lambda,i}^{+} \delta (x_i t y/z) \left(\begin{smallmatrix}
      \rho_{u_1}
  (\Xi^{\xi_{\pm}}(z))\Psi^{*}_{\lambda+\xi_{\mp}\cdot 1_i}(y)\\
      \phantom{\rho_{u_1}
  (\psi^{-} ((q/t)^{1/4}z))} \otimes\\
      \Psi_{\lambda + \xi_{\pm}\cdot 1_i}(y)\rho_{u_2}
        (\Xi^{-\xi_{\mp}}(z))
\end{smallmatrix}\right) = \\
= (\omega/q)^{\xi_{\mp}}\sum_{\lambda} \sum_{i=1}^{l(\lambda)+1} \left(
  \frac{q A_{\lambda+1_i,i}^{-}}{||M_{\lambda+1_i}||^2} +
  \frac{A_{\lambda,i}^{+}}{||M_{\lambda}||^2} \right) \delta (x_i t y/z) \left(\begin{smallmatrix}
      \rho_{u_1}
  (\Xi^{\xi_{\pm}}(z))\Psi^{*}_{\lambda+\xi_{\mp}\cdot 1_i}(y)\\
      \phantom{\rho_{u_1}
  (\psi^{-} ((q/t)^{1/4}z))} \otimes\\
      \Psi_{\lambda + \xi_{\pm}\cdot 1_i}(y)\rho_{u_2}
  (\Xi^{-\xi_{\mp}}(z))
\end{smallmatrix}\right) = 0
\end{multline}
where
\begin{align}
  \label{eq:81}
  A_{\lambda,i}^{+} &= (1-t) \prod_{j=1}^{i-1} \frac{\left( 1 - t
      \frac{x_i}{x_j} \right) \left( 1 - \frac{q}{t} \frac{x_i}{x_j}
    \right)}{\left( 1 - \frac{x_i}{x_j} \right) \left( 1 - q
      \frac{x_i}{x_j} \right)}\\
  A_{\lambda,i}^{-} & =(1-t^{-1}) \prod_{j=i+1}^{\infty} \frac{\left(
      1 - \frac{t}{q} \frac{x_i}{x_j} \right) \left( 1 - \frac{1}{t}
      \frac{x_i}{x_j} \right)}{\left( 1 - \frac{1}{q} \frac{x_i}{x_j}
    \right) \left( 1 - \frac{x_i}{x_j} \right)}
\end{align}
\be
\xi_+=1,\ \ \ \ \ \xi_-=0,\ \ \ \ \ \ \Xi^{\pm 1}(y)=\psi^{\mp}(\omega^{1/2}y),\ \ \ \ \ \ \Xi^0(y)=1,\ \ \ \ \ \ x_i = q^{\lambda_i} t^{-i}
\ee
and we remind that $\omega=\sqrt{q/t}$. The last line in
Eq.~\eqref{eq:80} vanishes because of a particular sum
rule for the norms of Macdonald polynomials. Commutation with
$\psi^{\pm}(z)$ can also be explicitly verified:
\begin{multline}
  \label{eq:87}
\rho_{u_1} \otimes \rho_{u_2} \Delta (\psi^{\pm}(y))  \sum_{\lambda} ||M_{\lambda}||^{-2} \left( \begin{smallmatrix}
       \Psi^{*}_{\lambda}(z)\\
      \otimes  \\
      \Psi_{\lambda}(z)
\end{smallmatrix}\right) =
\sum_{\lambda}  ||M_{\lambda}||^{-2} \left( \begin{smallmatrix}
      \rho_{u_1}
  (\psi^{\pm} (\omega^{\mp 1/2}y)) \Psi^{*}_{\lambda }(z)\\
      \phantom{\rho_{u_1}
  (\psi^{\pm} (y))}\otimes  \\
      \rho_{u_2}
  (\psi^{\pm} (\omega^{\pm 1/2} y)) \Psi_{\lambda}(z)
\end{smallmatrix}\right)=\\
=\sum_{\lambda} ||M_{\lambda}||^{-2} \left( \begin{smallmatrix}
    \Psi^{*}_{\lambda }(z)\rho_{u_1}
    (\psi^{\pm} (\omega^{\mp 1/2} y))\\
    \otimes \phantom{\rho_{u_1}
      (\psi^{\pm} (y))} \\
    \Psi_{\lambda}(z)\rho_{u_2} (\psi^{\pm} (\omega^{\pm 1/2}y))
\end{smallmatrix}\right) =   \sum_{\lambda} ||M_{\lambda}||^{-2} \left( \begin{smallmatrix}
       \Psi^{*}_{\lambda}(z)\\
      \otimes  \\
      \Psi_{\lambda}(z)
\end{smallmatrix}\right)  \rho_{u_1} \otimes \rho_{u_2} \Delta (\psi^{\pm}(y))
\end{multline}
All the commutation calculations above work by a similar mechanism,
summarized schematically in Fig.~\ref{fig:2},~b). The action of the
DIM element on the two horizontal representations is first transformed
into its action on the intermediate vertical segment and finally the
other side of the dashed line is also pulled through the vertex to get
the commutation.

The action of the element $t(z)$ gives the Ward identities of the
corresponding matrix model. This can be seen directly by computing the
operator product expansion of this current with the screening charges. For example in the
Virasoro case (see \cite{MMZdim} for details):
\begin{equation}
  \label{eq:82}
  \rho_{u_1}\otimes \rho_{u_2} (\Delta(t(z))) S(y) = \frac{1 - t
    \frac{y}{z}}{1 - \frac{y}{z}} :u_1 \Lambda_1(z)
  S(y): + t  \frac{1 - \frac{q}{t} \frac{y}{z}}{1 - q \frac{y}{z}} :u_2\Lambda_2(z) S(y):
\end{equation}
Since $t(z)$ commutes with the screening charges, in any correlator with these latter it can be brought through to the vacuum, which is annihilated by the negative modes of $t(z)$ . Following this logic, one gets
the matrix model Ward identities, the regularity of certain
averages:
\begin{equation}
  \label{eq:83}
  \left\langle K_{+}(z) \prod_i\frac{1 - t
      \frac{x_i}{z}}{1 - \frac{x_i}{z}} + K_{-}(z) \prod_i\frac{1 - \frac{q}{t}
      \frac{x_i}{z}}{1 - q \frac{x_i}{z}} \right\rangle = \mathrm{Regular}(z)
\end{equation}
where $K_{\pm}(z)$ are certain polynomials. This equation is the
appropriate $(q,t)$-deformations of the familiar Ward
identity~\eqref{WI3}.

\section{Vertical action of DIM\label{sec:vert-acti-viras}}

As was mentioned earlier, the vertical representation of the DIM
algebra has a combinatorial description in terms of Young diagrams, \cite{FT,FFJMM1}. We
have
\begin{align}
  \rho^{(0,1)}_u(x^{+}(z)) |M_{\lambda}^{(q,t)} \rangle &=
  \sum_{i=1}^{l(\lambda) + 1} A_{\lambda,i}^{+} \delta (x_i tu/z)
  |M_{\lambda+1_i}^{(q,t)} \rangle\\
    \rho^{(0,1)}_u(x^{-}(z)) |M_{\lambda}^{(q,t)} \rangle &=
  \omega \sum_{i=1}^{l(\lambda)} A_{\lambda,i}^{-} \delta (x_i tu/(qz)) |M_{\lambda-1_i}^{(q,t)} \rangle\\
    \rho^{(0,1)}_u(\psi^{+}(z)) |M_{\lambda}^{(q,t)} \rangle &=
  \omega B_{\lambda}^{+}(u/z)
  |M_{\lambda}^{(q,t)} \rangle\\
  \rho^{(0,1)}_u(\psi^{-}(z)) |M_{\lambda}^{(q,t)} \rangle &=
  \omega^{-1} B_{\lambda}^{-}(z/u) |M_{\lambda}^{(q,t)} \rangle
\end{align}
where $A^{\pm}_{\lambda,i}$ were defined in Eq.~(\ref{eq:81}) and
\begin{align}
  \label{eq:88}
  B_{\lambda}^{+}(z) &= \prod_{i=1}^{\infty} \frac{1 - zx_i}{1 -
    tzx_i}\frac{1 - \frac{t^2}{q} zx_i}{1 - \frac{t}{q} zx_i}\\
  B_{\lambda}^{-}(z) &= \prod_{i=1}^{\infty} \frac{1 -
    \frac{z}{x_i}}{1 - \frac{z}{tx_i}} \frac{1 -
    \frac{q}{t^2} \frac{z}{x_i}}{1 - \frac{q}{t} \frac{z}{x_i}}
\end{align}
where $x_i = q^{\lambda_i} t^{-i}$.

Similar action in the Yangian limit $q,t \to 1$ has been considered in
\cite{Mat1}, where the matrix model Ward identities or regularity condition for
$qq$-characters~(\ref{eq:83}) were derived from the intertwining
property with the Virasoro vertex operators. We should note here that the
interpretation of the vertex operators in this work was
\emph{spectrally dual} to our present consideration, i.e.\ the $SU(N)$
gauge theory corresponded there to the four-point conformal block of
the $W_N$ algebra as prescribed by the AGT relation.

In our formalism such an intertwining relation is natural: the vertex
operator is build out of the DIM intertwiners, which combinations
\emph{commute} with elements of DIM. However, this is only true unless
there are external legs. If we consider a horizontal section of the
web diagram, and try to move it between the ``layers'' of the diagram,
we necessarily encounter the external legs, or vertex operators in the
language of~\cite{Mat1}. The DIM generators \emph{do not} commute with the
intertwiners having external legs, since one should consider the
additional terms due to the action of DIM element on these
legs. However, these terms turn out to be tractable. Indeed, they
precisely reproduce the $qq$-character insertion into the matrix model
average. The matrix model arises from the sum over diagrams residing
on the legs intersected by the horizontal section.

\subsection*{Spectral duality and change of basis}
\label{sec:spectr-dual-change}

In \cite{MZ} it was shown that the change of preferred direction in
refined topological string is a nontrivial change of basis. The change
of basis is nontrivial in the sense that for the states on several
parallel legs the matrix of this transformation does not factorize
into a tensor product of matrices acting on each leg. Indeed, the
transformation is given by the spectral duality, and the two basis sets
are the standard Schur (or Macdonald) symmetric functions and
the generalized Macdonald polynomials \cite{gMac,AGTa}. This matrix was called generalized
Kostka function in \cite{MZ}.

In this subsection, we show how the spectral duality shows up in our present
algebraic approach. Let us consider the ``vertical'' basis in the
tensor product of vertical representations $\rho_{u_1, \ldots,
  u_K}^{(0,1)} = \rho_{u_1}^{(0,1)} \otimes \cdots \otimes
\rho_{u_K}^{(0,1)}$, which we define as the set of eigenvalues of the
particular DIM element $\psi^{+}_{-1}$:
\begin{equation}
  \label{eq:89}
\rho_{u_1, \ldots, u_K}^{(0,1)} \Delta^{K-1}  \psi^{+}_{-1}
|M^{(q,t)}_{\lambda_1} \rangle \otimes \cdots \otimes
|M^{(q,t)}_{\lambda_K} \rangle = (1-t)(1-t/q)\sum_{k=1}^K \left( u_k
\sum_{i=1}^{\infty} q^{\lambda_{k,i}}t^{-i} \right) |M^{(q,t)}_{\lambda_1} \rangle \otimes \cdots \otimes
|M^{(q,t)}_{\lambda_K} \rangle
\end{equation}
where $\psi_{-1}^{+} = \oint \psi^{+}(z) dz$. For generic $u_i$,
all eigenvalues are distinct, so this property defines the basis
uniquely. This basis is certainly very simple: it is a tensor product
of Macdonald polynomials.

Let us now perform the spectral duality. In the DIM algebra, this
corresponds to the $S$-transformation from $SL(2,\mathbb{Z})$ acting
on the generators of the algebra. Under this transformation,
$\psi_{-1}^{+}$ transforms into $x_0^{+} = \oint x^{+}(z) dz/z$. The
basis~(\ref{eq:89}) transforms into the basis of $x^{+}_0$ with the
same eigenvalues. However, the operator $x_0^{+}$ should be taken in
the new representation $\rho_{u_1, \ldots,
  u_M}^{(1,0)} = \rho_{u_1}^{(1,0)} \otimes \cdots \otimes
\rho_{u_K}^{(1,0)}$:
\begin{equation}
  \label{eq:90}
  \rho_{u_1, \ldots, u_K}^{(1,0)} \Delta^{K-1}  x^{+}_0
  |M^{(q,t)}_{\lambda_1,\ldots ,\lambda_K}(u_1, \ldots, u_K) \rangle = (1-t)(1-t/q)\sum_{k=1}^K \left( u_k
    \sum_{i=1}^{\infty} q^{\lambda_{k,i}}t^{-i} \right) |M^{(q,t)}_{\lambda_1,\ldots ,\lambda_K}(u_1, \ldots, u_K) \rangle
\end{equation}
We recognize the operator $\rho_{u_1, \ldots, u_K}^{(1,0)}
\Delta^{K-1} x^{+}_0$: this is just the generalized Macdonald
Hamiltonian. The eigenvalues also match, so the new basis
$|M^{(q,t)}_{\lambda_1,\ldots ,\lambda_K}(u_1, \ldots, u_K) \rangle$
is the basis of \emph{generalized} Macdonald polynomials. The generalized
Kostka functions are just representations of this $SL(2,\mathbb{Z})$
transformation.

\section{Conclusion}

In this paper, we presented technical details on evaluation of the Nekrasov functions
and their symmetries (including the $qq$-character correlators)
from the free field formalism for the DIM($gl_1$) algebra.
This is a very powerful method, but it is only at the first stage of development.
There are several restrictions which should be consequently lifted at the next stages.
If one considers them as a consequent specification of representation types, the list
should be read in inverse order.

\begin{itemize}
\item
The construction that admits fermionization of intertwiners $\Psi$ and $\Psi^*$ at the level one of DIM is much similar to that for the level $k=1$ Kac-Moody algebras. Hence, one could expect a straightforward generalization to arbitrary level
{\it a la} \cite{GMMOS} involving analogues of the $b,c$-systems.
Note, however, that the requirement on the level does not restrict the value of
the Virasoro central charge regulated by $\beta$: all matrix models and $\beta$-ensembles
and, hence, the generic Liouville and $W_K$-conformal blocks are already handled
by the existing formalism.
Also, at this level the difference disappears between the vertex operators (in particular, the screening charges) and
the stress tensors (including the W-operators):
all these are described by exponentials of the free fields,
the differences emerge only in the limit $q,t\longrightarrow 1$.

\item The formalism is best developed for the intertwiners, which act as operators
between the two "horizontal" Fock modules ${\cal F}^{(1,L)}$ and ${\cal F}^{(1,L\pm 1)}$,
while the third representation is the "vertical" leg associated with ${\cal F}^{(0,\pm 1)}$.
Such a non-symmetricity is inevitable since the resulting topological
vertex of \cite{IKV} is still asymmetric and remembers about the distinguished vertical
direction.
Technically this restricts consideration to the {\it balanced} networks, what makes
many important models, including the quiver ones, treatable only via additional
application of the spectral duality.

\item A better treatment should involve infinitely many free fields, giving rise
to MacMahon type modules, what should also allow one to define skew intertwiners,
where all the three legs are non-vertical.
An existing description of the MacMahon modules is pure combinatorial, in terms
of $3d$ Young diagrams (plane partitions).
A naive free field formalism would involve fields depending on two coordinates
instead of one, and this requires a far-going generalization of holomorphic
fields used in the ordinary $2d$ CFT.
Such a formalism is now developing, also with the motivation coming from MHV amplitudes,
but its incorporation into the DIM representation theory is a matter of future.
Still, it seems important for a full understanding of the spectral dualities
and of generic networks, including the sophisticated ones from \cite{NfNc}.
They can be treated by the existing formalism, but it leaves the underlying symmetries
well hidden: they show up only in answers, but not at any of the intermediate stages.

\item A further challenge is further generalization from DIM($\mathfrak{gl}_1$) to DIM($\mathfrak{gl}_n$)
and the triple-Pagoda algebras DIM($\widehat{\mathfrak{gl}_1})$ and DIM($\widehat{\mathfrak{gl}_n})$.
An intriguing problem (see Appendix A3) is that already DIM($\mathfrak{gl}_n$)
is built from the {\it affine} Dynkin diagram of $\widehat{\mathfrak{gl}_n}$, thus, the triple-affine
generalization should involve more sophisticated Dynkin diagrams.
\end{itemize}

\bigskip

We hope that the present text can serve as a good introduction in the DIM-based
generalization of conformal theories, where the conformal blocks are the
generic Nekrasov functions and the Ward identities are the associated regularity conditions for $qq$-characters.
We hope that it will help to attract more attention
to emerging challenging problems, which we have just enumerated.
Technical means for this seem to be already at hand.

\newpage

\section*{Appendix. Properties of the DIM algebras and their limits}
\label{sec:abstr-algebr-descr}

In this Appendix, we describe the algebraic structures of DIM algebras and their degenerations.

\subsection*{A1. Constructing DIM($\mathfrak{gl}_1$) from $W_{1+\infty}$ algebra}

Let us discuss how one can construct DIM($\mathfrak{gl}_1$) starting from the algebra of difference operators, \cite{M,Aw}.

\paragraph{Algebra $W_{1+\infty}$.}

Consider the algebra $W_{1+\infty}$  (as usual, $1+\infty$ refers here to adding the Heisenberg algebra to $W_\infty$) given by the generators $W^k_n = W(z^n{\cal D}^k)$, $n\in \mathbb{Z},\ k\in \mathbb{Z}_{\geq 0}$, where ${\cal D}=z\partial_z$. One can consider the central extension of this algebra:
\be
\l[W(z^n{\cal D}^k),W(z^m{\cal D}^l)] =W\Big([z^n{\cal D}^k,z^m{\cal D}^l]\Big) + c\delta_{n+m,0} \cdot \psi_{n,kl}, \nn \\
\psi_{n,kl} = \left\{\begin{array}{ccc} \sum_{j=1}^n (-j)^k(n-j)^l, && n>0  \\
0 && n=0 \end{array}\right.
\ee
or, in the different basis of $W^k_n = W(z^nD^k)$ with $D\equiv \mathfrak{t}^{\cal D}$ (see (\ref{D})),
\be\label{eq:167}
\l[W(z^nD^k),W(z^mD^l)] =(\mathfrak{t}^{mk}-\mathfrak{t}^{nl})W\Big(z^{n+m}D^{k+l}\Big)-c\delta_{n+m,0}{\mathfrak{t}^{mk}-\mathfrak{t}^{nl}\over \mathfrak{t}^{k+l}-1}
\ee
Note that, if $ k+l \neq 0 $,
the second term in the right hand side of (\ref{eq:167}) can be absorbed into
the first term by redefining the generators $W({\cal D}^k)$ with $k\neq 0$: $W(D^k) \to W(D^k) - \displaystyle{{c \over \mathfrak{t}^k - 1 }}$,   $k\neq 0$. However, at $k+l=0$ this term can not be absorbed and is equal to $ n c \mathfrak{t}^{-nk} \delta_{n+m,0} \delta_{k+l,0} $, see (\ref{wbar}).

\paragraph{Algebra $\overline{W_{1+\infty}}$.}

The next step is to consider the algebra $\overline{W_{1+\infty}} = \Big\{W(z^n\mathfrak{t}^{k{\cal D}}), \ n,k \in \mathbb{Z}\Big\}$, which is a double of the $W_{1+\infty}$ and may have {\it two} central extensions:
\be\label{wbar}
\l[W(z^nD^k),W(z^mD^l)] =(\mathfrak{t}^{mk}-\mathfrak{t}^{nl})W\Big(z^{n+m}D^{k+l}\Big) +
\mathfrak{t}^{-nk}(nc_1+kc_2)\delta_{m+n,0}\delta_{k+l,0}
\ee

\paragraph{Automorphisms.}

The algebra $\overline{W_{1+\infty}}$, (\ref{wbar}) has the evident automorphisms $\sigma, \tilde\sigma$ and
$\tau$ defined by
\begin{align}
&\sigma(W_n^k)= \mathfrak{t}^{-nk} W_k^{-n}\,,
&&\sigma(c_1)=-c_2\,,
&&\sigma(c_2)=c_1\,,
\nonumber \\
&\tilde\sigma(W_n^k)= -W_{k}^{n}\,,
&&\tilde\sigma(c_1)=c_2\,,
&&\tilde\sigma(c_2)=c_1\,,
\nonumber \\
&\tau(W_n^k)= \mathfrak{t}^{\frac{1}{2}n^2} W_{n}^{k+n}\,,
&&\tau(c_1)=c_1+c_2\,,
&&\tau(c_2)=c_2\,.
\end{align}
In particular, $\sigma$ and $\tau$ form $SL(2,\mathbb{Z})$ acting
on two central charges $c_1$ and $c_2$\,.

\paragraph{Heisenberg subalgebras.}

By the commutation relations \eqref{wbar}, it is easy to see that it contains
a Heisenberg subalgebra generated by $\{W_n^0, c_1\}_{n\in \mathbb{Z}}$
satisfying
\begin{align}
[W_n^0, W_m^0] = n c_1 \delta_{n+m,0}\,.
\end{align}
From the viewpoint of the root lattice of $\overline{W_{1+\infty}}$\,,
this can be seen as the {\it vertical} embedding of the Heisenberg algebra.
By using the automorphisms $\sigma$ and $\tau$ in the above,
it is easy to find the {\it horizontal} and the embedding with arbitrary {\it slope}
$\alpha\in \mathbb{Z}$ as follows;
\begin{align}
[W_0^{n}, W_0^{m}]
&= n c_2 \delta_{n+m,0}\,, \nonumber \\
[W_n^{\alpha n}, W_m^{\alpha m}]
&= n \mathfrak{t}^{-\alpha n^2} (c_1+\alpha c_2) \delta_{n+m,0}\,.
\end{align}

\paragraph{Chevalley generators and Serre relations.}

The generators $W_n^{\pm,0}=W(z^nD^{\pm 1,0})$ form a closed subalgebra:
\be\label{preDIM}
\left[W_n^+,W_m^-\right]=(\mathfrak{t}^m-\mathfrak{t}^{-n})W_{m+n}^0+(nc_1+c_2)\mathfrak{t}^{-n}\delta_{n+m,0}\nn\\
\left[W_n^0,W_m^{\pm}\right]=(1-\mathfrak{t}^{\pm n})W_{m+n}^{\pm}\nn\\
\left[W_n^0,W_m^0\right]=nc_1\delta_{n+m,0}
\ee
One can generate the whole algebra from this subalgebra provided the Serre relations are added:
\be\label{preDIMSerre}
\Big[W_n^\pm,[W_{n+1}^\pm,W_{n-1}^\pm]\Big]=0
\ee

\paragraph{Quantization: from $\overline{W_{1+\infty}}$ to DIM($\mathfrak{gl}_1$).}

This algebra can be deformed with the deformation parameter $\mathfrak{q}$. Let us denote the deformed (properly rescaled) generators through $W_n^0\to x^0_n$, $W_n^\pm\to x_n^\pm$. Then,
\be
\left[x^0_n,x^\pm_m\right]=\mp {\kappa_n\over n}\mathfrak{q}^{(n\pm |n|)c_1/2}x^\pm_{n+m}\nn\\
\left[x^0_n,x^0_m\right]=-{\kappa_n\over n}{\mathfrak{q}^{c_1n}-\mathfrak{q}^{-c_1n}\over \mathfrak{q}-\mathfrak{q}^{-1}}\delta_{n+m,0}\nn\\
\left[x^+_n,x^-_m\right]={1\over\kappa_1}(\mathfrak{q}^{c_2+nc_1}\psi^+_{n+m}-\mathfrak{q}^{-nc_1-c_2}\psi^-_{n+m})
\ee
where
\be
\kappa_n\equiv (q_1^n-1)(q_2^n-1)(q_3^n-1),\ \ \ \ \ \ \ \ q_1=\mathfrak{t}^2, \ \ \ q_2=\mathfrak{q}^{-2}\mathfrak{t}^{-2}, \ \ \ q_3=\mathfrak{q}^{2}\ \ \ \ (q_1q_2q_3=1)
\ee
and
\be
\sum_{k=0}^\infty \psi_k^\pm z^{\mp k}\equiv \mathfrak{q}^{\mp c_2}\exp\left(\pm\sum_{n=1}^\infty x^0_{\pm n}z^{\mp n}\right)
\ee
Introducing the series of generators,
\be
\psi^\pm (z)={(1-\mathfrak{q}^2)(1-\mathfrak{q}^{-2})\over\kappa_1^2}\ \sum_{k=0}^\infty \psi_k^\pm \mathfrak{q}^{-c_1k/2}z^{\mp k},\ \ \ \ \ \ \ x^\pm (z)=\sum_{n\in\mathbb{Z}}x^\pm_nz^{-n}
\ee
we immediately come to the DIM($\mathfrak{gl}_1$) algebra of s.\ref{sec:dim-algebra} upon identification $q_1= q$, $q_2= t^{-1}$.

\paragraph{Free field realization.}

At the values of central charges $(c_1,c_2)=(1,0)$, the constructed DIM algebra has the deformed affine $U(1)$ subalgebra so that the generators are realized in its terms as
\be
x^+(z) = \exp\left(\sum_{n>0}\frac{1-t^{-n}}{n}\cdot z^n\ttau_n\right)\cdot
\exp\left(\sum_{n>0} (q^n-1)z^{-n}\p_{\ttau_n}\right), \nn \\
x^-(z) = \exp\left(-\sum_{n>0}\frac{1-t^{-n}}{n}\cdot \omega^{-n}z^n\ttau_n\right)\cdot
\exp\left(-\sum_{n>0} (q^n-1)\omega^{-n}z^{-n}\p_{p_n}\right), \nn \\
\psi^+(z) =
\exp\left(\sum_{n>0} (q^n-1)(1-\omega^{-2n})z^{-n}\omega^{n/2}\p_{\ttau_n}\right), \nn \\
\psi^-(z) = \exp\left(\sum_{n>0}\frac{1-t^{-n}}{n}(1-\omega^{-2n})\omega^{n/2}\cdot  z^n\ttau_n\right)
\ee
After the Miwa transform of variables  $\ttau_n=\sum_{i=1}^N z_i^n$,
these expressions reduce to the Macdonald operators
\begin{align}
&(t^{\pm 1}-1)
\sum_{i=1}^N \prod_{j(\neq i)} \frac{t^{\pm 1}z_i-z_j}{z_i-z_j} \cdot
z_i^n q^{\pm {\cal D}_i}
\nn\\
&=
\oint \frac{dz}{2\pi i z} z^{-n}
\left\{ t^{\pm N}
\exp\left(\sum_{n>0} \frac{1-t^{\mp n}}{n} z^{ n} \ttau_{ n} \right)
-
\exp\left(\sum_{n>0} \frac{1-t^{\pm n}}{n} z^{-n} \ttau_{-n} \right)
\right\}
\exp\left(\sum_{n>0} (q^{\pm n}-1) z^{-n}\p_{\ttau_n}\right)
\label{eq:MacOpBoson}
\end{align}
with ${\cal D}_i:=z_i \frac{\partial}{\partial z_i}$.
Here we use the identity
\be
(t-1)
\sum_{i=1}^N \prod_{j(\neq i)} \frac{tz_i-z_j}{z_i-z_j} \cdot
\delta(z_i z)
=
t^{N}
\exp\left(\sum_{n>0} \frac{1-t^{-n}}{n} z^{ n} \ttau_{ n} \right)
-
\exp\left(\sum_{n>0} \frac{1-t^{n}}{n} z^{-n} \ttau_{-n} \right)
\nn
\ee
with $\delta(x):=\sum_{n\in\mathbb Z}x^n$.
Note that the second term in the r.h.s. of (\ref{eq:MacOpBoson}) 
vanishes for $n>0$
so that
\be
(t^{\pm 1}-1)
\sum_{i=1}^N \prod_{j(\neq i)} \frac{t^{\pm 1}z_i-z_j}{z_i-z_j} \cdot
z_i^n q^{\pm {\cal D}_i}
=
(qt)^{\frac{n\mp n}{4}}
t^{\pm N} x^{\pm}_{-n} -\delta_{n,0}
\ee
with $n\geq 0$. Similarly, at the values of central charges $(c_1,c_2)=(2,0)$ this DIM algebra contains a $q$-deformed subalgebra (Virasoro $\otimes\ \widehat{U(1)})$  (and is realized by two free fields), at $(c_1,c_2)=(3,0)$ it contains a $q$-deformed subalgebra ($W^{(3)}\otimes\widehat{U(1)})$  (and is realized by three free fields), etc.

\subsection*{A2. Elliptic DIM($\mathfrak{gl}_1$) algebra}
\label{sec:dim-algebra-1}

Elliptic version of DIM algebra is generated by the same set of
operators as the ordinary DIM: $x^{\pm}(z)$, $\psi^{\pm}(z)$ and the
central element $\gamma$. The relations are a copy of
Eq.~\eqref{eq:24}, except for the $[x^{+}, x^{-}]$ relation, which
changes to
\begin{equation}
  \label{eq:2}
  \l[x^+(z),\, x^-(w)]\ =\
  \frac{\Theta_{q'} (q; q') \Theta_{q'} (t^{-1};
    q')}{(q';q')_{\infty}^3 \Theta_{q'} (q/t ; q')}\,\Big(\delta(\gamma^{-1}z/w)\,\psi^+(\gamma^{1/2}w)
  \ -\ \delta(\gamma z/w)\,\psi^-(\gamma^{-1/2}w)\Big)
\end{equation}
where $\Theta_p(z) = (p;p)_{\infty} (z;p)_{\infty} (p/z;p)_{\infty}$
is the theta-function. Also, most importantly, the structure function
$G^{\pm}(z)$ is now not trigonometric, but elliptic:
\begin{equation}
  \label{eq:49}
  G^{\pm}_{\mathrm{ell}}(z) = \Theta_p(q^{\pm 1}z)\Theta_p(t^{\mp 1}z)
    \Theta_p(q^{\mp 1} t^{\pm 1} z),
\end{equation}
The comultiplication $\Delta$ is exactly the same as in the
trigonometric case, given by Eqs.~\eqref{eq:23}. The essential difference with the
trigonometric case appears when one tries to build Fock representation
of elliptic DIM: one set of bosons turns out not to be enough. One needs
at least \emph{two} sets of Heisenberg generators $\hat{a}_n$ and
$\hat{b}_n$ to reproduce the commutation relations of the elliptic
algebra. Concretely, we have for the level one representation:
\begin{gather}
  \rho_u (x^{+}(z)) = u \eta(z) = u\ : \exp \left( - \sum_{n \neq 0}
    \frac{(1-t^n) z^{-n}}{n (1 - q'^{|n|})} \hat{\mathfrak{a}}_n \right)
  \exp \left( - \sum_{n \neq 0}
    \frac{(1-t^{-n}) q'^{|n|} z^n}{n (1 - q'^{|n|})} \hat{\mathfrak{b}}_n \right): \notag\\
  \rho_u (x^{-}(z)) = u^{-1} \xi(z) = u^{-1} : \exp \left( \sum_{n
      \neq 0} \frac{(1-t^n) \omega^{-|n|} z^{-n}}{n (1 - q'^{|n|})}
    \hat{\mathfrak{a}}_n \right) \exp \left( \sum_{n \neq 0}
    \frac{(1-t^{-n}) \omega^{|n|} q'^{|n|} z^n}{n (1 - q'^{|n|})} \hat{\mathfrak{b}}_n \right):\notag\\
  \rho_u (\psi^{+}(z)) = \varphi^{+}(z) = \exp \left( \sum_{n > 0}
    \frac{(1-t^n) (\omega^{-n} - \omega^{n}) \omega^{-n/2} }{n (1 - q'^n)} \left(
      z^{-n} \hat{\mathfrak{a}}_n - \omega^{n} q'^n z^n
      \hat{\mathfrak{b}}_n\right)
  \right)   \label{eq:50}\\
  \rho_u (\psi^{-}(z)) = \varphi^{-}(z)= \exp \left( -\sum_{n > 0}
    \frac{(1-t^{-n}) (\omega^{-n} - \omega^{n}) \omega^{-n/2} }{n (1 - q'^n)}
    \left( z^n \hat{\mathfrak{a}}_{-n} - \omega^{n} q'^n z^{-n}
      \hat{\mathfrak{b}}_{-n} \right)
  \right)  \ \ \ \ \ \ \
  \rho_u(\gamma) = \left( t/q \right)^{1/2}
\end{gather}
where the bosons $\hat{\mathfrak{a}}_n$ and
$\hat{\mathfrak{b}}_n$ satisfy the following commutation relations:
\be\label{eq:8}
  [\hat{\mathfrak{a}}_m , \hat{\mathfrak{a}}_n] = m \frac{(1 - q'^{|m|})(1 -
    q^{|m|})}{1 - t^{|m|}} \delta_{m+n,0},\notag\ \ \ \ \ \ \
  [\hat{\mathfrak{b}}_m , \hat{\mathfrak{b}}_n] = m \frac{(1 - q'^{|m|})(1 -
    q^{|m|})}{(p q')^{|m|} (1 - t^{|m|})} \delta_{m+n,0},\ \ \ \ \  \ \
  [\hat{\mathfrak{a}}_m , \hat{\mathfrak{b}}_n] = 0.\notag
\ee

\bigskip

The \emph{dressed} current $t(z) = \alpha(z) x^{+}(z) \beta(z)$, corresponding
to the stress energy tensor is given by exactly the same
expression~\eqref{eq:48}, as in the ordinary DIM case. Moreover, the
dressing operators $\alpha(z)$ and $\beta(z)$ are constructed from the
$\psi^{\pm}$ generators of the elliptic DIM algebra using the same
formulas~\eqref{eq:27} as give above. In the level two representation
$\rho_{u_1, u_2}^{(2)}$ the element $t(z)$ produces the elliptic
Virasoro stress-energy tensor
\begin{equation}
  \label{eq:12}
  \mathcal{T}(z) =\ : e^{\hat{\Phi}(z)} e^{-\hat{\Phi}(t^{-1}z)}: +\ t :
e^{-\hat{\Phi}(tz/q)} e^{\hat{\Phi}(z/q)}:
\end{equation}
where
\begin{equation}
  \label{eq:11}
  \hat{\Phi}(z) = \sum_{n \neq 0} \frac{z^n}{n (1 - q'^{|n|})}
  {\hat{\mathfrak{a}}_{-n}\over\sqrt{1+\omega^{|n|}}} - \sum_{n \neq 0} \frac{z^{-n}}{n (1 - q'^{|n|})}(\omega^2q')^{|n|/2} \hat{\mathfrak{b}}_{-n}
\end{equation}
Let us also mention that the \emph{undressed} elliptic DIM charge
$\oint x^{+}(z) dz/z$ also leads to several very interesting
objects. In the level one representation it gives elliptic Ruijsenaars
Hamiltonian, while in the second level representation it is the
difference version of the intermediate long-wave
Hamiltonian~\cite{ILW}, which itself is a generalization of the
Benjamin-Ono system.

\subsection*{A3. Rank $>1$:\  DIM($\mathfrak{gl}_n$) = quantum toroidal algebra of type $\mathfrak{gl}_n$}
\label{sec:dimN-algebra}

In complete parallel with the previous consideration, DIM($\mathfrak{gl}_n$) emerge as a deformation of the universal enveloping algebra of
the Lie algebra $A_n = {\rm Mat}_n\otimes C[z^{\pm 1},D^{\pm 1}]$
with
\be\label{D}
D=q_1^{z\frac{\p}{\p z}}
\ee
i.e. of $n\times n$ matrices with entries being elements of the algebra of functions on the quantum torus, $zD=q_1Dz$. The deformation of $A_{\cal N}$ introduces another parameter, $q_2$. Providing this deformed algebra with two-dimensional central extension, one arrives at DIM($\mathfrak{gl}_n$).

The set of generators of DIM($\mathfrak{gl}_n$) is $E_{ik},F_{ik},H_{ir},K_{i0}^\pm,q^{\pm c}$
with $k\in \mathbb{Z}$, $r\in \mathbb{Z}/\{0\}$, $0\leq i \leq n-1$.
The generating functions (currents) are:
\be
E_i(z) = \sum_{k\in \mathbb{Z}} E_{ik}z^{-k}, \nn \\
F_i(z) = \sum_{k\in \mathbb{Z}} F_{ik} z^{-k}, \nn \\
K^{\pm}_i(z) = K_{i0}^{\pm 1} \exp\left(\pm(\mathfrak{q}-\mathfrak{q}^{-1})\sum_{r=1}^\infty H_{i,\pm r} z^{\mp r}\right)
\ee
The two centers are $\mathfrak{q}^c$ and $\kappa = \prod_{i=0}^{n-1} K_{i0}$.

The commutation relations are
\be
d_{ij}G_{ij}(z,w) \,E_i(z) E_j(w) + G_{ji}(w,z)\, E_j(w)E_i(z) = 0, \nn \\
d_{ij}G_{ij}(z,w) \,K_i^\pm\Big(\mathfrak{q}^{(1\mp 1)c/2}z\Big) E_j(w)
+ G_{ji}(w,z)\, E_j(w)K^\pm_i\Big(\mathfrak{q}^{(1\mp c)/2}z\Big) = 0, \nn \\
d_{ji}G_{ji}(z,w) \,F_i(z) F_j(w) + G_{ij}(w,z)\, F_j(w)F_i(z) = 0, \nn \\
d_{ji}G_{ji}(z,w) \,K_i^\pm\Big(\mathfrak{q}^{(1\pm 1)c/2}z\Big) F_j(w)
+ G_{ij}(w,z)\, F_j(w)K^\pm_i\Big(\mathfrak{q}^{(1\pm c)/2}z\Big) = 0, \nn \\ \nn \\
\Big[ E_i(z), \, F_j(w)\Big] = \frac{\delta_{ij}}{\mathfrak{q}-\mathfrak{q}^{-1}}\left(
\delta\left(\frac{\mathfrak{q}^cw}{z}\right)K^+_i(z) - \delta\left(\frac{\mathfrak{q}^cz}{w}\right)K^-_i(w)\right)\nn\\
{G_{ij}(\mathfrak{q}^{-c}z,w)\over G_{ij}(\mathfrak{q}^cz,w)}K^-_i(z)K^+_j(w)={G_{ji}(w,\mathfrak{q}^{-c}z)\over G_{ji}(w,\mathfrak{q}^cz)}K^-_i(z)K^+_j(w)\nn\\
\left[K_i^{\pm}(z),K^{\pm}_j(w)\right]=0
\ee
where, in variance with the DIM($\mathfrak{gl}_1$)-case,
\be
q_1=\mathfrak{t}\mathfrak{q}^{-1}, \ \ \ q_2=\mathfrak{q}^2, \ \ \ q_3=\mathfrak{t}^{-1}\mathfrak{q}^{-1}
\ee
and powers of $\mathfrak{q}$ are made from entries of the Cartan matrix.
The commutation relations can be added with the Serre relations
\be
{\rm for}\ n\geq 3 & {\rm sym}_{z_1,z_2} \
\left[E_i(z_1), \ \Big[E_i(z_2),E_{i\pm 1}(w)\Big]_\mathfrak{q}\right]_{\mathfrak{q}^{-1}}=0\nn\\
{\rm for}\ n=2 & {\rm sym}_{z_1,z_2,z_3} \
\left[E_i(z_1), \left[E_i(z_2), \Big[E_i(z_3),E_{i\pm 1}(w)\Big]_{\mathfrak{q}^2}\right]_{\mathfrak{q}^0=1}\right]_{\mathfrak{q}^{-2}}=0
\ee
and similarly for $F$.
The $q$-commutator is $[A,B]_\mathfrak{q}=AB-\mathfrak{q}BA$.

The comultiplication is the same as for DIM($\mathfrak{gl}_1$).

The structure functions are build from the {\it affine} Dynkin diagrams and for $\mathfrak{gl}_n$-case are defined
as follows:
\begin{itemize}
\item
for the simply laced case $n\geq 3$
\be
\parbox{5cm}{\includegraphics[width=4cm]{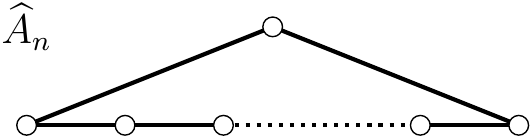}}
G_{ij}(z,w) = \left\{\begin{array}{ccc}
(z-q_1w) & {\rm for} & i=j-1 \\
(z-q_2w) & {\rm for} & i=j \\
(z-q_3w) & {\rm for} & i=j+1 \\
(z-w) & {\rm for} & i\ne j,j\pm 1
\end{array}\right. \nn \\
d_{ij} = \left\{\begin{array}{ccc}
\mathfrak{t}^{\pm 1} & {\rm for} & i=j\pm 1,\ n\ge 3 \\
1 & & {\rm otherwise}
\end{array}\right.
\ee
\item
The affine Dynkin diagram for $n=2$ is not simply laced, and in this case
\be
G_{00}^{\mathfrak{gl}_2}(z,w) =  G_{11}^{\mathfrak{gl}_2}(z,w) = (z-q_2 w) \nn \\
\parbox{3.5cm}{\includegraphics[width=2cm]{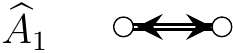}} G_{01}^{\mathfrak{gl}_2}(z,w) = G_{10}^{\mathfrak{gl}_2}(z,w) = (z-q_1w)(z-q_3w)\nn \\
d_{00}=d_{11}=1, \ \ \ d_{01}=d_{10}=-1
\ee
\item
For $n=1$ we return to sec.\ref{sec:dim-algebra}, i.e.
\be
\parbox{3.5cm}{\includegraphics[width=2cm]{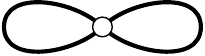}} G_{00}^{\mathfrak{gl}_1}(z,w) = (z-q_1w)(z-q_2w)(z-q_3w),\ \ \ \ \ \
d_{00}=1
\ee
\item
One expects in the Pagoda (triple-affine) case DIM($\widehat{\mathfrak{gl}_1}$) (or $U_{q,t,\widetilde{t}}(\widehat{\widehat{\widehat{\mathfrak{gl}}}}_1)$, hence, the name Pagoda) the Dynkin diagram of the form:
$\parbox{1cm}{\includegraphics[width=2cm]{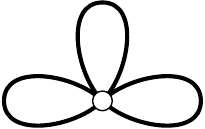}}$
\end{itemize}

\subsection*{A4. Affine Yangian of $\mathfrak{gl}_1$ \cite{Pro}}

One can consider a "quasiclassical" limit of the DIM($\mathfrak{gl}_1$) algebra, $q=e^{\hbar h_1}$, $t^{-1}=e^{\hbar h_2}$, $t/q = e^{\hbar h_3}$ with properly rescaled generators. We also use another parameterizations:
\be
\sigma_1 = h_1+h_2+h_3=0, \\
\sigma_2 = h_1h_2+h_1h_3+h_2h_3, \nn \\
\sigma_3 = h_1h_2h_3
\ee
In the limit of $\hbar\to 0$, one obtains the affine Yangian, which, on the gauge theory side, describes the $4d$ theories/Nekrasov functions. It is given by the commutation relations:
\be
\l[e_i,f_j]=\psi_{i+j}
\ee
\be
\l[\psi_i,\psi_j]=0
\ee
\be\label{er}
\l[e_{i+3},e_j] - 3[e_{i+2},e_{j+1}]+3[e_{i+1},e_{j+2}]-[e_i,e_{j+3}]
+ \sigma_2 \Big([e_{i+1},e_j]-[e_i,e_{j+1}]\Big) - \sigma_3(e_ie_j+e_je_i) = 0 \nn \\
\l[\psi_{i+3},e_j] - 3[\psi_{i+2},e_{j+1}]+3[\psi_{i+1},e_{j+2}]-[\psi_i,e_{j+3}]
+ \sigma_2 \Big([\psi_{i+1},e_j]-[\psi_i,e_{j+1}]\Big) - \sigma_3(\psi_ie_j+\psi_je_i) = 0
\ee
and two more relations similar to (\ref{er}) with $e_i$ substituted by $f_i$ and $\sigma_3$ substituted by $-\sigma_3$. These commutation relations should be added by the Serre relations
\be
sym_{i_1,i_2,i_3} \ \Big[e_{i_1},[e_{i_2},e_{i_3+1}]\Big] = 0
\ee
and similarly for $f_i$.

The commutation relations should be supplemented with the "initial conditions":
\begin{itemize}
\item $\psi_{0,1}$ are the central elements, i.e. commute with everything all generators
\item $\psi_2$ is the grading element, i.e.
\be
\ \ [\psi_2, e_j] = 2e_j,\ \ \ \ \ \ [\psi_2, f_j] = -2f_j,\ \ \ \ \ \ [\psi_2, \psi_j] = 0
\ee
\end{itemize}

Note that, introducing the generator functions
\be
e(u) = \sum_{i=0}^\infty e_iu^{-i-1}, \nn \\
f(u) = \sum_{i=0}^\infty f_iu^{-i-1}, \nn \\
\psi(u) = 1+\sigma_3 \sum_{i=0}^\infty \psi_iu^{-i-1}
\ee
one can rewrite the commutation relations as
\be
e(u)e(v) \sim \Phi(u-v)\,e(v)e(u), \nn \\
f(u)f(v) \sim \Phi(v-u)\,f(v)f(u), \nn \\
\psi(u)e(v) \sim \Phi(u-v)\, e(v)\psi(u), \nn \\
\psi(u)f(v) \sim \Phi(v-u)\, f(v)\psi(u), \nn \\
e(u)f(v) -f(v)e(u)\sim -{1\over\sigma_3}{\psi(u)-\psi(v)\over u-v}, \nn \\
\psi(u)\psi(v) \sim \psi(v)\psi(u)
\ee
with $\Phi(u) = \frac{(u+h_1)(u+h_2)(u+h_3)}{(u-h_1)(u-h_2)(u-h_3)}$.

\subsubsection*{Virasoro $\oplus$ Heisenberg subalgebra}

The commutation relations of the Virasoro algebra with extended $\widehat{U(1)}$-algebra,
\be
\l[J_m,J_n] = k m\delta_{m+n} \nn \\
\l[L_m,J_n] = -nJ_{m+n} \nn \\
\l[L_m,L_n] = (m-n)L_{m+n} + \frac{c}{12}n(n^2-1)\delta_{m+n}
\ee
can be realized with identification:
\be
J_{-1} = e_0, \ \ \ J_1 = -f_0, \nn \\
L_{-1} = e_1 + \alpha e_0, \ \ \ L_1 = f_1-\alpha f_0, \ \ \ \Longrightarrow
\ \ L_0 = \psi_2+2\alpha\psi_1+\alpha^2\psi_0 \nn \\
L_{-2} = \frac{1}{2}[e_2,e_0]-\frac{1}{2}\beta\sigma_3\psi_0[e_1,e_0], \ \ \
L_{2} = \frac{1}{2}[f_2,f_0]+\frac{1}{2}\beta\sigma_3\psi_0[f_1,f_0]
\ee
From the first line it follows that $k =\psi_0$. The other current mode are constructed by repeated commutators:
$J_{-2} = [e_1,e_0]$, $J_{2} =-[f_1,f_0]$ etc. Consistency conditions (e.g. $J_{-3} \sim [L_{-1},J_{-2}]\sim [L_{-2},J_{-1}]$) require
$2\alpha = (1-\beta)\sigma_3\psi_0$ (the dependence on $h$-parameters comes from relation with $[e_0,\psi_3]$, which does not involve
$e_3$, because $[e_3,\psi_0]=0$). Thus, there remains a free parameter $\beta$.

The central charge is $c = -\sigma_2\psi_0 - \sigma_3\psi_0^3 = 1-(1-\lambda_1)(1-\lambda_2)(1-\lambda_3)$, where $\lambda_a = -\psi_0h_bh_c$ with $(abc)$ is a cyclic permutation of $(123)$.

\subsubsection*{Representations: plane partitions}

The basis of a quasi-finite representation of this affine Yangian\footnote{Such representations are labeled by a triple of ordinary Young diagrams:
"minimal" plane partitions are labeled by boundary conditions, \cite{FJMMrep,Pro}.} can be described by plane partitions ($3d$ Young diagrams). The generators of algebra act on the plane partition as follows:
\be
e(u) \sim \text{adding a box to $3d$ Young diagram} \nn \\
f(u) \sim \text{removing a box to $3d$ Young diagram} \nn \\
\psi(u) \sim \text{diagonal action}
\ee
More precisely,
\begin{itemize}
\item the diagonal action is
\be\label{psil}
\psi|\Lambda> \ = \psi_\Lambda(u)|\Lambda> \nn \\
\psi_\Lambda(u) = \psi_\emptyset(u) \prod_{\Box \in \Lambda}\Phi\Big(u-u_0-h(\Box)\Big)
\ee
where $h(\Box) = xh_1+yh_2+zh_3$ and $(x,y,z)$ are the coordinates of the box within the plane partition;
\item the raising (lowering) action is
\be\label{EF}
e(u)|\Lambda> \ = \sum_{\Box\in\Lambda_+\backslash\Lambda}
\frac{E(\Lambda\longrightarrow \Lambda_+)}{u-u_0-h(\Box)}|\Lambda_+>, \nn \\
f(u)|\Lambda> \ = \sum_{\Box\in\Lambda\backslash\Lambda_-}
\frac{F(\Lambda\longrightarrow \Lambda_-)}{u-u_0-h(\Box)}|\Lambda_->
\ee
where $\Lambda_+$ ($\Lambda_-$) denotes arbitrary plane partition with one additional (one subtracted) box as compared to $\Lambda$.
\end{itemize}

Here $F$ and $E$ are coefficients which have to be defined from the commutation relations of the algebra and are
some residues of $\psi_\Lambda(u)$, $u_0$ is a constant shift, a counterpart of inhomogeneity in the standard spin chain.

Formula (\ref{psil}) is derived by acting with the both sides of the commutation relation $\psi(u)e(v) \sim \Phi(u-v)\, e(v)\psi(u)$ on $|\Lambda>$, using (\ref{EF}) and then taking the residue at $v=h(\Box)$

\paragraph{Constraints on the coefficients $E$ and $F$.}

Constraints on functions $E(\Lambda\longrightarrow \Lambda_+)$
and $F(\Lambda \longrightarrow \Lambda_-)$ can be derived from the commutation relations
$[e_i,f_j]=\psi_{i+j}$. For the generating functions it looks like
\be
\psi_\Lambda(u)=1+\sigma_3\sum_{\Box} \frac{E(\Lambda_-\longrightarrow \Lambda)
F(\Lambda\longrightarrow \Lambda_-)}{u-h(\Box)} -
\sigma_3\sum_{\Box} \frac{F(\Lambda_+\longrightarrow \Lambda)
E(\Lambda\longrightarrow \Lambda_+)}{u-h(\Box)}
\ee
where the second-order pole does not contribute. This relation does not fix $E$ and $F$ completely.
Imposing an additional requirement of unitarity $E(\Lambda\longrightarrow \Lambda_+)=
F(\Lambda_+\longrightarrow \Lambda)$, one immediately obtains \cite{Pro}
\be
\sigma_3E(\Lambda\longrightarrow \Lambda_+)^2 = -{\rm res}_{u\longrightarrow h(\Box)}\psi_\Lambda(u)\nn \\
\sigma_3E(\Lambda_-\longrightarrow \Lambda)^2 = {\rm res}_{u\longrightarrow h(\Box)}\psi_\Lambda(u)
\ee
One still has to fix the sign (after taking the square root).

The commutation relation $e(u)e(v)\sim \varphi(u-v)e(v)e(u)$ relates adding two boxes in different order:
\be
\frac{E(\Lambda\longrightarrow \Lambda+\Box_A)E(\Lambda+\Box_A\longrightarrow \Lambda+\Box_A+\Box_B)}
{E(\Lambda\longrightarrow \Lambda+\Box_B)E(\Lambda+\Box_B\longrightarrow \Lambda+\Box_A+\Box_B)} =
\Phi\Big(h_{\Box_B}-h_{\Box_A}\Big)
\ee
To check that it is satisfied, calculate the square of the l.h.s.:
\be
{{\rm res}_{u\longrightarrow h(\Box_A)}\psi_\Lambda(u)\cdot {\rm res}_{u\longrightarrow h(\Box_B)}\psi_{\Lambda+\Box_A}(u)\over {\rm res}_{u\longrightarrow h(\Box_B)}\psi_\Lambda(u)\cdot {\rm res}_{u\longrightarrow h(\Box_A)}\psi_{\Lambda+\Box_B}(u)}={{\rm res}_{u\longrightarrow h(\Box_A)}\psi_\Lambda(u)\cdot {\rm res}_{u\longrightarrow h(\Box_B)}\left\{\psi_\Lambda (u)\Phi \Big(u-h(\Box_A)\Big)\right\}\over {\rm res}_{u\longrightarrow h(\Box_B)}\psi_\Lambda(u)\cdot {\rm res}_{u\longrightarrow h(\Box_A)}\left\{\psi_\Lambda (u)\Phi \Big(u-h(\Box_B)\Big)\right\}}=\nn\\
={\Phi \Big(h(\Box_B)-h(\Box_A)\Big)\over \Phi \Big(h(\Box_A)-h(\Box_B)\Big)}=\Phi\Big(h(\Box_B)-h(\Box_A)\Big)^2
\ee
Similarly one can check the Serre relations by adding three boxes:
\be
\sum_{\pi \in S_3} \left[h(\Box_{A_{\pi(1)}})-2h(\Box_{A_{\pi(2)}})+h(\Box_{A_{\pi(3)}})\right]
E\Big(\Lambda\longrightarrow \Lambda+\Box_{A_{\pi(1)}}\Big)\times\nn\\ \times E\Big(\Lambda\longrightarrow \Lambda+\Box_{A_{\pi(1)}}+\Box_{A_{\pi(2)}}\Big)E\Big(\Lambda\longrightarrow \Lambda+\Box_{A_{\pi(1)}}+\Box_{A_{\pi(2)}}+\Box_{A_{\pi(3)}}\Big) = 0
\ee

\paragraph{A simplest example of the highest-weight representation.}

Consider a representation with the highest weight $|\Lambda>$:
\be
\psi_j|\Lambda>=\psi_{j,\Lambda}|\Lambda>,\ \ \ \ \ \ \
f_j|\Lambda>=0
\ee
Since we consider the quasi-finite representations, there should be linear relations among $e_i|\Lambda>$.
Consider vectors in the representation at the first level with finitely many, $r-1$ independent vectors. This means that the Shapovalov matrix at the level one, which is
\be
B_\Lambda(e_i,e_j)<\Lambda|\Lambda> = <\Lambda|f_ie_j|\Lambda> = \psi_{i+j,\Lambda}<\Lambda|\Lambda>
\ee
should have only $r-1$ independent lines, i.e. there is a relation
\be
\sum_{i=0}^{r-1}\alpha_i\psi_{i+k,\lambda} = 0, \ \ \ k\geq 0
\ee
Then, the generating function of eigenvalues
\be
\psi_\Lambda(u) = 1+\sigma_3 \sum_{j=0}^\infty \psi_{j,\Lambda} u^{-j-1}= \frac{f(u)}{g(u)}
\ee
where $f(u)$ and $g(u)$ are polynomials of degree $r-1$.

Consider the case of $r=2$, i.e. a single state at the level one and linear functions $f(u)$ and $g(u)$:
\be
\psi_\lambda(u) = \frac{u+\sigma_3\psi_{0,\lambda}}{u} = 1+\frac{\sigma_3\psi_{0,\lambda}}{u}
\ee
Then, the commutation relations and the Serre relations implies that there are 3 states at the second level (this is since the function $\Phi (u)$ is a ratio of cubic polynomials) and 6 states at the third level. These particular numbers are equal to the number of $3d$ Young diagrams with a given number of boxes. This means that the highest weight is associated with the trivial plane partition $|\lambda>=\emptyset$, and the single first level vector is associated with the only one box plane partition $|\Box>$:
\be
e_i|\emptyset>=0,\ i>0;\ \ \ \ \ \ \ \ e_0|\emptyset>\sim |\Box>; \ \ \ \ \ \ \ \psi_i|\emptyset>=0,\ i>0
\ee
Since $\psi_1$ is a center and $[\psi_2,e_0]=2e_0$ one immediately obtains
\be
\psi_1|\Box>=0,\ \ \ \ \ \ \ \psi_2|\Box>=2|\Box>
\ee
Using these formulas, from the Serre relations that involve $\psi_j$ and $e_{0,1,2,3}$, one gets
\be
\psi(u)|\Box>\ =   \frac{u+\sigma_3\psi_{0,\emptyset}}{u}\varphi(u)|\Box>
\ee

\section*{Acknowledgements}

A.M.'s and Y.Z. are grateful for remarkable hospitality at Nagoya University during the work on this project.

Our work is supported in part by Grant-in-Aid for Scientific Research (\# 24540210) (H.A.),
(\# 15H05738) (H.K.), for JSPS Fellow (\# 26-10187) (Y.O.), JSPS Grant-in-Aid for Young Scientists (B) \# 16K17567 (T.M.) and JSPS Bilateral Joint Projects (JSPS-RFBR collaboration)
\lq\lq Exploration of Quantum Geometry via Symmetry and Duality\rq\rq\
from MEXT, Japan. It is also partly supported by grants 15-31-20832-Mol-a-ved (A.Mor.),
15-31-20484-Mol-a-ved (Y.Z.), mol-a-dk 16-32-60047 (And.Mor), by RFBR grants 16-01-00291 (A.Mir.)  and
16-02-01021 (A.Mor.\ and Y.Z.), by joint grants 15-51-50034-YaF,
15-51-52031-NSC-a, 16-51-53034-GFEN.

\end{document}